\documentclass[12pt,letterpaper]{article}
\usepackage{bbm}
\usepackage[multiple]{footmisc}
\usepackage{floatpag,amsmath,amsthm,amssymb}
\newtheorem{proposition}{Proposition}
\numberwithin{equation}{section}
\newtheorem{nono-prop}{Proposition}[]

\newcommand{\mobilitypath}{.}
\newcommand{\mortalitypath}{.}
\newcommand{\figpath}{.}

\usepackage[latin1]{inputenc}
\usepackage{setspace}
\usepackage{amsmath}
\usepackage{amsthm}
\usepackage{amsfonts}
\usepackage{longtable}
\usepackage{fullpage}
\usepackage{amssymb}
\usepackage{graphicx}
\usepackage{epstopdf}
\usepackage{multirow}
\usepackage{array}
\usepackage{harvard}
\usepackage{tabularx}
\citationmode{abbr}

\usepackage{float}
\usepackage{lscape}
\usepackage{verbatim}
\usepackage{pdflscape}
\usepackage{chngcntr}
\usepackage{appendix}
\usepackage{booktabs,calc}
\usepackage{ulem}

\usepackage{color,colortbl}
\usepackage{soul}
\definecolor{Yellow}{rgb}{.88,1,.65}
\definecolor{Green}{rgb}{.65,1,.65}
\definecolor{Red}{rgb}{1,.65,.65}

\citationstyle{dcu}

\usepackage[labelfont=bf,center,large,labelsep=newline]{caption}
\usepackage{subfigure}

\newcommand{\superscript}[1]{\ensuremath{^{\textrm{#1}}}}

\newcommand{\cnewline}{\hspace{\linewidth}}

\usepackage[right=1in,left=1in,top=1in,bottom=1in]{geometry}
\usepackage[all=normal,title,sections]{savetrees}

\usepackage{nohyperref} 

\title{Partial Identification of Expectations with Interval
  Data\footnote{We are thankful for useful discussions with Emily
    Blanchard, Raj Chetty, Eric Edmonds, Shahe Emran, Francisco
    Ferreira, Nate Hilger, David Laibson, Ethan Ligon, Erzo Luttmer,
    Nina Pavcnik, Bruce Sacerdote, Na'ama Shenhav, Forhad Shilpi, Gary
    Solon, Doug Staiger, Chris Snyder and Elie Tamer. Toby Lunt, Ryu
    Matsuura, and Taewan Roh provided excellent research
    assistance. We are grateful to the authors of Bratberg et
    al. (2017) for sharing data. All errors are our own.}}
\author{Sam Asher\thanks{World Bank, sasher@worldbank.org} \\ \\ Paul
  Novosad\thanks{Dartmouth College, paul.novosad@dartmouth.edu,
    corresponding author} \\ \\ Charlie Rafkin\thanks{National Bureau
    of Economic Research, crafkin@nber.org} \\ \\ }

\begin{document}
\date{January 2018}
\maketitle

\begin{abstract}
A conditional expectation function (CEF) can at best be partially
identified when the conditioning variable is interval censored. When
the number of bins is small, existing methods often yield minimally
informative bounds. We propose three innovations that make meaningful
inference possible in interval data contexts. First, we prove novel
nonparametric bounds for contexts where the distribution of the
censored variable is known. Second, we show that a class of measures
that describe the conditional mean across a fixed interval of the
conditioning space can often be bounded tightly even when the CEF
itself cannot. Third, we show that a constraint on CEF curvature can
either tighten bounds or can substitute for the monotonicity
assumption often made in interval data applications. We derive
analytical bounds that use the first two innovations, and develop a
numerical method to calculate bounds under the third. We show the
performance of the method in simulations and then present two
applications. First, we resolve a known problem in the estimation of
mortality as a function of education: because individuals with high
school or less are a smaller and thus more negatively selected group
over time, estimates of their mortality change are likely to be
biased. Our method makes it possible to hold education rank bins
constant over time, revealing that current estimates of rising
mortality for less educated women are biased upward in some cases by a
factor of three. The method is also applicable to the estimation of
education gradients in patterns of fertility, marriage and disability,
among others, where similar compositional problems arise. Second, we
apply the method to the estimation of intergenerational mobility,
where researchers frequently use coarsely measured education data in
the many contexts where matched parent-child income data are
unavailable. We show that conventional measures like the rank-rank
correlation may be uninformative once interval censoring is taken into
account, but CEF interval-based measures of mobility are bounded
tightly.

\end{abstract}

\newpage
\clearpage
\doublespacing
\section{Introduction}
\label{sec:intro}

\nocite{Bratberg2017}

The value of a conditional expectation function (CEF) can at best be
partially identified when the conditioning variable is interval
censored \cite{Manski2002}. When the observed intervals are coarse or
the CEF slope is large in magnitude, existing methods may yield bounds
that are minimally informative. In this paper, we develop three
innovations that can yield narrower bounds on parameters of interest,
and we develop analytical and numerical methods to calculate these
bounds.  We apply the methods in two policy-relevant settings: the
estimation of mortality as a function of education
\cite{Meara2008,Case2015}, and the estimation of intergenerational
mobility \cite{Solon1999,Guell2013,Chetty2014b}.

First, we show that using information on the distribution of the
conditioning variable leads to tighter bounds on the CEF. We prove
sharp analytical bounds on the value of the CEF when the latent
conditioning variable has a known distribution but is
interval-censored. This approach is broadly applicable, because
distributions are known or commonly assumed for many economic
variables. For some conditioning variables (such as ranks), no
additional assumptions are required; for example, ranks are uniform by
construction. For others (such as income), distributional assumptions
on the variable of interest are common and reasonable, and results
under alternative assumptions can be tested.

Second, we derive a class of measures that describe the CEF mean
across a fixed interval of the conditioning variable. Such interval
means can in practice be bounded tightly in many cases and point
estimated for some intervals. This makes meaningful inference possible
for policy-relevant parameters even when the bounds on the CEF itself
are very wide. For example, when the bottom bin is large, the mean
value of the CEF in the bottom quintile of the conditioning variable
may be bounded more tightly than the value of the CEF at any point in
the bottom quintile.

Third, we show that a curvature constraint on the CEF can be
implemented in a nonparametric setup using numerical constrained
optimization, further narrowing bounds on the CEF and functions of the
CEF. The assumption of limited curvature can also substitute for the
monotonicity assumption that earlier approaches to this problem relied
upon. In our applications, conservative curvature limits generate
identified sets of similar size to those under assumptions of
monotonicity. This result provides a tractable framework for
nonparametric inference with interval-censored data even in contexts
without monotonicity.

In practice, we find that to obtain informative bounds, the first and
second innovations (known distribution and interval means) are
required. Further, even a weak curvature constraint can tighten the
bounds significantly. Although this paper focuses on conditional
expectation functions, the method can be directly applied to any
function or moment of the variable of interest. For example, the
method can bound any percentile of the conditional distribution of $y$
given an interval-censored variable $x$.

This paper contributes to a growing literature focused on partial
identification of solutions to problems where point identification is
difficult without excessively restrictive assumptions
\cite{Manski2003,Tamer2010,Ho2015a}. This paper is most closely
related to \citeasnoun{Manski2002}, who calculate analytical bounds on
a CEF with an interval-censored conditioning variable from an unknown
distribution. The \citeasnoun{Manski2002} bounds are sharp---we can
only improve upon them by making additional assumptions, but the
assumptions we make are weak and reasonable in many contexts, and
tighten bounds in some cases by an order of magnitude or more. We also
provide a tractable numerical framework for calculating nonparametric
bounds under more complex constraints. Our analysis is limited to
conditional expectation functions with a single parameter. As we show
below, this setup nevertheless describes a broad class of problems and
the innovations are more broadly applicable; extensions to more
complex models are a subject for future research.\footnote{Other work
  on partial identification in contexts with interval data include
  \citeasnoun{Magnac2008}, who focus on cases with binary dependent
  variables. For this case, they show that bounds are tighter under
  known distributions and reduce to points under the uniform
  distribution. \citeasnoun{Bontemps2012} focus primarily on cases
  where the $y$ variable is interval-censored. Our focus is on
  continuous dependent variables with interval-censored conditioning
  variables.}

Here, we briefly describe the two applications that we will focus on
below.

\vspace{4mm}
\noindent \underline{Application 1: Mortality as a Function of Education}

\noindent In the first application, we resolve a long-standing problem
in the estimation of mortality as a function of education. Researchers
have noted recent increases in the mortality of less-educated
individuals in the
U.S. \cite{Meara2008,Cutler2010,Cutler2011,Olshansky2012,Case2015,Case2017}.
For example, mortality among women aged 50--54 with high school
education or less (LEHS) has risen from 459 deaths per 100,000 people
in 1992 to 587 deaths in 2015. A known concern with these estimates is
that rising education levels over time, particularly among women, make
these numbers difficult to interpret. 

Figure~\ref{fig:mort_scatter} shows mortality for 50--54 year old
U.S. women as a function of the median education rank in each of three
educational categories, illustrating the simultaneous changes in
mortality and in the distribution of education. Women with a high
school degree or less represented 64\% of women in 1992 and only 39\%
of women in 2015. If mortality is a decreasing function of the latent
education rank, then the increasing negative selection of LEHS women
could explain some or all of the mortality change for this group, even
if the underlying mortality-education rank relationship is unchanged.
Whether and how to adjust for these compositional changes is an
important debate in the mortality literature. Some studies have argued
that the bias is close to zero, while others have suggested that it
may explain all of the recent mortality increases.\footnote{Recent
  high profile work by Case and Deaton (2015, 2017) focuses on
  unadjusted estimates for non-Hispanic whites with high school
  education or less (rather than dropouts), arguing that their average
  school completion has not substantially changed over the sample
  period they study. For our sample period (1992 to the present, all
  races) LEHS men have gone from 54\% to 44\% of the population in
  2015 and LEHS women have gone from 64\% of the population to
  39\%. Changes are even larger for other age groups over other time
  periods. \citeasnoun{Dowd2014} and \citeasnoun{Currie2018} argue
  that the bias may be so large that estimates of mortality change
  among LEHS people are effectively uninformative.} Estimates of
mortality within fixed education quantiles would solve the problem,
but there is no established method to generate quantiles when
intervals in the data do not correspond to quantile
boundaries.\footnote{Mortality data typically report education in a
  small number of coarse categories. Most studies on mortality and
  education use only two or three categories of
  education.}\superscript{,}\footnote{\citeasnoun{Bound2015} generate
  quantile point estimates, but only under the implicit assumption
  that the latent mortality-education gradient has zero slope within
  each education bin. This is a strong assumption given the important
  gradient across education bins. The partial identification approach
  that we propose lets us avoid making strong assumptions about
  censored data.}

To make progress on this problem, we make two assumptions. First, we
assume that the observed education rank represents a latent,
continuous rank that is observed only in coarse intervals, a common
assumption in this literature \cite{Goldring2016}.\footnote{The latent
  variable can be interpreted as the total net benefit of pursuing a
  given quantity of education. Those individuals at the high end of a
  latent education rank bin are the ones who would move to a higher
  education bin if their net benefit of education marginally increased.}
Second, we assume that mortality is decreasing in latent educational
rank. This assumption holds across bins for every year between 1992
and 2015 for both men and women, and across every income ventile
\cite{Chetty2016b}. Under just these assumptions, our method can
generate bounds on the expectation of mortality at any rank or in any
rank interval or quantile.

We focus on women age 50-54, because their increasing education over
time means the selection bias for this group may be large. We bound
the mortality rate of the bottom 64\% of the education distribution --
a fixed share of the population representing LEHS in 1992.  The bounds
are tight and informative. The mortality increase for women from
1992--2015 in this part of the education distribution is between 29
and 38 additional deaths per 100,000. The unadjusted estimate (which
compares the bottom 64\% in 1992 to the bottom 39\% in 2015) suggests
an increase of 128 deaths, more than three times higher than the upper
bound from our calculation.  For some population groups, the mortality
increases noted in the recent literature are sustained when we use our
method to study constant rank groups, while for others, the unadjusted
estimates are substantially biased or have the incorrect sign.

This application focuses on mortality, but similar compositional
issues arise in any context where the researcher is interested in
changes in the relationship between education and some outcome
variable over time. For example, our method could resolve bias due to
changing composition in studies on education gradients in birth
outcomes, marriage patterns or disability
\cite{Cutler2010a,Aizer2014,Bertrand2016}.\footnote{In a context where
  education is strictly considered as an input to the production
  function (such as estimating the returns to education), unadjusted
  estimates may be preferred. But if education and the outcome are
  correlated with any omitted variable, then adjusting for population
  share and education rank will be a useful exercise. We take no stand
  on the health production function or whether the mortality-education
  relationship should be treated causally.}

\vspace{4mm}
\noindent \underline{Application 2: Intergenerational Educational Mobility}

\noindent The methods presented here can also resolve several
challenges in the estimation of intergenerational 
mobility.  The object of interest in many studies of intergenerational
mobility is the CEF of child education given parent education, in part
because data on educational attainment are widely available and may be
less subject to measurement error than parent income data
\cite{Black2003,Guell2013}. 

The coarse binning of education data poses a key problem in this
context. Many mobility measures require observation of the child CEF
at a specific point in the parent rank distribution. Absolute upward
mobility, for instance, is defined by \citeasnoun{Chetty2014b} as the
expected outcome of a child who is born to a family at the
25\superscript{th} percentile of the parent rank distribution. Binned
education data make this challenging to estimate.  In older
generations in India, for example, over 50\% of parents report having
less than two years of education, the lowest recorded category in many
datasets. In such a context, the 25\superscript{th} percentile parent
is not directly observed, so absolute upward mobility can at best be
partially identified.\footnote{We assume that absolute upward mobility
  is measured in terms of the continuous latent educational rank
  rather than the directly observed rank bin. Treating the bin mean as
  the true value in the rank bin (the approach of
  \citeasnoun{Bound2015} to mortality) has the undesirable property
  that more granular measures of education will lead to lower measures
  of absolute upward mobility.}  Expected child outcomes at constant
parent ranks are also required for meaningful cross-group mobility
comparisons \cite{Hertz2005}.\footnote{The rank-rank gradient and
  other linear estimators of the parent-child outcome function are not
  informative about subgroup mobility, because they compare children
  of low-ranked parents with children of high-ranked parents from the
  same subgroup, which can be misleading \cite{Aaronson2008}.} With
education rank boundaries that change over time, subgroup educational
mobility estimates are difficult to compare over time.

The CEF bounds proposed above are a direct solution to these problems,
as they bound the expected outcome of a child born at arbitrary points
or intervals in the parent rank distribution. We propose a new measure
of mobility, \textit{upward interval mobility}, which is the mean
value of the child CEF in the bottom half of the parent rank
distribution. Applying our method to Indian data, we show that
conventional mobility measures are biased or uninformative about the
mobility of older cohorts once we account for interval censoring, but
upward interval mobility can be tightly bounded.\footnote{Upward
  interval mobility has very similar policy relevance to absolute
  upward mobility. Absolute upward mobility measures the expected
  outcome of the \textit{median} child in the bottom half of the
  parent distribution, whereas upward interval mobility measures the
  \textit{mean}.}

The interval problem for educational mobility is most severe in
developing countries, but is important in other contexts as well. In
wealthier countries, it is common for a large share of the population
to be in a topcoded education bin.\footnote{In one mobility study from
  Sweden, for example, 40\% of adoptive parents were topcoded with 15
  or more years of education \cite{Bjorklund2006}. Studies on the
  persistence of occupation across generations also frequently use a
  small number of categories and face a similar challenge when the
  occupational structure changes significantly over time, as it has
  with farm work in the United States. See, for example,
  \citeasnoun{Long2013}, \citeasnoun{Xie2013} and
  \citeasnoun{Guest1989}.} Internationally comparable censuses also
frequently report education in as few as four categories; our method
is thus particularly relevant for cross-country comparison.

\vspace{6mm} In the next section, we describe the setup, prove the new
bounds and present the numerical solution framework. In
Section~\ref{sec:sim}, we explore properties of the bounds in a
simulation. Sections~\ref{sec:mort} and \ref{sec:mob} present the
applications to the measurement of mortality and of intergenerational
mobility in more detail. Section~\ref{sec:conc} concludes. Stata and
Matlab code to implement all methods in the paper are available on the
corresponding author's web site.\footnote{Code can be downloaded at
  https://github.com/paulnov/anr-bounds.}

\section{Bounds on CEFs with a Known
  Conditioning Distribution}
\label{sec:method}

This section describes the main contribution of the paper. We
calculate analytical and numerical bounds on a CEF where the
conditioning variable is interval censored but has a known
distribution.  The bounds are sharp and depend either on the
assumption of a weakly monotonic CEF or on the assumption that the CEF
has limited curvature.  The method can also bound any statistic that
can be derived from the CEF, such as the mean over an arbitrary
interval, or the best linear approximator to the CEF.

We describe the method by working through an example motivated by
Figure~\ref{fig:mort_scatter}, which plots total mortality against
education, where education is only observed in one of three
education bins: (i) less than or
equal to high school; (ii) some college; or (iii) bachelor's degree or
higher.\footnote{Points are plotted at the midpoint of the education
  rank bins.} We focus on women aged 50--54, because (i) this age
group has been highlighted in other recent research, and (ii) the
change in education for this group has been large over the sample
period. We wish to estimate some statistic that describes mortality in
1992 and in 2015 for a group of people occupying the same set of
education ranks in the population. This is challenging because the rank bin
boundaries change between 1992 and 2015. In 1992, 64\% of women had
less than or equal to a high school education, while in 2015, this
number was 39\%. 

Our approach is to estimate the conditional expectation function of
mortality given education in each year, which would allow us to
partially identify mortality at any rank in any year. We implicitly
assume that there exists a latent, continuous education rank that we
only observe in discrete intervals. This section focuses on the
problem of identifying a CEF given interval data, and
Section~\ref{sec:mort} explores the findings on mortality in more
detail.

Figure~\ref{fig:cef_sample} depicts the setup for 2015. The points
show mortality at the midpoints of three education bins and the
vertical lines show the rank bin boundaries. The lines plot two (of
many) possible nonparametric CEFs, each of which fit the sample means
with zero error. These two functions have the same mean in each bin,
even if they do not cross the mean at the bin midpoint.\footnote{A
  naive polynomial fit to the midpoints in the graph would be a biased
  fit to the data because of Jensen's Inequality.} These are the
functions we aim to bound.  We begin with the assumption that
mortality is weakly decreasing in latent rank, and then show how a
curvature constraint can supplement or substitute for this assumption.

\subsection{Nonparametric Inference with Interval Data}

Define the outcome as $y$ and the conditioning variable as $x$; the
conditional expectation function is $Y(x)=E(y|x)$. Let the function
$Y(x)$ be defined on $x \in [0,100]$, and assume $Y(x)$ is
integrable. We also assume throughout that $\underline{Y} \leq Y(x)
\leq \overline{Y}$, that is, the function is bounded
absolutely.\footnote{In most applications, parameters of interest are
  likely to have upper and lower bounds either in theory or in
  practice. Loosening the absolute upper and lower bound restriction
  would result in wider bounds for the CEF in the bottom or top
  intervals, but informative inference is still possible even in these
  outer bins. In the case of mortality, we will impose that the upper
  bound is a mortality rate of 100\%.}

With interval data, we do not observe $x$ directly, but only that it
lies in one of $K$ bins. Let $f_k(x)$ be the probability density
function of $x$ in bin $k$. Define the expected outcome in the
$k^{th}$ bin as $$r_k = E \left(y|x \in [x_k,x_{k+1}] \right) =
\int_{x_k}^{x_{k+1}} Y(x)f_k(x)dx,$$ where $x_k$ and $x_{k+1}$ define
the bin boundaries of bin $k$. This expression holds due to the law of
iterated expectations. The limits of the conditioning variable are
assumed to be known, and are denoted by $x_1$ and $x_{K+1}$.  Further
define the expected outcomes in the intervals directly above and below
the intervals of interest as $r_{k+1}=E \left(y|x \in
[x_{k+1},x_{k+2}] \right)$ and $r_{k-1}=E\left(y|x \in [x_{k-1},x_{k}]
\right)$, if they exist. Define $r_0 = \underline{Y}$ and $r_{K+1} =
\overline{Y}$. The sample analog to $r_k$ is the observed mean outcome
in bin $k$, which we denote $\overline{r}_k$.

Sharp bounds on $E(y|x)$ given interval
measurement of $x$ are derived by \citeasnoun{Manski2002}, when the
distribution of $x$ is unknown. The essential structural assumption
that constrains the CEF is Monotonicity (M): 
\begin{equation}
  E(y|x) \text{ must be weakly increasing in } x.
  \tag{Assumption M} 
\end{equation} 
Note that we apply this assumption to the survival rate, which is one
minus the mortality rate; however, our graphs show the mortality rate
which is the parameter of interest. The CEF in the monotonic graphs is
thus monotonically decreasing.\footnote{Mortality is decreasing in
  educational attainment for every group and time period in the CDC
  data; it is also a monotonically decreasing function of income
  \cite{Chetty2016b}.}  \citeasnoun{Manski2002} also introduce the
following Interval (I) and Mean Independence (MI) assumptions. For $x$
which appears in the data as lying in bin $k$,
\begin{align}
  \tag{Assumption I} 
  &x  \text{ in bin } k \implies P(x \in [x_{k}, x_{k+1}]) = 1. \\  
  \tag{Assumption MI} 
  &E( y \vert x, \text{ } x \text{ lies in bin } k) = E(y
  \vert x).
\end{align}

\noindent Assumption $I$ states that the rank of all people who report education
ranks in category $k$ are actually in bin $k$. Assumption $MI$
states that censored observations are not different from uncensored
observations. These always hold in our context because all of the data
are interval censored.

If all observations of $x$ are interval censored, the \citeasnoun{Manski2002}
bounds are:
\begin{equation}                     
  r_{k-1} \leq E(y | x) \leq r_{k+1} 
  \tag{Manski-Tamer bounds} 
\end{equation}                       

\noindent The value of the CEF in each bin is bounded by the
means in the previous and next bins.

We can improve upon these bounds if the distribution of $x$ is
known. In some cases, as with ranks, the distribution is given by the
definition of the variable. In other cases, conventional distributions
are frequently assumed (such as lognormal or Pareto for income
data). Alternatively, data could be transformed into a known
distribution, for example, by transforming the conditioning variable
into ranks. We first show bounds under the assumption that $x$ has a
uniform distribution because the analytical results are particularly
parsimonious, but we derive all of our results under a general known
distribution. We therefore consider the following assumption (U):
\begin{equation}
x \sim U(x_0,x_{K+1}) 
  \tag{Assumption U} 
\end{equation}
\noindent where $U$ is the uniform distribution. 

If $x$ is uniformly distributed, we
know that:
\begin{equation}
E\left(x \vert x \in [x_{k},x_{k+1}]\right) = \frac{1}{2} \left( x_{k+1} -
x_k \right). 
\end{equation} 

We derive the following proposition. 

\begin{proposition} 
  \label{eq:cef_bound}
  Let $x$ be in bin $k$. Under assumptions IMMI and
  U, and without additional information, the
  following bounds on $E(y \vert x)$ are sharp:
  $$
  \begin{cases}
    r_{k-1} \leq E(y \vert x) \leq \frac{1}{x_{k+1} - x} \left(
    \left(x_{k+1} - x_k\right) r_k - \left(x - x_k\right) r_{k-1} \right), & x < x_k^* \\
    \frac{1}{x - x_k} \left( \left(x_{k+1} - x_k\right) r_k -
    \left(x_{k+1} - x\right) r_{k+1} \right) \leq E(y \vert x) \leq r_{k+1}, & x \geq x_k^* 
  \end{cases}
  $$
  where $$x_k^* = \frac{x_{k+1} r_{k+1}
    - \left(x_{k+1} - x_k\right) r_k -
    x_k r_{k-1}  }{r_{k+1} - r_{k-1} }.$$ 
\end{proposition} 

The proposition is obtained from the insight that the value of
$E(y|x=i)$ at a point $i$ in bin $k$ (below the midpoint) will only be
minimized if all points in bin $k$ to the left of $i$ have the same
value. Since all points to the right of $i$ are constrained by the
outcome value in the subsequent bin $k+1$, $E(y|x=i)$ will need to
rise above the \citeasnoun{Manski2002} lower bound as $i$ increases,
in order to meet the bin mean. Intuitively, consider the point
$E(y|x=x_{k+1}-\varepsilon)$. In order for this point to take on a value
below the bin mean $r_k$, it needs to be the case that virtually
all of the density in bin $k$ lies between $x_{k+1}-\varepsilon$
and $x_{k+1}$. This is ruled out by the uniform distribution, and
indeed by most distributions; for many distributions, therefore, the
\citeasnoun{Manski2002} bounds are too conservative. We prove the
proposition and provide additional intuition in
Appendix~\ref{app:proofs}.

We generalize the proposition to obtain the following result for an arbitrary
known distribution of $x$:
\begin{proposition} 
\label{eq:cef_distrib}
Let $x$ be in bin $k$. Let $f_k(x)$ be the probability density function
of $x$ in bin $k$. Under assumptions IMMI, and
without additional information, the
following bounds on $E(y \vert x)$ are sharp:
$$
\begin{cases}                                                                                                                          
r_{k-1} \leq E(y \vert x) \leq \frac{r_k - r_{k-1} \int_{x_{k}}^x
  f_k(s)ds}{\int_{x}^{x_{k+1}}f_k(s)ds}, & x < x_k^*      \\           
\frac{r_k - r_{k+1}\int_x^{x_{k+1}} f_k(s)ds }{\int_{x_k}^x f_k(s)ds}  \leq E(y \vert x)  \leq                                                                         
r_{k+1} , & x \geq x_k^*                                                                                                             
\end{cases}
$$
where $x_k^*$ satisfies: 
$$r_k = r_{k-1} \int_{x_k}^{x_k^*} f_k(s) ds + r_{k+1}
\int_{x_k^*}^{x_{k+1}} f_k(s) ds.$$ 
\end{proposition} 

\noindent A proof of the proposition is in Appendix B. 

Figure~\ref{fig:analytic_bounds} compares \citeasnoun{Manski2002}
bounds to those obtained under the additional assumption of
uniformity, using the mortality data. The new bounds are a significant
improvement, especially where the data are particularly
coarse and near the bin boundaries. For example, without using
information on the distribution type, one could not reject that
mortality for people in the first bin is 100,000 per 100,000 until just
before the first bin boundary. The improvements in the other bins are
less extreme but still substantial.

In addition to bounding the value of $Y=E(y|x)$ at any given
point, we can also bound many functions of the CEF, which
we represent in the form $M(Y)$. One function of interest is the
slope of the best linear approximation to the CEF; this is difficult to
bound analytically, but we bound this numerically in
Section~\ref{sec:numeric}. 

Here, we highlight a function that describes the average value of the CEF
over an arbitrary interval of the conditioning space, or $\mu_a^b =
E(y|x \in [a, b])$.  This function has several desirable
properties. First, it can be bounded analytically. Second, it is
frequently bounded more tightly than $E(y|x)$. Third, it has a similar
interpretation to $E(y|x)$ and is thus likely to be
policy-relevant. We show in Sections~\ref{sec:mort} and \ref{sec:mob}
that for our applications, $\mu_a^b$ can be bounded considerably more
tightly than $E(y|x)$.

Let $f(x)$ represent the probability density function of $x$. 
Define $\mu_a^b$ as
\begin{align}
  \label{eq:mu_bounds}
  \mu_a^{b} = \frac{\int_a^{b} E(y | x) f(x) dx} { \int_a^b
      f(x) dx} . 
\end{align}
We now state analytical bounds on $\mu_a^b$ given uniformity. Let
$Y_x^{max}$ be the analytical upper bound on $E(y \vert x)$, given by
Proposition~\ref{eq:cef_bound}. Let $Y_x^{min}$ be the analytical
lower bound on $E(y \vert x)$. The following proposition defines sharp
bounds on $\mu_a^b$ under the assumption that $x$ is uniformly
distributed:
\begin{proposition}
  Let $a \in [x_h, x_{h+1}]$ and $b \in [x_k, x_{k+1}]$, with $a<b$. Let
  assumptions IMMI and U hold. Then, if no
  additional information is available, the
  following bounds are sharp: 
  \label{eq:bound_mu} 
$$ 
  \begin{cases} 
    Y_b^{min} \leq \mu_a^b \leq Y_a^{max} & h = k \\
    \frac{r_h (x_k - a) + Y_b^{min}(b - x_k)}{b-a} \leq
    \mu_a^b \leq \frac{Y_a^{max} (x_k - a) + r_k
      (b-x_k)}{b-a} & h +
    1 = k \\
    \frac{r_h (x_{h+1} - a) + \sum_{\lambda = h+1}^{k-1} r_{\lambda}
      (x_{\lambda+1} - x_{\lambda}) + Y_b^{min}(b - x_k)}{b-a} \leq
    \mu_a^b 
    \leq \frac{Y_a^{max} (x_{h+1} - a) + \sum_{\lambda = h+1}^{k-1} r_{\lambda}
      (x_{\lambda+1} - x_{\lambda}) + r_k (b-x_k)}{b-a} & h +
    1 < k. 
  \end{cases} 
$$ 
\end{proposition} 

\noindent We prove this proposition under uniformity and under an
arbitrary known conditioning distribution in Appendix~\ref{app:proofs}.

We note two special cases. First, if $a=b$, then $\mu_a^b =
E(y|x=a)$. Second, if $a$ and $b$ correspond exactly to bin
boundaries, then the bounds on $\mu_a^b$ collapse to a point: in this
case, $\mu_a^b$ is just a weighted average of the bin means between
$a$ and $b$.

In fact, $\mu_a^b$ can be very tightly bounded whenever $a$ and $b$
are close to bin boundaries.  For intuition, consider the following
examples.  If $\delta \in [a,b]$, $\mu_a^b$ can be written as a
weighted mean of the two subintervals $\frac{\delta - a}{b-a}
\mu_a^{\delta} + \frac{b - \delta}{b-a} \mu_\delta^b$.\footnote{The
  weights on each subcomponent here assume that $x$ is uniformly
  distributed. A different distribution would use different weights.}
If $\mu_a^{\delta}$ is known (because there are bin boundaries at $a$
and $\delta$), then any uncertainty about the value of the CEF in the
range $[a, \delta]$ is not consequential for the bounds on
$\mu_a^b$. If $b$ is close to $\delta$, the weight on the unknown
value $\mu_\delta^b$ is very small, and $\mu_a^b$ can be tightly
bounded. Similarly, if instead $\mu_a^b$ is known, and $b$ is again
close to $\delta$, then $\mu_a^\delta$ can be tightly estimated even
if $\mu_\delta^b$ has wide bounds.

Bounds on other functions of the CEF may be difficult to calculate
analytically, but can be defined as the set of solutions to a pair of
minimization and maximization problems that take the following
structure. We write the conditional expectation function in the form
$Y(x) = s(x,\gamma)$, where $\gamma$ is a finite-dimensional vector
that lies in parameter space $G$ and serves to parameterize the CEF
through the function $s$. For example, we could estimate the
parameters of a linear approximation to the CEF by defining
$s(x,\gamma)=\gamma_0+\gamma_1*x$. We can approximate an arbitrary
nonparametric CEF by defining $\gamma$ as a vector of discrete values
that give the value of the CEF in each of $N$ partitions; we take this
approach in our numerical optimizations, setting $N$ to
100.\footnote{For example, $s(x,\gamma_{50})$ would represent $E(y|x
  \in [49,50])$.}  Any statistic $m$ that is a single-valued function
of the CEF, such as the average value of the CEF in an interval
$(\mu_a^b)$, or the slope of the best fit line to the CEF, can be
defined as $m(\gamma)=M(s(x,\gamma))$.

Let $f(x)$ again represent the probability distribution
of $x$. Define $\Gamma$ as the set of parameterizations of
the CEF that obey monotonicity and minimize mean squared error with
respect to the observed interval data:
\begin{align}
  \label{eq:opt1}
  \Gamma = \underset{g \in G}{\text{argmin}}  \sum_{k=1}^K \Bigg\{
    \int_{x_k}^{x_{k+1}} f(x) dx \left( \left(
    \frac{1}{\int_{x_k}^{x_{k+1}} f(x)dx } \int_{x_k}^{x_{k+1}} s(x,g) f(x) dx \right) - \overline{r}_k
    \right)^2 \Bigg\} \\ \nonumber \text{such that}
  \\ \tag{Monotonicity} E(y \vert x) \text{ is weakly increasing in }
  x.
\end{align}
\noindent Decomposing this expression, $\frac{1}{\int_{x_k}^{x_{k+1}}
  f(x)dx } \int_{x_k}^{x_{k+1}} s(x,g) f(x) dx$ is the mean value of
$s(x,g)$ in bin $k$, and $\int_{x_k}^{x_{k+1}} f(x) dx$ is the width
of bin $k$. The minimand is thus a bin-weighted MSE.\footnote{While we
  choose to use a weighted mean squared error penalty, in principle
  $\Gamma$ could use other penalties.} Recall that for the rank
distribution, $x_1=0$ and $x_{K+1}=100$.

The bounds on $m(\gamma)$ are
therefore:

\begin{equation}
  \begin{aligned}
    \label{eq:m_bounds}
    m^{min} &= \inf\{m(\gamma) \ \vert \ \gamma \in \Gamma \} \\
    m^{max} &= \sup\{m(\gamma) \ \vert \ \gamma \in \Gamma \}. 
  \end{aligned}
\end{equation}

For example, bounds on the best linear approximation to the CEF can be
defined by the following process. First, consider the set of all CEFs
that satisfy monotonicity and minimize mean-squared error with respect
to the observed bin means.\footnote{In many cases, and in all of our
  applications, there will exist many such CEFs that exactly match the
  observed data and the minimum mean-squared error will be zero.}
Next, compute the slope of the best linear approximation to each
CEF. The largest and smallest slope constitute $m^{min}$ and
$m^{max}$. Stata code to generate bounds on the CEF and on $\mu_a^b$,
and Matlab code to run these numerical optimizations for more complex
functions (as well as with the curvature constraints described below)
are posted on the corresponding author's web site.


\vspace{8mm}

\noindent \underline{CEF Bounds Under Constrained Curvature}
\label{sec:curv}

\noindent The candidate CEFs that underlie the bounds in
Proposition~\ref{eq:cef_bound} are step functions with substantial
discontinuities. If such functions are implausible
descriptions of the data, then the researcher may wish to impose an
additional constraint on the curvature of the CEF, which will generate
tighter bounds. For example,
examination of the mortality-income relationship (which can be
estimated at each of 100 income ranks, displayed in Figure
\ref{fig:mort_splines}) suggests no such discontinuities.\footnote{More
  complex structural restrictions can also be imposed. For example,
  the CEF might be continuous within education bins, but there could
  be large discontinuities due to sheepskin effects at the education
  bin boundaries \cite{Hungerford1987}.} Alternately, in a context
where continuity has a strong theoretical underpinning but
monotonicity does not, a curvature constraint can substitute for a
monotonicity constraint and in many cases deliver useful bounds.

We consider a curvature restriction with the following structure:
\begin{align}
  \label{eq:c_bar}
  \tag{Curvature Constraint}  
  s(x,\gamma) \text{ is twice-differentiable and } \vert
  s''(x,\gamma) \vert \leq \overline{C}.
\end{align}

\noindent This is analogous to imposing that the first derivative is
Lipshitz.\footnote{Let $X, Y$ be metric spaces with metrics $d_X, d_Y$
  respectively. The 
  function $f:X \to Y$ is \textit{Lipschitz continuous} if
  there exists $K \geq 0$ such that for all $x_1,x_2 \in X$,
  $$d_Y(f(x_1),f(x_2)) \leq K d_X(x_1,x_2).$$} Depending on the value
of $\overline{C}$, this constraint may or may not bind. 

The most restrictive curvature constraint, $\overline{C}=0$, is
analogous to the assumption that the CEF is linear. Note that the
default practice in many studies of mortality is to estimate the best
linear approximation to the CEF of mortality given education (e.g.,
\citeasnoun{Cutler2011} and \citeasnoun{Goldring2016}). In the study
of intergenerational mobility (Section~\ref{sec:mob}), the best linear
approximation to the parent-child CEF is the canonical estimator. A
moderate curvature constraint is therefore a \textit{less} restrictive
assumption than the approach in many studies. We discuss the choice of
curvature restriction below.

In the rest of this section, we show results under a range of
curvature restrictions to shed light on how these additional
assumptions affect bounds in an empirical application. In our
applications in Sections~\ref{sec:mort} and \ref{sec:mob}, we show all
results under the most conservative approach of $\overline{C}=\infty$.

\subsection{Numerical Calculation of CEF Bounds}
\label{sec:numeric}

This section describes a method to numerically solve the constrained
optimization problem suggested by Equations~\ref{eq:opt1} and
\ref{eq:m_bounds}. We take a nonparametric approach for generality:
explicitly parameterizing an unknown CEF with limited data is
unsatisfying and could yield inaccurate results if the interval
censoring conceals a non-linear within-bin CEF. In the context of
mortality (and mobility, Section~\ref{sec:mob}), many CEFs of interest
do not appear to obey a familiar parametric form (see Figures
\ref{fig:mort_splines} and \ref{fig:mob_splines}).

To make the problem numerically tractable, we solve the discrete
problem of identifying the feasible mean value taken by $E(y|x)$ in
each of $N$ discrete partitions of $x$. We thus assume $E(y|x) =
s(x,\gamma)$, where $\gamma$ is a vector that defines the mean value
of the CEF in each of the $N$ partitions. We use $N=100$ in our
analysis, corresponding to integer rank bins, but other values may be
useful depending on the application. Given continuity in the latent
function, the discretized CEF will be a very close approximation of
the continuous CEF; in our applications, increasing the value of $N$
increases computation time but does not change any of our results.

We solve the problem through a two-step process. Define a $N$-valued
vector $\hat{\gamma}$ as a candidate CEF. First, we
calculate the minimum MSE from the constrained optimization problem
given by Equation~\ref{eq:opt1}. We then run a second pair of
constrained optimization problems that respectively minimize and
maximize the value of $m(\hat{\gamma})$, with the
additional constraint that the MSE is equal to the value obtained in
the first step, denoted $\underbar{MSE}$. Equation \ref{eq:opt2} shows
the second stage setup to calculate the lower bound on
$m(\hat{\gamma})$. Note that this particular setup is specific to the
uniform rank distribution, but setups with other distributions would
be similar.
\begin{align}
  \label{eq:opt2}
  m^{min} &= \underset{ \hat{\gamma} \in
    [0,100]^{N} }{ \text{min} } m(\hat{\gamma}) \\
  &\nonumber \text{such that} \\ 
  \tag{Monotonicity}  s(x, \hat{\gamma}) \text{ is
    weakly increasing in } x \\ 
  \tag{Curvature}  \lvert s''(x, \hat{\gamma}) \rvert \leq \overline{C}, \\
  \tag{MSE Minimization} \sum_{k=1}^K \left[ \frac{\Vert X_k \Vert}{100}
    \left( \left( \frac{1}{\Vert X_k \Vert} \sum_{x \in X_k} 
   s(x,\hat{\gamma}) \right) -
    \overline{r}_k \right)^2  \right] &= \underbar{MSE}
\end{align}

\noindent
$X_k$ is the set of discrete values of $x$ between $x_{k}$ and
$x_{k+1}$ and $\Vert X_k \Vert$ is the width of bin $k$. The
complementary maximization problem obtains the upper bound on
$m(\hat{\gamma})$.

Note that setting $m(\gamma) = \gamma_x$ (the x\superscript{th}
element of $\gamma$) obtains bounds on the value of the CEF at point
$x$. Calculating this for all ranks $x$ from 1 to 100 generates
analogous bounds to those derived in proposition \ref{eq:cef_bound},
but satisfying the additional curvature constraint. Similarly
$m(\gamma) = \frac{1}{b-a} \sum_{x=a}^{b}\gamma_x$ obtains bounds on
$\mu_a^b$. 

\subsection{Example with Sample Data}
\label{sec:example}

In this section, we demonstrate the bounding method using data from
mortality in the United States, continuing with the mortality of
50--54 year-old women in 2015. We focus here on the properties of the
bounds under different assumptions. We explore mortality change in
more detail in Section~\ref{sec:mort}.

Panel A of Figure \ref{fig:cef_f2_mort} graphs the analytical upper
and lower bounds on $E(y|x)$ at each value of $x$ under just the
assumption of monotonicity. These bounds do not reflect statistical
uncertainty but uncertainty about the CEF in the unobserved parts of
the latent rank distribution.\footnote{We do not present standard
  errors because we are working with the universe of deaths in a large
  country and statistical imprecision is very small in this
  context. We discuss and present bootstrap confidence sets in
  Section~\ref{sec:mob} where statistical imprecision is more
  important.}

We next consider a curvature-constrained CEF.\footnote{With neither
  the monotonicity nor the curvature constraint, the CEF cannot be
  bounded except by the maximum possible value of the variable of
  interest.} The mortality-education data are not in themselves
informative regarding which curvature restriction to choose. To
identify a conservative curvature constraint, we examine the curvature
of a closely related conditional expectation function that is not
interval censored: the CEF of mortality given income rank. We show
this CEF in Figure \ref{fig:mort_splines}, using data from
\citeasnoun{Chetty2016b}. Using a spline approximation to income rank
data for 52-year-old women in 2015, we calculate a maximum
$\overline{C}$ of 1.6; we use a constraint approximately twice as high
as a conservative starting point. Panel B of
Figure~\ref{fig:cef_f2_mort} shows the bounds obtained under curvature
constraints of 2, 3 and 5, but \textit{without} the assumption of
monotonicity.
Relative to
those under monotonicity, the curvature-constrained bounds
are less informative at the tails of the distribution,
and more informative close to the bin midpoints.

In Panel C, we impose the monotonicity and curvature constraints
simultaneously. Panel D shows the limit case with $\overline{C}=0$;
the CEF in this figure is identical to the predicted values from a
regression of mortality on median education rank. Note that while 
stricter curvature restrictions can tighten the bounds, this
may come at the expense of ruling out a plausible CEF, even if the
MSE remains zero. In Figure~\ref{fig:cef_f2_mort}, only Panel D has a
non-zero MSE.

Table~\ref{tab:ests_mortality} presents estimates of $p_x=E(y|x)$
and $\mu_a^b=E(y|x \in (a, b))$ for women ages 50--54 in 2015, for various
values of $x$, $a$ and $b$, under different constraints.  We first
highlight the statistics $p_{32}$ and $\mu_0^{64}$. In 1992, 64\% of
women had high school education or less, and thus occupied the bottom
rank bin in the education distribution. $p_{32}$ and $\mu_0^{64}$
respectively describe the median and mean mortality of the comparably
ranked group of women in 2015. These statistics give us mortality estimates for
constant ranks in the education distribution, even though the
distribution of education levels is changing over time. 

We draw attention to two features of the table. First, the interval
mean estimates ($\mu_a^b$) are in most cases considerably more tightly
bounded than estimates of the CEF value at the midpoint of the
interval ($p_x$). $\mu_0^{64}$ is nearly point identified in 1992
because $0$ and $64$ are very close to bin boundaries in 1992, and it
is tightly bounded in 2015 as well, regardless of the constraint
set.\footnote{We have used integer approximations to these parameters
  for convenience; if we used the average mortality for the precise
  proportion of women with less than or equal to a high school degree
  ($\mu_0^{63.658}$), then the parameter would be precisely point
  identified.}  These two statistics are both useful summaries of
mortality among the less educated, but $\mu_0^{64}$ is estimated with
at least 22 times more precision than $p_{32}$.  Similarly,
$\mu_0^{39}$ is effectively point estimated in 2015, where 39\% of
women had attained high school or less.  The advantage of $\mu_a^b$
over $p_x$ (where $x=\frac{a+b}{2}$) is greatest when $a$ and $b$ are
close to boundaries in the data.

Second, $\mu_0^{39}$ and $\mu_0^{64}$ are very robust to
different bounding assumptions. Inference is more difficult on a
parameter like $\mu_0^{20}$ with boundaries far from any in the data,
and the width of the bounds depends strongly on the assumptions being
made. Mortality in the bottom 20\% of the education distribution may
be of policy interest, but our method shows that it cannot be
precisely estimated with these data.

Because it is a frequently estimated parameter, in Column 5 we show
the predicted values from the best linear approximation to the
mortality-education CEF. This parameter is point estimated, but
implicitly assumes away large increases in mortality at the bottom of
the distribution, increases that are consistent with the data and in fact
suggested by Figure~\ref{fig:mort_splines}. In contrast, our method
allows researchers to generate consistent bounds on mortality across
the education distribution under considerably less restrictive
assumptions.

\section{Simulation: Bounds on the U.S. Mortality-Income CEF}
\label{sec:sim} 
\label{sec:denmark}

In this section, we validate our method in a simulation by taking data
from the fully supported U.S. mortality-income CEF \cite{Chetty2016b},
interval censoring that data, and then recovering bounds on the true
CEF from the interval censored data. The exercise shows that our
approach works in practice. It also illustrates that studying
partially identified bounds permits the researcher to recover
important features of the CEF that she might miss if she attempted
simply to fit a parametric form to the observed bin means.  We use
data on mortality by income percentile, gender, age and year
\cite{Chetty2016b}. We focus on women aged 52 in 2014, the group most
comparable what we have examined so far.

First, we estimate the true CEF from the
mortality-income data by fitting a cubic spline with four knots to the
data, the same spline used to obtain an estimate of
$\overline{C}$. We plot this in Appendix Figure
\ref{fig:mort_splines}.\footnote{We use a spline approximation rather
  than the raw data because the variation across neighboring rank bins
  is most likely idiosyncratic given the small number of deaths in an
  age bin defined by a single year. By using information from
  neighboring points, the spline is a better estimate of
  mortality risk than the individual rank bin means.}

Next, we simulate interval censoring by obtaining the mean of the true
CEF within income rank bins that cover the same ranks as the education
bins observed in our 2015 mortality-education data. In this
simulation, there are 39\% of people in the bottom bin, 29\% of people
in the middle bin, and 33\% of people in the top bin.\footnote{We
  round to the nearest integer, since we only observe integer
  percentiles in the data from \citeasnoun{Chetty2016b}.} After
interval censoring, we have a dataset with average mortality in each
of three bins, comparable to the data from
Section~\ref{sec:method}. We compute bounds on the CEF using only the
binned data.

Panels A--D of Figure~\ref{fig:mort_bounds} present CEF bounds
generated from the binned data, under monotonicity and curvature
limits that vary from $\infty$ (unconstrained) to $1$. The dashed
lines show the underlying data. The solid circles show the constructed
bin means of the censored data; these are the only data that we use
for the optimization. The solid lines show the upper and lower
envelopes that we calculate for the nonparametric CEF.

The suggested curvature constraint ($\overline{C}=3$) yields bounds
that contain the true CEF at every point; but when we impose
$\overline{C}=1$, the constraint is excessive and the bounds do
not contain the true CEF. The true CEF
is not always centered within the bounds; from ranks 25 to 40, the
true CEF is near the bottom bound, and from ranks 90 to 100, it is
nearer the upper bound.

The exercise also illustrates that assuming a parametric form for the
underlying CEF can yield misleading results. A quadratic or
linear fit to the data would fail to identify the convexity at the
bottom of the distribution. The strength of our method is that 
it makes transparent how the structural assumptions affect the CEF bounds.

Table \ref{tab:sims_stats} shows bounds on a range of statistics of
interest under different curvatures, as well as the true estimate.  We
highlight three results. First, the interval mean measures ($\mu_a^b$)
generate tighter bounds than the CEF values $p_x$, with no greater
propensity for error. Second, $\overline{C}$ is consequential for
$p_x$, but considerably less important for $\mu_a^b$. Third, the
linear estimates generated with $\overline{C} = 0$ are biased by as
much as 25\% relative to the true estimates, and sometimes produce
estimates outside the bounds even of CEFs with unconstrained
curvature.

\section{Application: Estimating U.S. Mortality in Constant Education Rank Bins}
\label{sec:mort}

In this section, we apply our methodology to study changes in
U.S. mortality for individuals at constant ranks in the education
distribution. Many researchers have noted that mortality is rising for
individuals in less educated groups; however, the changing composition
of these groups over time has made this finding difficult to
interpret. For example, women with a high school education or less
(LEHS) represented the least educated 64\% of the population in 1992,
and the least educated 39\% of the population in 2015. Those with LEHS
are thus more negatively selected in 2015 than they were in 1992; the
changing size and composition of this group may account for at least
some of the mortality increase. This bias has been frequently noted in
the literature, but different authors have reached widely different
conclusions regarding its size and
importance.\footnote{\citeasnoun{Cutler2011} adjust for compositional
  shifts by predicting propensity to attend college using region,
  marital status and income, and then using this propensity as a
  conditioning variable. They argue that compositional shifts are not
  important for mortality changes from the 1970s to the 1990s. This
  approach is limited by the extent to which these variables can
  predict education, and in many cases (e.g. with vital statistics
  data), these additional variables are
  unavailable. \citeasnoun{Case2015} and \citeasnoun{Case2017} argue
  that changes in the proportion of middle-aged whites with LEHS from
  the 1990s to the present are too small to influence mortality
  rates. In contrast, \citeasnoun{Dowd2014} and \citeasnoun{Bound2015}
  perform analytical exercises that suggest that compositional shifts
  can explain most or all of recent mortality
  changes. \citeasnoun{Bound2015} estimate mortality for the bottom
  quartile of the education distribution, implicitly assuming that
  mortality is constant within each interval-censored mortality rank
  bin. \citeasnoun{Currie2018} suggests that studying mortality for
  the least educated is entirely misleading because of the shrinking
  size of this group. \citeasnoun{Goldring2016} derive a one-tailed
  test for changes in the mortality-education gradient, but they do do
  not calculate the bias in existing mortality estimates or estimate
  mortality in constant rank bins. Our method requires no additional
  covariates, and bounds mortality at an arbitrary education rank
  under only the assumption of monotonicity.} In this section, we use
the methods above to bound the value of the mortality CEF at constant
education ranks, even if these ranks are not directly observed in the
data. We can then study a group with constant size and education rank
over time, and thus study any subset of the education rank
distribution without bias from changing education levels over time.

Mortality by education records come from the U.S. Center for Disease
Control's WONDER database and total population by age, gender and
education come from the Current Population Survey, as in
\citeasnoun{Case2017}. Additional details on data construction are
available in Appendix~\ref{sec:app_mort_data}.

As above, we assume that the observed mortality data describe a
monotonic relationship between mortality and latent education rank,
the latter of which is observed only in coarse bins. Results are
virtually identical if we constrain curvature using the parameter
suggested in Section~\ref{sec:method} and forgo the monotonicity
constraint.  We focus in this section on women aged 50--54, because
this is a group whose education composition has shifted substantially
over time.\footnote{We use 5-year bins for ages rather than larger
  bins to ensure that the average age in the bin does not change over
  time \cite{Gelman2016}.}

Panel A of Figure~\ref{fig:mort_overlay} plots mean total mortality
for women age 50-54 in each education group in 1992 and in 2015, along
with analytical bounds on CEFs with unconstrained curvature. The
bounds are largely overlapping across the entire education
distribution, and too wide to infer very much about changes in
mortality. In Panel B, we restrict curvature to approximately twice
the maximum curvature from the income-mortality data, as discussed in
Section~\ref{sec:method}. Panel B identifies a clear decline in
mortality at the top of the education distribution, but the bounds
remain minimally informative at the bottom.

The interval mean measures ($\mu_a^b$) are more informative.  We focus
on mean mortality in the bottom 64\%, denoted by
$\mu_0^{64}$.\footnote{Specifically, we calculate $\mu_0^{63.658}$ for
  analytical monotonic bounds and $\mu_0^{64}$ for numerical curvature
  constrained bounds.} This measure describes mortality for the set of
women who occupied positions in the rank distribution that would give
them high school education or less in 1992. In 2015, the bottom 64\%
includes all women with high school or less, and some women with some
college education, but none with bachelor's degrees or higher. For
men, we focus on the bottom 54\%, which is the population share with
high school or less in 1992; by 2015, 44\% of men have LEHS, so the
bottom 54\% again includes some men with two-year college degrees. As
in \citeasnoun{Chetty2016b}, we rank men and women against members of
their own gender, estimating mortality for a given percentile group of
men or women; that is, the least educated group can be interpreted as
the ``the 64\% of least educated women,'' rather than ``women in the
bottom 64\% of the population education distribution.''  We chose
own-gender reference points because women's and men's labor market
opportunities and choices are often different and because women and
men often share households and incomes, making population ranks
misleading. However, alternate choices could be considered and
estimated with the same method.

Panel A of Figure~\ref{fig:mort_change} shows bounds on total
mortality for women aged 50-54 in the bottom 64\%. Mortality in 1992
can be point estimated, because the 0-64 rank bin interval is exactly
observed in the data. As education levels diverge from those in 1992,
the bounds progressively widen. The ``x'' markers in the figure plot
the unadjusted estimates of mortality among women with less than or
equal to high school education; these mortality estimates describe a
group occupying a shrinking and more negatively selected share of the
population over time. The unadjusted estimates, which are the object
of study in most earlier work on the mortality-education relationship,
significantly overstate mortality increases relative to the constant
rank group. The upper bound on mortality gain for the bottom 64\%
is 8.5\%, compared to the unadjusted estimate of 28\%. Panel B shows
the same figure for men. The unadjusted estimates are closer to the
bounds here because men have gained less education than women over
this period. We can bound the mortality change for men in the interval
$[-7.1\%, +0.3\%]$, compared with the unadjusted estimate of
$+1.2\%$.\footnote{This result is not directly comparable to Case and
  Deaton (2015, 2017), who focus on \textit{white} men and women,
  whose unadjusted mortality is rising more substantially among the
  less educated. Estimating separate bounds for different racial
  groups requires additional assumptions about the relative positions
  of these groups in the unobserved part of the latent education
  distribution, and we leave this exercise for future work. The
  increases in mortality for less educated women that we identify here
  are still a cause for concern even if they are lower than
  previous estimates.} Panels C and D present analogous results for
combined deaths from suicide, poisoning and liver disease, described
by \citeasnoun{Case2017} as ``deaths of despair.'' The unadjusted
mortality estimates continue to overstate the constant rank mortality
changes, but the difference is small here because (i) deaths of
despair have increased substantially among all groups; and (ii) the
education gradient in deaths of despair was small in 1992. Appendix
Figure~\ref{fig:mort_change_nomon} shows the same plots, but removes
the monotonicity assumption and instead imposes the curvature
restriction of $\overline{C}=3$ suggested above. The plots are highly
similar; as discussed in Section~\ref{sec:method}, when $a$ and $b$
are close to bin boundaries, the bounds on $\mu_a^b$ are very robust
to alternate bounding assumptions.

Table~\ref{tab:mort_changes} shows unadjusted and constant-rank
estimates of women's mortality changes from 1992-2015 for age groups
from 20 to 69, for all education categories. We fix education rank
bins based on the 1992 rank bin divisions; results are very similar if
we fix estimates at the 2015 boundaries. The unadjusted estimates
systematically overstate mortality increases for all groups,
because the mean rank in each group has declined over this period. 

The extent of the bias on the naive estimates is increasing in the
magnitude of the mortality-education gradient, and in the magnitude of
the shift in bin boundaries. Given the significant variation across
age groups and genders, blanket assumptions about the existence or
lack of bias in unadjusted mortality differences are therefore
unlikely to be useful. Unadjusted estimates of men's mortality changes
from 1992-2015 are close to the constant rank bounds, as are
unadjusted estimates of deaths of despair for both men and women. For
women's total mortality, however, the naive estimates overstate
mortality increases in many cases by a factor of three or more, and in
some cases they have the wrong sign.

\section{Application: Intergenerational Educational Mobility}
\label{sec:mob}

The study of intergenerational mobility is another research context
where the conditional expectation function of interest in many cases
has an interval-censored conditioning variable.\footnote{For a review
  of intergenerational mobility, see \citeasnoun{Solon1999},
  \citeasnoun{Hertz2008}, \citeasnoun{Corak2013a},
  \citeasnoun{Black2011}, and \citeasnoun{Roemer2016}.}  Studies of
intergenerational mobility typically rely upon some measure of rank in
the social hierarchy which can be observed for both parents and
children \cite{Chetty2014c,Chetty2017}. In many contexts, the only
measure of social rank available for parents is their level of
education. In richer countries, this arises for studies of mobility in
eras that predate the availability of administrative income
data.\footnote{See, for example, \citeasnoun{Black2003} and
  \citeasnoun{Guell2013}.} In developing countries, matched
parent-child data are considerably more rare, and educational mobility
is often the only feasible object of study.\footnote{See, for example,
  \citeasnoun{Wantchekon2015}, \citeasnoun{Hnatkovska2013} or
  \citeasnoun{Emran2015}.}  Interval-censored parent education data is
ubiquitous in studies of intergenerational educational
mobility. Table~\ref{tab:mob_ests} reports the number of parent
education bins used in a set of recent studies of intergenerational
mobility from several rich and poor countries. Several of the studies
observe education in fewer than ten bins, the population share in the
bottom bin is often above 20\%, and sometimes it is above
50\%.\footnote{We specifically selected a set of studies where coarse
  data is likely to be an important factor. Note that internationally
  comparable censuses often report education in as few as four or five
  categories.}

Studies on educational mobility typically focus on linear estimators
of the parent-child outcome relationship, such as the slope of the
best linear approximator to the CEF of child education rank given
parent education rank, i.e., the rank-rank gradient. This is a useful
mobility statistic but it has two important limitations. First, it is
not useful for cross-group comparison.  The within-group rank-rank
gradient measures children's outcomes against better off members of
their own group; a subgroup can therefore have a lower gradient
(suggesting more mobility) in spite of having worse outcomes than
other groups at every point in the parent distribution.\footnote{An
  extreme example makes this clear. Suppose children in some
  population subgroup A all end up at the 10th percentile of the
  outcome distribution with certainty. The rank-rank gradient for this
  group would be zero (assuming some variation in parent outcomes),
  implying perfect mobility. But in fact the group would have
  virtually no upward mobility.} Second, the rank-rank gradient
aggregates information about mobility at the top and at the bottom of
the parent distribution; it is not directly informative about upward
mobility in the bottom half of the distribution. 

Because of these limitations, recent studies have focused on measures
based on the value of the parent-child CEF at a point in the parent
distribution, termed absolute mobility at percentile $x$ by
\citeasnoun{Chetty2014c} and denoted $p_x$. For example,
\citeasnoun{Chetty2014c} focus on $p_{25}$, which describes the
expected outcome of the child born to the median family in the bottom
half of the rank distribution. Unlike the rank-rank gradient, these
measures are both informative about child outcomes at arbitrary points
in the parent rank distribution and can be meaningfully compared
across population subgroups. These measures are central to current
research on mobility, but there is no established method for
calculating such measures with education data, where any given
percentile in the parent distribution lies within some larger bin. The
problem is most stark when the bins are very large, so we focus our
application on measuring intergenerational mobility in India, where
over 50\% of older generation parents are in the bottom education bin
($<2$ years of education). Appendix Table~\ref{tab:trans_matrices}
shows the complete education transition matrices for decadal birth
cohorts from 1950 to 1989. A tempting but misleading approach would be
to simply assume that the expected child outcome is exactly the same
at all ranks within a given rank bin. In this case, the value of
$p_{25}$ will change when education is measured with a different
degree of granularity. In contrast, our bounds will widen when the
granularity of the measure decreases, but they will contain the bounds
generated from more granular data.

We take the following approach. We assume that the latent parent-child
rank CEF can be described by an increasing monotonic function; this
relationship is monotonic in virtually every country
\cite{Dardanoni2012}, as well across every rank bin in every year of
our data on India. We use the CEF bounding method derived in
Section~\ref{sec:method} to obtain bounds on (i) the value of the
parent-child CEF at arbitrary parent percentiles ($p_x$); and (ii) the
average value of the parent-child CEF across arbitrary percentile
ranges of the parent distribution.\footnote{Our method is loosely
  related to \citeasnoun{Chetty2016}, who use a numerical procedure
  with similar constraints to bound absolute mobility at the 25th
  percentile, given just the marginal distributions of children's and
  parents' incomes and no information on the joint
  distribution. However, the substantive problem they solve is very
  different from ours.} We call the latter statistic, which
corresponds to $\mu_a^b$ from Section~\ref{sec:method},
\textit{interval mobility}.  We show below that, (i) the rank-rank
gradient may be biased or uninformative when estimated from interval
data; and (ii) interval mobility ($\mu_a^b$) can be bounded
considerably more tightly than the other measures that we consider. We
combine data from two sources, including administrative data on the
education of every person in India in 2012, to obtain a representative
sample of every father-son pair in India.\footnote{We are restricted
  to the study of fathers and sons because the data do not match
  daughters to parents or children to mothers when they do not live in
  the same household.} The details of data construction are described
in Appendix~\ref{sec:app_mob_data}.

We observe education for both fathers and sons in seven
categories.\footnote{The categories are (i) less than two years of
  education; (ii) at least two years but no primary; (iii) primary;
  (iv) middle school; (v) secondary; (vi) senior secondary; and (vii)
  post-secondary or higher.}  Because sons' education levels are also
reported categorically, we do not directly observe the expected child
outcome in each parent education bin. In this section, we instead
assign to children the midpoint of their rank bin. We show in
Appendix~\ref{app:impute} that data on son wages (for which the rank
distribution is uncensored) suggests that the midpoint is a very close
approximation to the true expected rank, because the residual
correlation of father education and son wages is very small once son's
education is controlled for. Note that such an exercise is impossible
for interval censoring of parent data, because no additional data on
parents is available, as is typical in studies of intergenerational
mobility.\footnote{Because parental education is often obtained by
  asking children, it is common to have data on many child outcomes,
  but only the education level of parents, as we do here.}
Appendix~\ref{app:impute} also provides a method that generates bounds
under joint censoring, which can be used in contexts where additional
data on sons is not available. An alternate approach would be to
estimate child rank directly using a socioeconomic measure for sons
that can be observed continuously.

Panel A of Figure~\ref{fig:mob_changes} shows the raw data for cohorts
born in the 1950s and in the 1980s. Each point plots the midpoint of a
father education rank bin against the expected child rank in that
bin. The vertical lines plot the boundary for the lowest education bin
for each cohort, which corresponds to fathers with less than two years
of education. In the 1950s birth cohort (solid line), this group
represents 60\% of the population; it represents 38\% for the 1980s
cohort (dashed line). The points in the figure suggest that the
rank-rank CEF has not changed in the bottom half of the parent
distribution over this period: the bottom point in the 1950s lies
almost directly between the bottom two points in the 1980s.  However,
when we estimate the rank-rank gradient directly on these bin means,
we find small but unambiguous mobility gains over this 30-year
period. The graph makes clear that the decrease in the gradient is
driven by changes in mobility in the top half of the
distribution. Alternately, if we treat the data as uncensored, such
that the expected child outcome is the same at all latent ranks within
each parent bin, we would conclude that absolute upward mobility
($p_{25}$, or the expected child outcome at the 25\superscript{th} parent
percentile) has unambiguously fallen from the 1950s to the
1980s. Neither of these conclusions appears to represent the true
change in mobility.\footnote{Note also that the CEF is evidently
  non-linear, so a naive nonlinear parametric fit to the bin midpoints
  would be biased due to Jensen's Inequality. It also assumes away
  concavity at the bottom of the distribution, which is observed in
  many other countries (see Appendix Figure~\ref{fig:mob_splines}).}
We therefore turn to estimating bounds on the CEF in each period.

Panel B of Figure~\ref{fig:mob_changes} shows the bounds on the
parent-child CEFs for these birth cohorts; we select a curvature
constraint of 0.1, which is approximately 1.5 times the maximum
curvature observed in uncensored parent-child \textit{income} data
from the United States, Denmark, Sweden and Norway.\footnote{We
  selected these countries because we were able to obtain precise
  uncensored parent-child income rank data for them from
  \citeasnoun{Chetty2014b}, \citeasnoun{Boserup2014} and
  \citeasnoun{Bratberg2017}. Graphs for the spline estimations used to
  calculate the curvature constraints are displayed in Appendix
  Figure~\ref{fig:mob_splines}. Results are substantively similar
  under different curvature constraints.}  The bounds on the CEF are
widest at the bottom of the distribution where interval censoring is
most severe, and are worse for the older generation with the larger
bottom rank bin.  The bounds in the bottom half of the distribution
are consistent with both large positive and large negative changes in
mobility, and thus uninformative. Absolute upward mobility can
evidently not be bounded informatively.\footnote{Note that a more
  restrictive curvature constraint would narrow the bounds, but at the
  expense of imposing excessive structure that would rule out
  plausible CEFs, especially given the evident nonlinearity in the
  data.}

We can make meaningful progress by focusing on an interval-based
measure such as $\mu_a^{b}=E(y|x \in [a, b])$. We call this measure
\textit{interval mobility}, and focus in particular on
$\mu_0^{50}=E(y|x \in [0, 50])$, which we call \textit{upward interval
  mobility}. This statistic is closely related to absolute upward
mobility ($p_{25}$). The latter describes the outcome of the median
child born to a parent in the bottom half of the parent distribution,
whereas upward interval mobility describes the \textit{mean} child
outcome in the bottom half of the parent distribution. These measures
are of similar economic importance, but we show here that upward
interval mobility can be bounded tightly in contexts with severe
interval censoring, while absolute upward mobility cannot.

Figure~\ref{fig:mob_time_stats} shows bounds on the three mobility
statistics discussed for each decadal cohort: the rank-rank gradient,
absolute upward mobility ($p_{25}$), and upward interval mobility
($\mu_0^{50}$).\footnote{The bounds on the rank-rank gradient describe
  the slopes of the set of best linear approximators to feasible
  CEFs.}  For reference, we plot recent estimates of similar
educational mobility measures from USA and Denmark.\footnote{Rank-rank
  correlations of education are from \citeasnoun{Hertz2008}, which are
  equal to the slope of the rank-rank regression coefficient if
  estimated on uncensored rank data. For absolute mobility, we
  calculate $p_{25}$ for the U.S. and Denmark from the distributions
  shown in Figure~\ref{fig:mob_splines}, with data from
  \citeasnoun{Chetty2014c}.}\superscript{,}\footnote{We calculate and
  show bootstrap confidence sets using 1,000 bootstrap samples from
  the underlying datasets, following methods described in
  \citeasnoun{m_Imbens2004} and \citeasnoun{Tamer2010}.}  Once we
allow expected child outcomes to vary within the bottom parent
education bin, both the rank-rank gradient and absolute upward
mobility have wide and minimally informative bounds. In contrast,
upward interval mobility is estimated with tight bounds in all
periods. According to this measure, upward mobility has changed very
little over the four decades studied; there is a small gain from the
1950s to the 1960s, followed by a small decline from the 1960s to the
1980s. On average, Indian mobility is as far below that in the United
States as mobility in the United States is below
Denmark. Table~\ref{tab:ests_india} reports the bounds for each
measure and cohort with bootstrap confidence sets under a range of
curvature restrictions. Moderate curvature restrictions generate
substantial improvements on the estimation of the value of the CEF
(e.g., $p_{25}$), but are considerably less important for the interval
mean measures (e.g., $\mu_{0}^{50}$), which are tightly bounded even
with unconstrained curvature.

In conclusion, the most widely used mobility estimator, the rank-rank
gradient, presents an incomplete and potentially biased picture of
intergenerational educational mobility.  Upward interval mobility, in
contrast, yields informative estimates even without a curvature
constraint, making it feasible to study upward mobility in the
lower-ranked parts of the distribution, even in a context with extreme
interval censoring. The advantage of upward interval mobility is
likely to be replicated in mobility studies in the many other
countries where older generations are clustered in less educated bins.

\section{Conclusion}
\label{sec:conc}

We propose a method that generates useful bounds on a
conditional expectation function when the conditioning variable is
interval-censored. Tight bounds on parameters of interest are possible
because of three innovations. First, we show that CEF bounds are
substantially improved when the distribution of the conditioning
variable is known, and many economic contexts have distributions that
are either known with certainty or are assumed by convention. Second,
we show that there are many intervals in which the conditional mean
can be bounded tightly, even when the bounds on the CEF itself are
wide across its domain. Third, bounds can be improved by imposing a
constraint on the curvature of the CEF, which is justified in many
empirical contexts. A curvature constraint can further substitute for
the assumption of monotonicity, making it possible to conduct
inference in interval data contexts where there is not a strong
theoretical basis for monotonicity. 

We also propose a traactable numerical framework for bounding CEFs and
functions of the CEF with arbitrary structural
restrictions. Simulations of interval censoring indicate that the
methods perform well in common empirical scenarios. In our
applications, the first two innovations prove sufficient to bound
parameters of interest informatively, but any of these three alone is
insufficient.

A useful thought experiment when working with interval data is to
explore how estimates are affected as intervals become more or less
granular. The bounds presented in this paper become wider when the
data become more coarse, as should be expected given that information
has been removed. In contrast, with conventional point estimation
approaches, the use of coarser intervals can lead to different point
estimates, thus obscuring the loss of information to the
researcher. Our method is transparent about what is known and what is
not known.

We have shown that our method can be used to solve known problems in
the study of mortality and of intergenerational mobility. Generating
bounds on outcome variables by education quantile is an
application with many other potential uses, given the large number of
contexts where education is of interest as a dependent variable but
available only in a small number of bins.  Other useful applications
may be found where the conditioning variable takes the form of
interval-censored income data, or Likert scale responses, among
others. 


\begin{appendix}

\newpage
\singlespace
\bibliographystyle{aer}
\bibliography{manual,mendeley}


\begin{figure}[H]
  \caption{Women's Total Mortality by Education Group, \cnewline Age
    50-54, 1992-2015} 
  \label{fig:mort_scatter}

  \begin{center}
    \begin{tabular}{c}
      \\
      \includegraphics[scale=0.85]{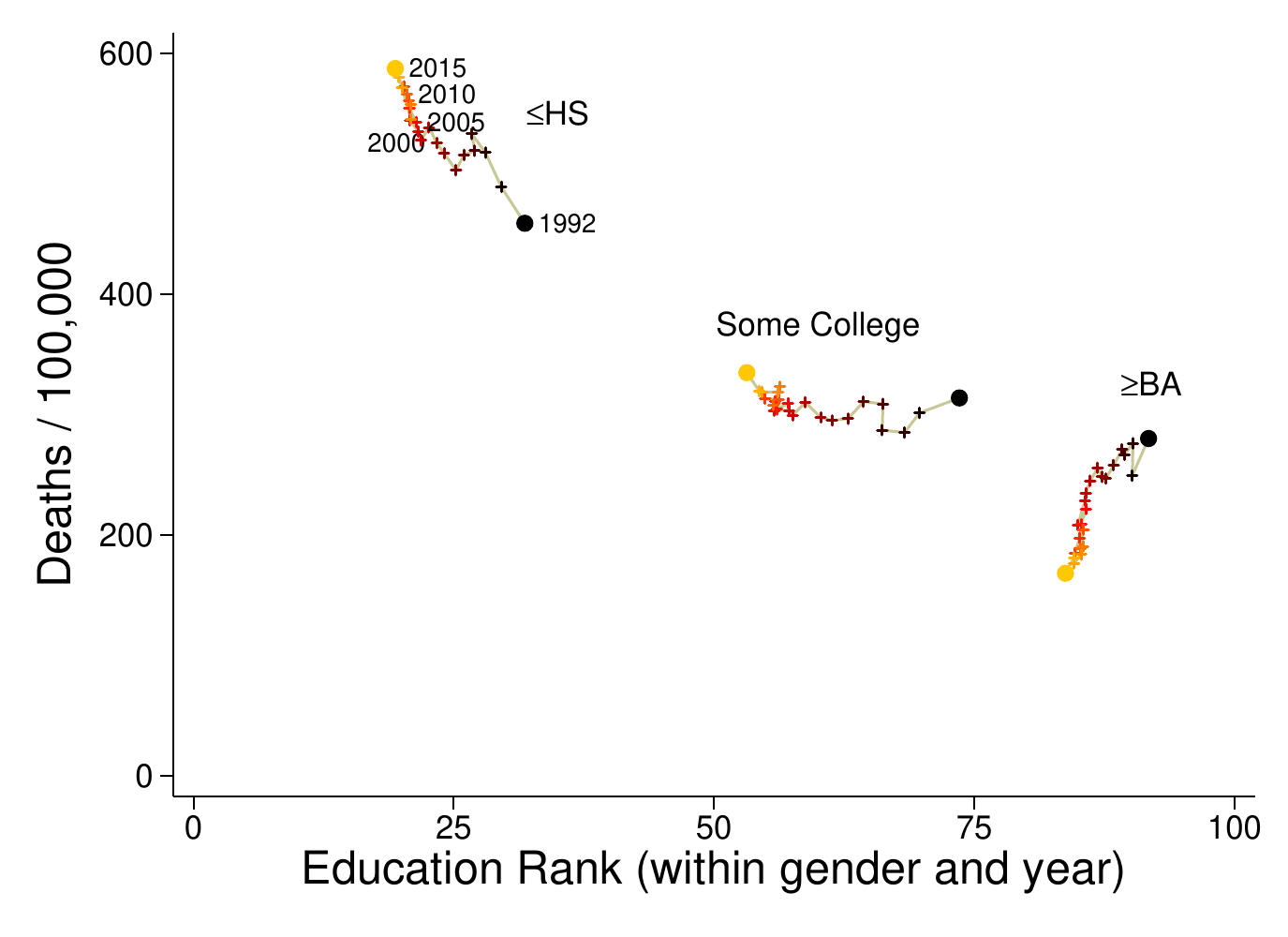} \\

      \hline
    \end{tabular}
  \end{center}

\footnotesize{Figure \ref{fig:mort_scatter} plots mortality rates vs. mean
education rank for three groups: women with less than or equal to
a high school degree, women with some college education, and
women with a BA or more. Each point represents a mortality rate (in
deaths per 100,000) within a
year and education group. The lighter colored points correspond to
later years. Ranks are calculated within gender and year.}
\end{figure}

\begin{figure}[H]
  \caption{Candidate Functions for Conditional Expectation of
    Mortality \cnewline given Education Rank} 
  \label{fig:cef_sample}

  \begin{center}
    \begin{tabular}{c}
      \\
      \includegraphics[scale=0.85]{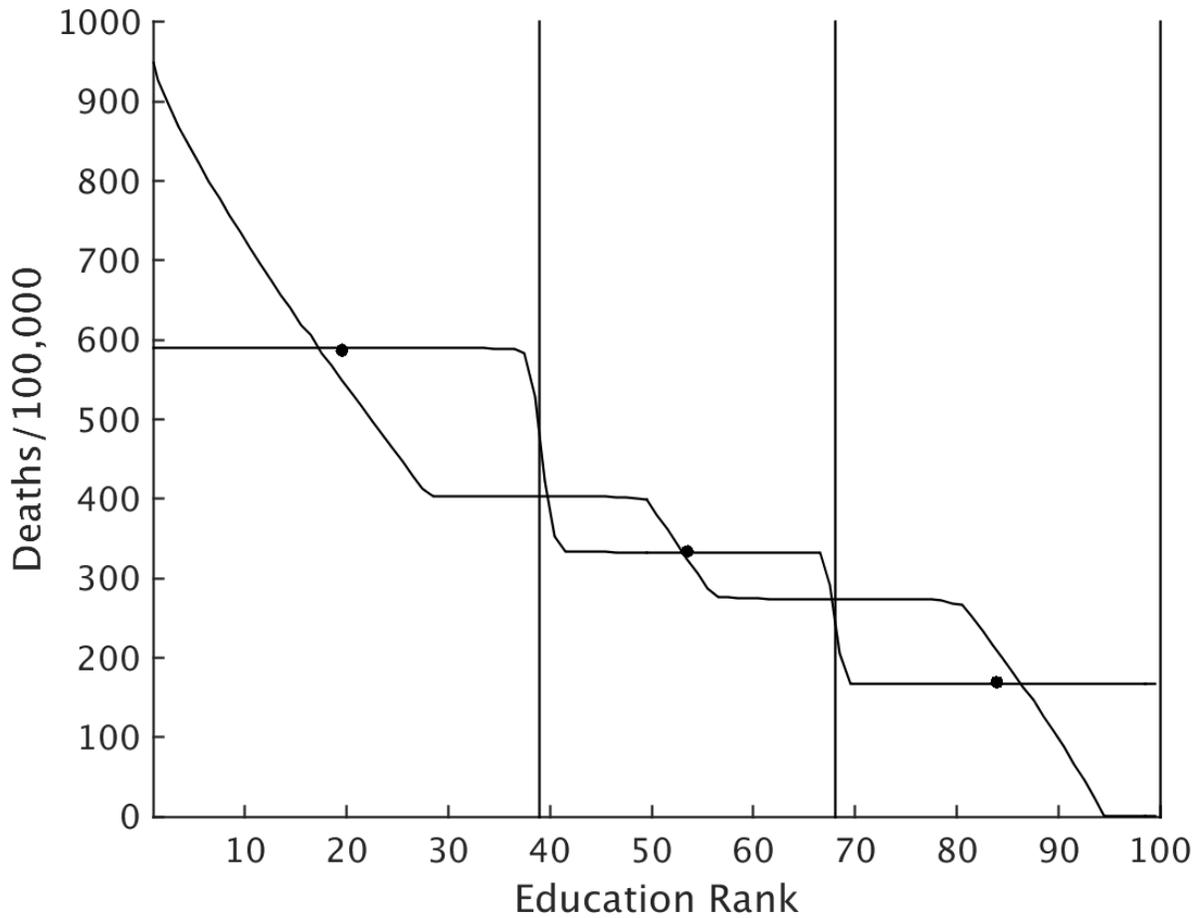} \\

      \hline
    \end{tabular}
  \end{center}

  \footnotesize{Figure \ref{fig:cef_sample} shows two candidate
    conditional expectation functions of mortality given education
    rank for women aged 50--54 in the United States in 2015. The
    vertical lines show the bin boundaries and the points show mean
    mortality and mean child rank in each bin.}

\end{figure}

\begin{figure}[H]
  \caption{Analytical Bounds on the CEF of Mortality given Education Rank} 
  \label{fig:analytic_bounds}
  \begin{center}
    \begin{tabular}{c}
      \\
      \includegraphics[scale=0.85]{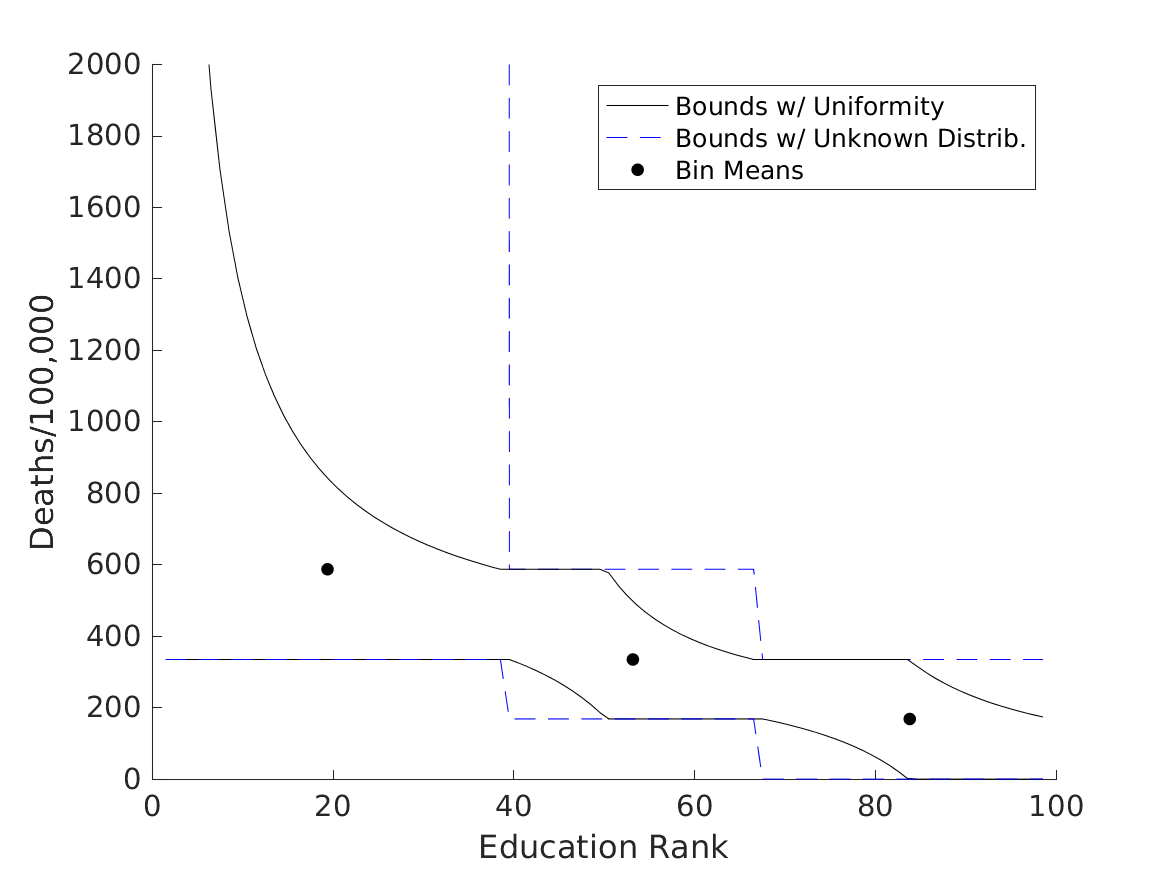} \\

      \hline
    \end{tabular}
  \end{center}

  \footnotesize{Figure \ref{fig:analytic_bounds} shows bounds on the conditional
  expectation of mortality given education rank for women aged
  50--54 in the United States in 2015. The vertical lines show
  the bin boundaries and the points show the mean total mortality and child rank in
  each bin. The dashed lines
  show analytical bounds when the distribution of the $x$
  variable is unknown \cite{Manski2002}. The solid line shows analytical
  bounds when the distribution of the $x$ variable is uniform. }

\end{figure}

\begin{figure}[H]
  \caption{CEF of Mortality given Education Rank:
\cnewline
Bounds Under
    Different Constraint Assumptions}
  \label{fig:cef_f2_mort}
  \begin{center}

    \begin{tabular}{cc}
      Panel A: Monotonicity Only $(\overline{C} = \infty) $ & Panel B:
      Curvature Only $(\overline{C} \in \{2,3,5\}) $ \\
      \includegraphics[scale=0.40]{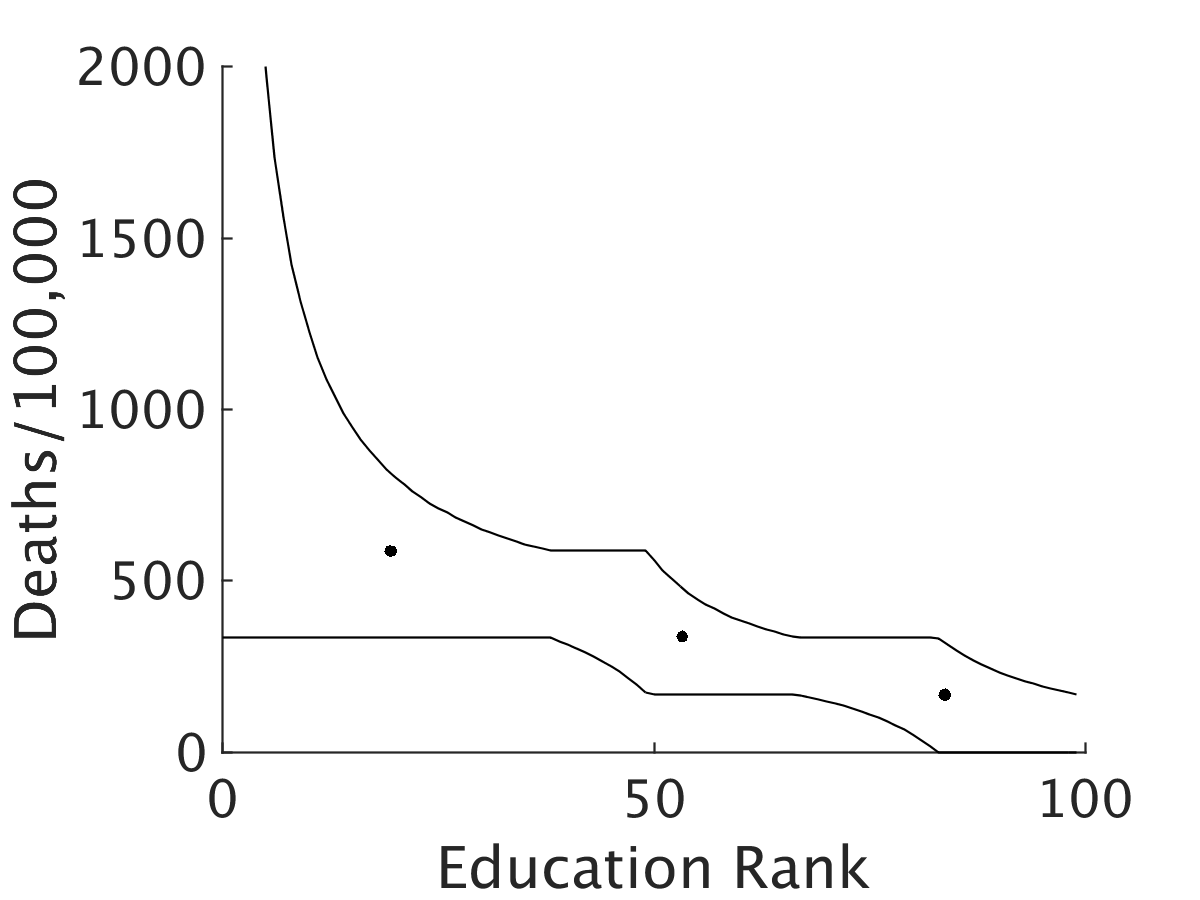} & \includegraphics[scale=0.40]{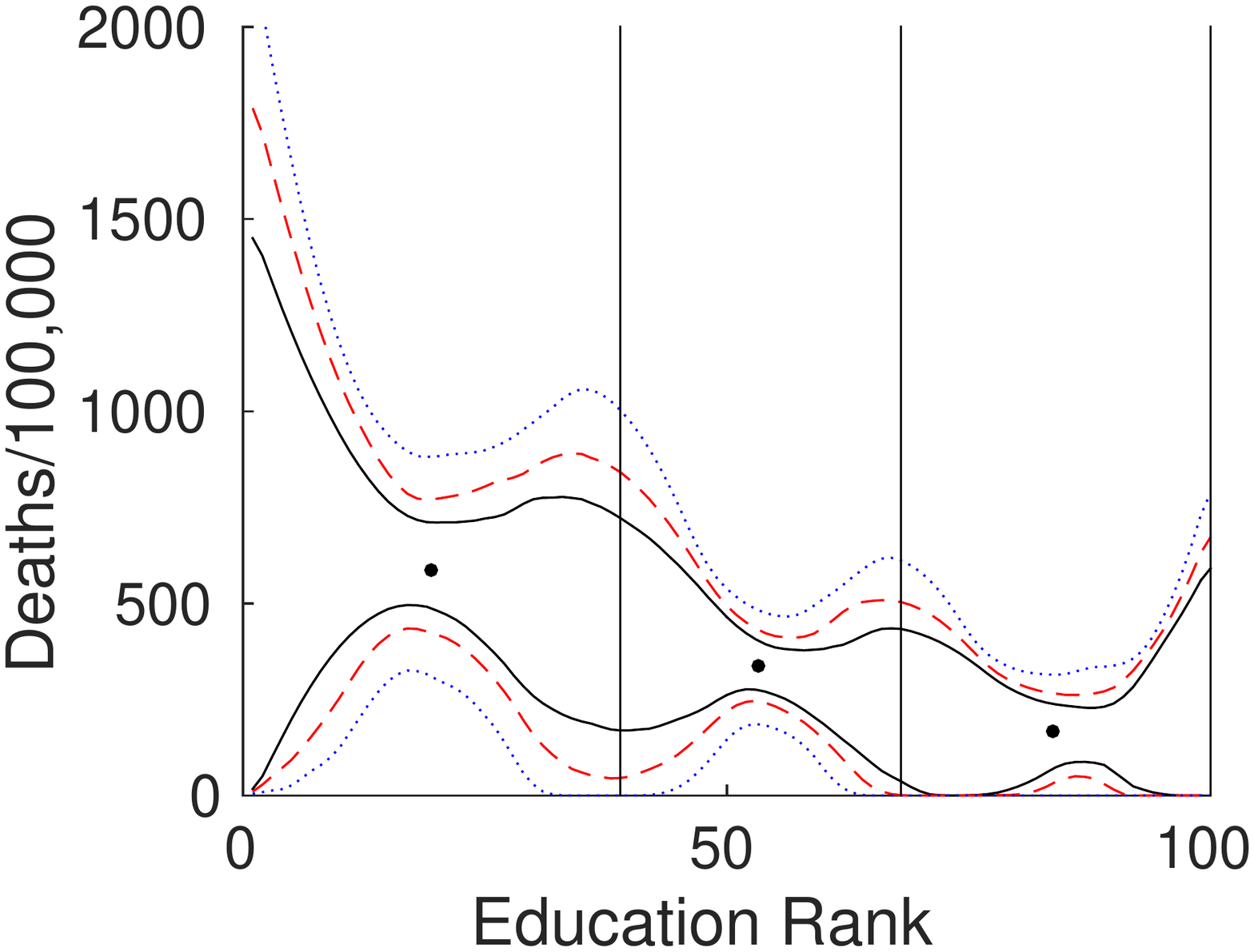} \\

      Panel C: Monotonicity and Curvature $(\overline{C} \in \{2,3,5\}) $ & Panel
      D: Linear Fit $(\overline{C} = 0)$ \\
      \includegraphics[scale=0.40]{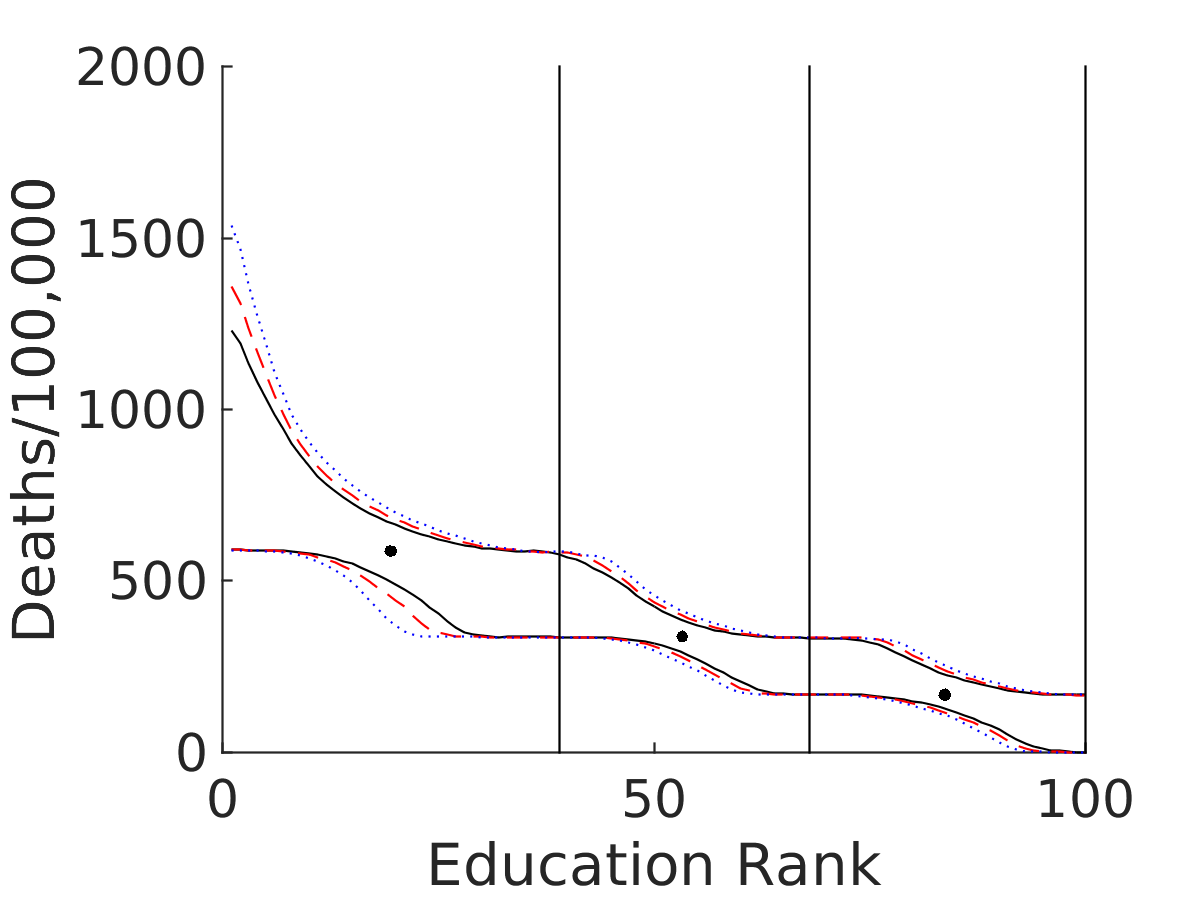} & \includegraphics[scale=0.40]{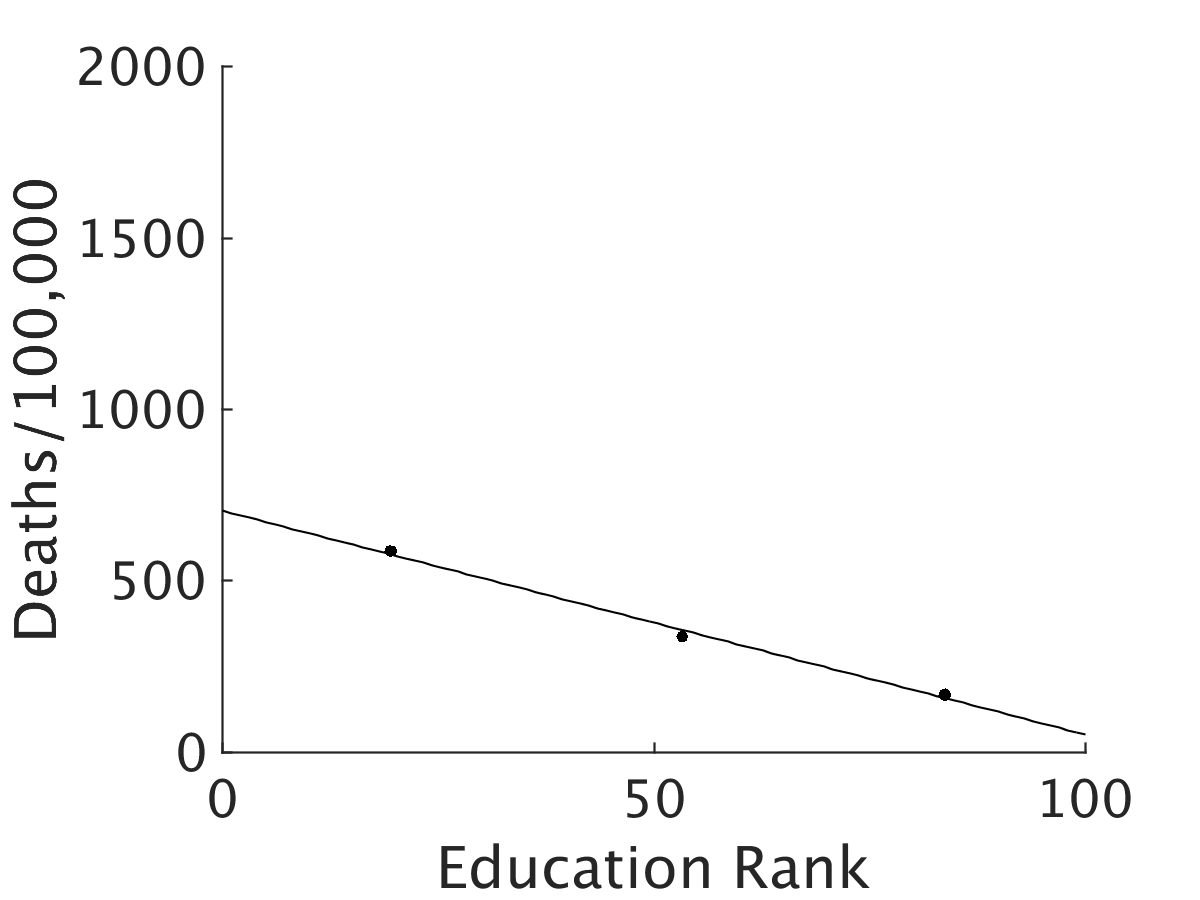} \\

      \hline

    \end{tabular}

  \end{center}
  \noindent
  \footnotesize{Figure \ref{fig:cef_f2_mort} presents bounds on
    conditional expectation functions of mortality given education
    ranks for women aged 50--54 in 2015 under different assumptions
    sets. Education rank is measured relative to the set of all women
    aged 50-54. The lines in each panel represent the upper and lower
    bounds on the CEF at each rank, obtained under different
    monotonicity or curvature restrictions. Panel A imposes
    monotonicity only. Panel B imposes curvature constraints only,
    with the solid, dashed and dotted lines respectively showing
    bounds with $\overline{C}=2$, $\overline{C}=3$ and
    $\overline{C}=5$. Panel C imposes monotonicity and curvature
    constraints, with the same limits as Panel B. Panel D imposes
    linearity, by setting $\overline{C}$ to zero.  The points show
    the mean mortality and education rank of women in each bin in the education
    distribution.}

\end{figure}

\newpage
\begin{figure}[H]
  \caption{Simulated Interval Censoring and Bounds using
    U.S. Mortality-Income Data}
  \label{fig:mort_bounds}
  \begin{center}
    
    \begin{tabular}{cc}
      Panel A: $\overline{C} = \infty$
      & Panel B: $\overline{C} = 20$
      \\
      \includegraphics[scale=0.40]{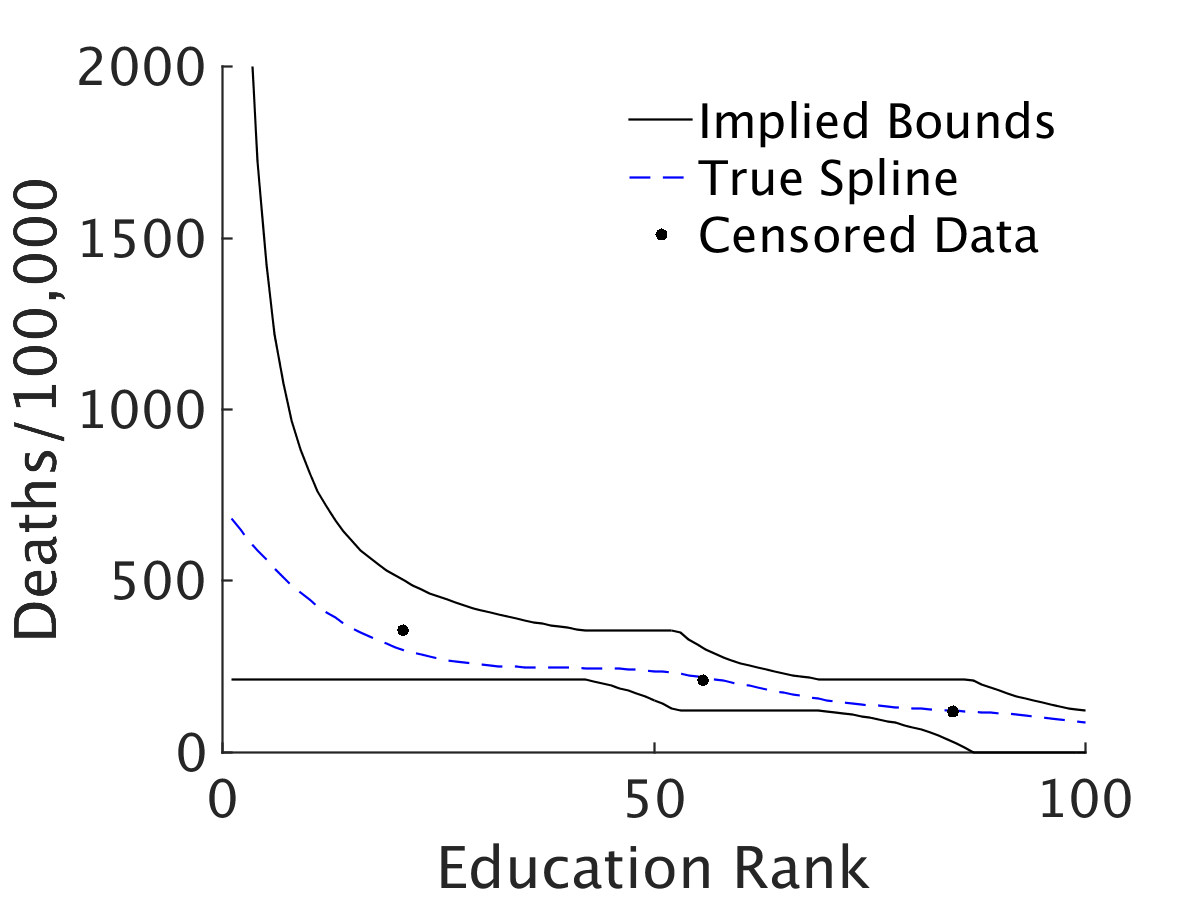}
      &
      \includegraphics[scale=0.40]{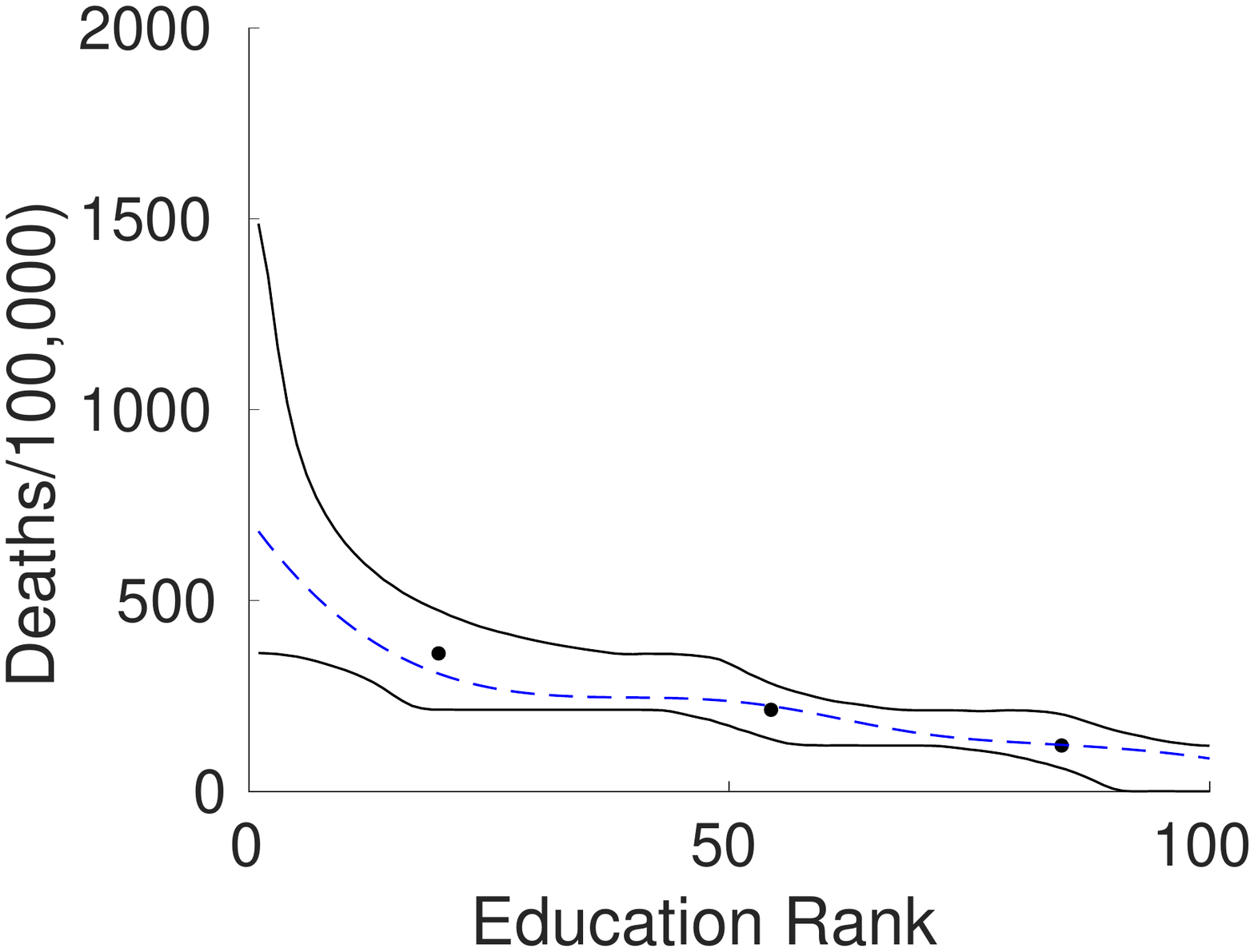}
      \\
      
      Panel C: $\overline{C} = 3 $
      & Panel D: $\overline{C} = 1 $
      \\
      \includegraphics[scale=0.40]{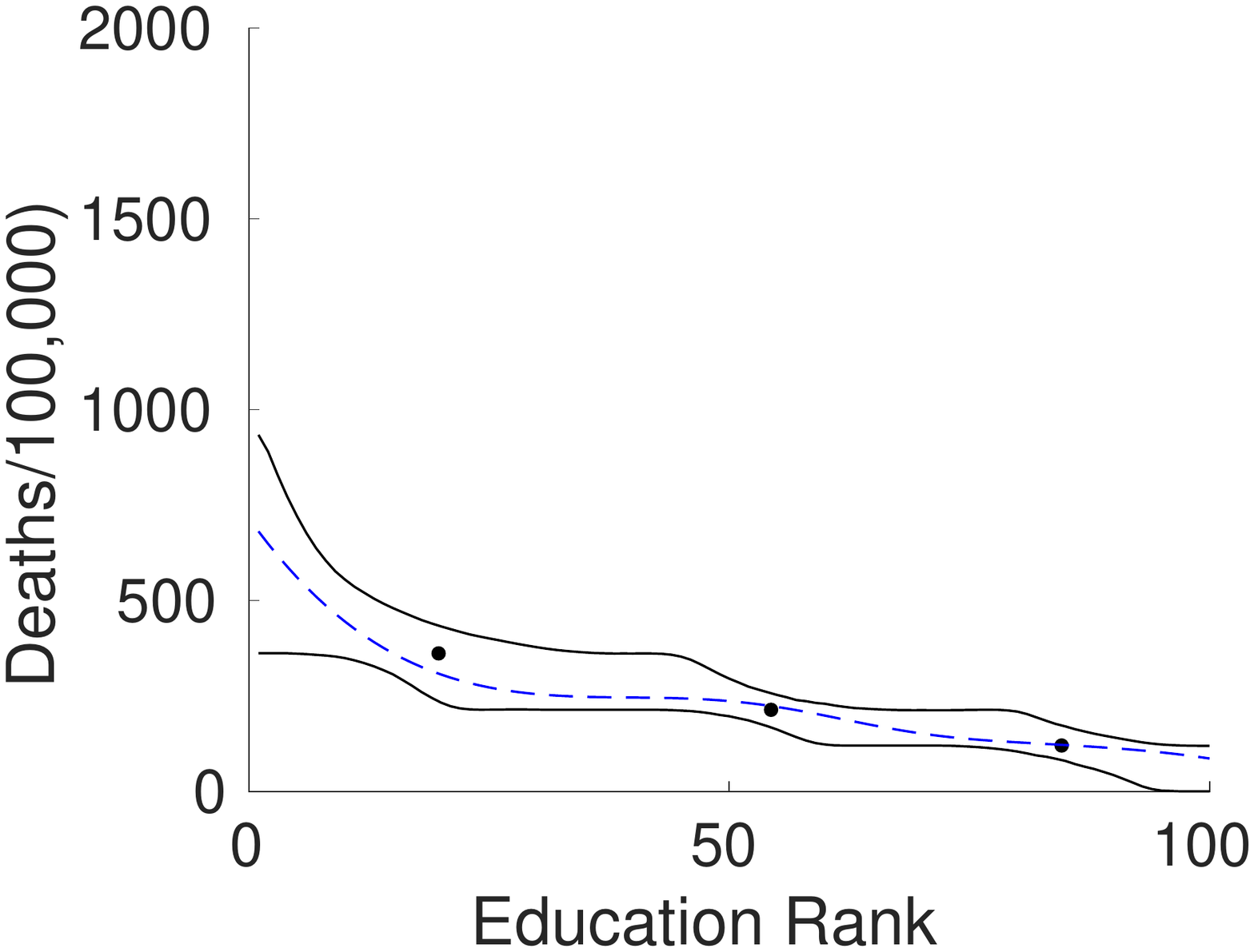}
      &
      \includegraphics[scale=0.40]{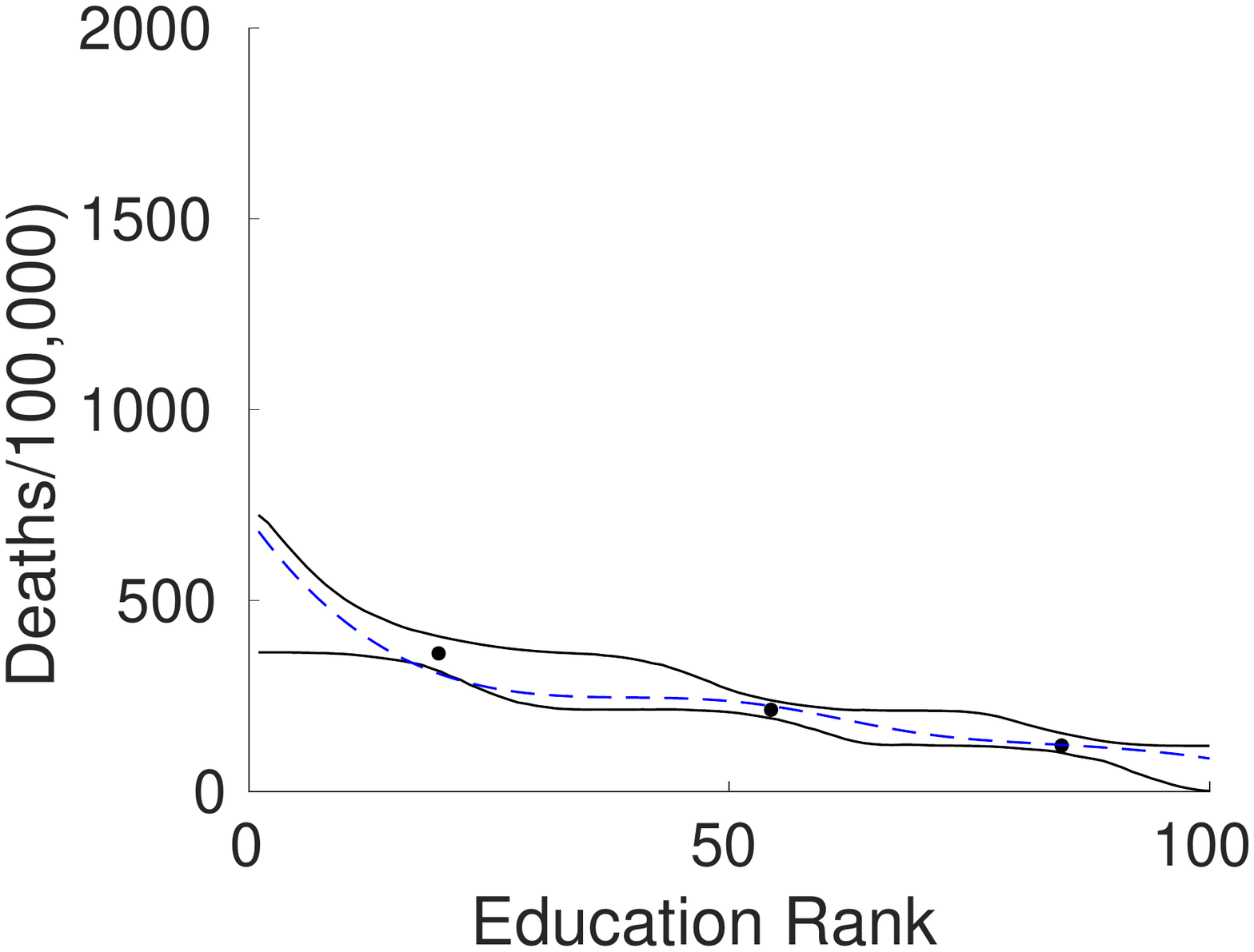}
      \\
      \hline
      
    \end{tabular}
  \end{center}
  \noindent
  
  \footnotesize{Figure \ref{fig:mort_bounds} shows results from a
    simulation using matched mortality-income rank data from
    \citeasnoun{Chetty2016b} in 2014 for women aged 52. We simulated
    interval censoring along the bin boundaries from the 2014
    education-mortality data, so that the only observable data were
    the points in the graphs, which show mean mortality and education rank
    in each
    education bin. We then calculated bounds under four
    different curvature constraints, indicated in the graph
    titles. The solid lines show the upper and lower bound of the CEF
    at each point in the parent distribution, and the dashed line
    shows the spline fit to the underlying data (described in
    Figure~\ref{fig:mort_splines}).}
\end{figure}

\begin{figure}[H]
\caption{Change in Total Mortality of U.S. Women, Age 50-54
  \cnewline Bounds on Conditional Expectation Functions}
\label{fig:mort_overlay}
\begin{center}

  \begin{tabular}{c}
    Panel A: Monotonicity Only \\ 

    \includegraphics[scale=.45]{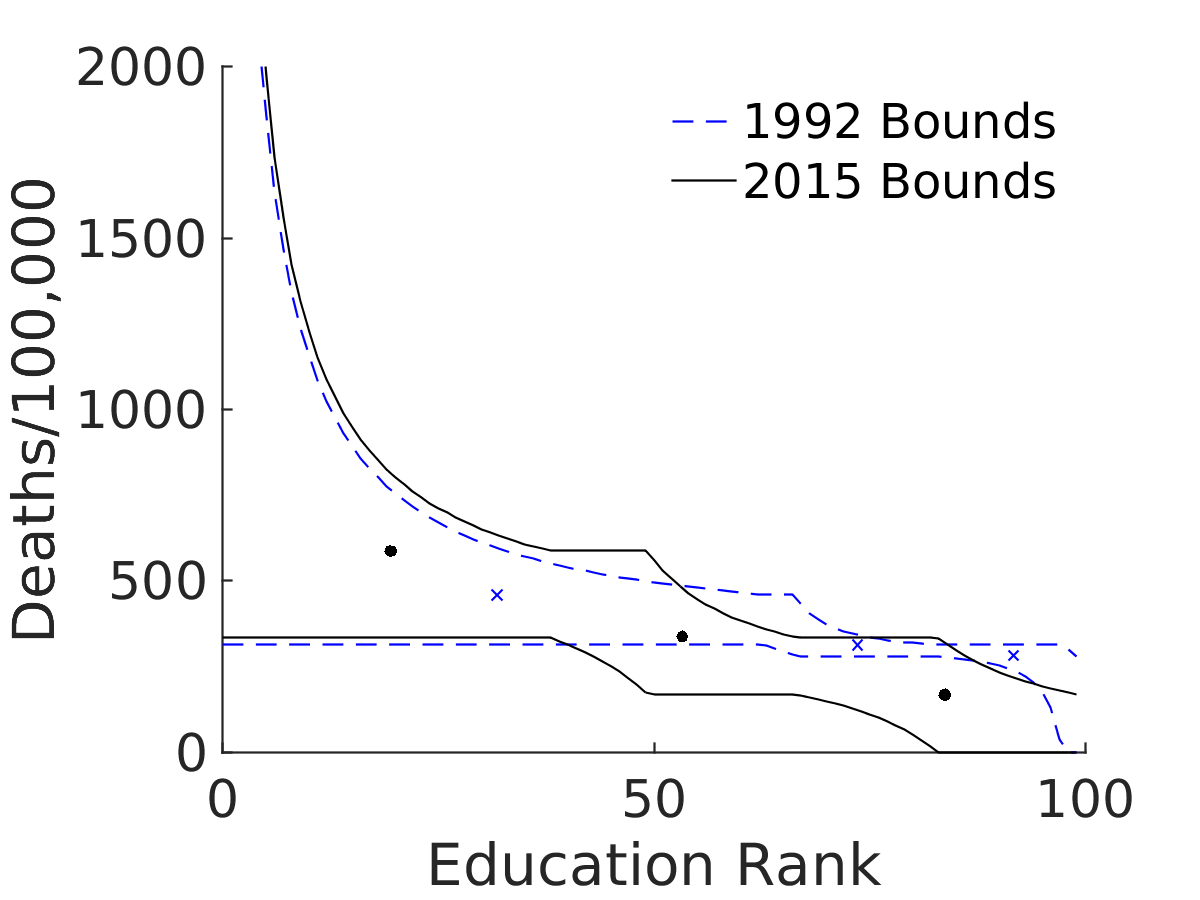}
    \\ 
    Panel B: Monotonicity and $\overline{C} = 3$
    \\
    \includegraphics[scale=.45]{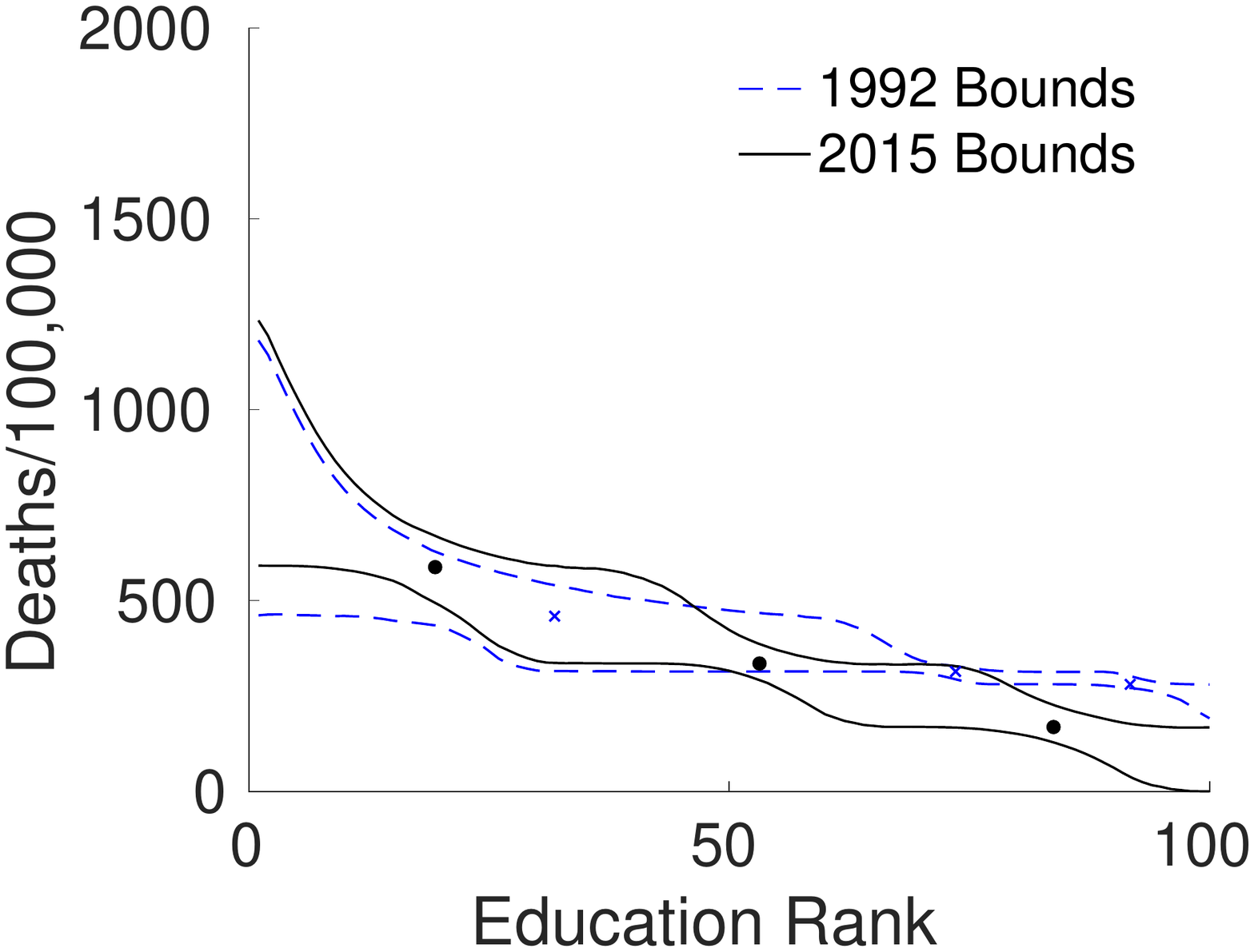} \\
    
    \hline

  \end{tabular}

\end{center}
\noindent
\footnotesize{Figure \ref{fig:mort_overlay} shows bounds on the conditional
  expectation function of mortality as a function of latent
  educational rank. The sample consists of U.S. women aged 50-54;
  mortality is measured in deaths per 100,000 women, and the graph
  shows the mean across the sample years 1992 and
  2015. Panel A shows analytical bounds with no curvature
  constraint. Panel B uses the curvature constraint
  suggested in Section~\ref{sec:method}. Education rank is measured
  relative to the set of all women aged 50-54.}

\end{figure}

\begin{figure}[H]
\caption{Changes in U.S. Mortality, Age 50-54, 1992-2015:
  \cnewline Constant Rank Interval Estimates (High School or Less in 1992)}
\label{fig:mort_change}
\begin{center}

  \begin{tabular}{cc}
    Panel A: Women (Total Mortality) & Panel B: Men (Total Mortality) \\

    \includegraphics[scale=.54]{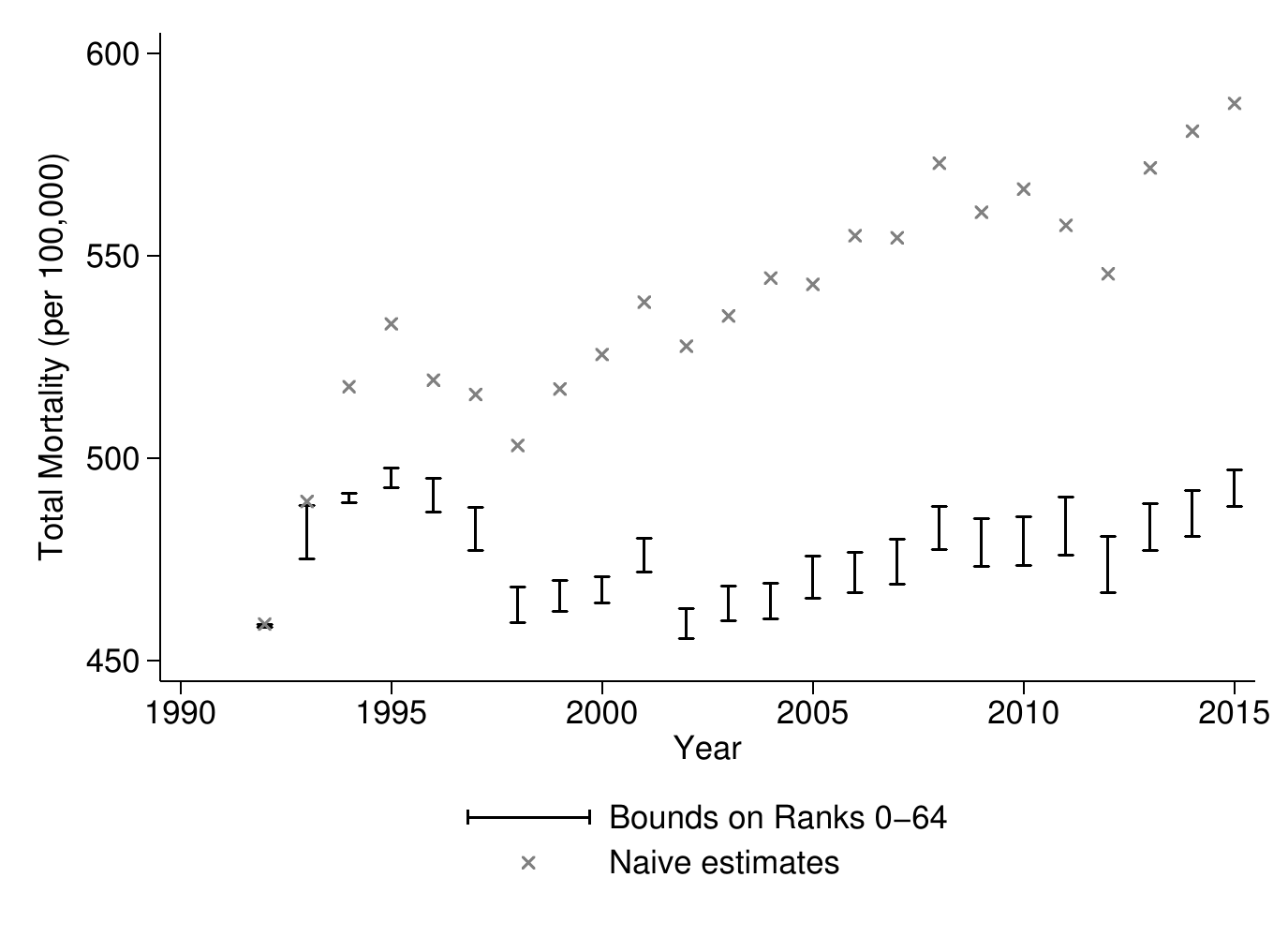}
    &
    \includegraphics[scale=.54]{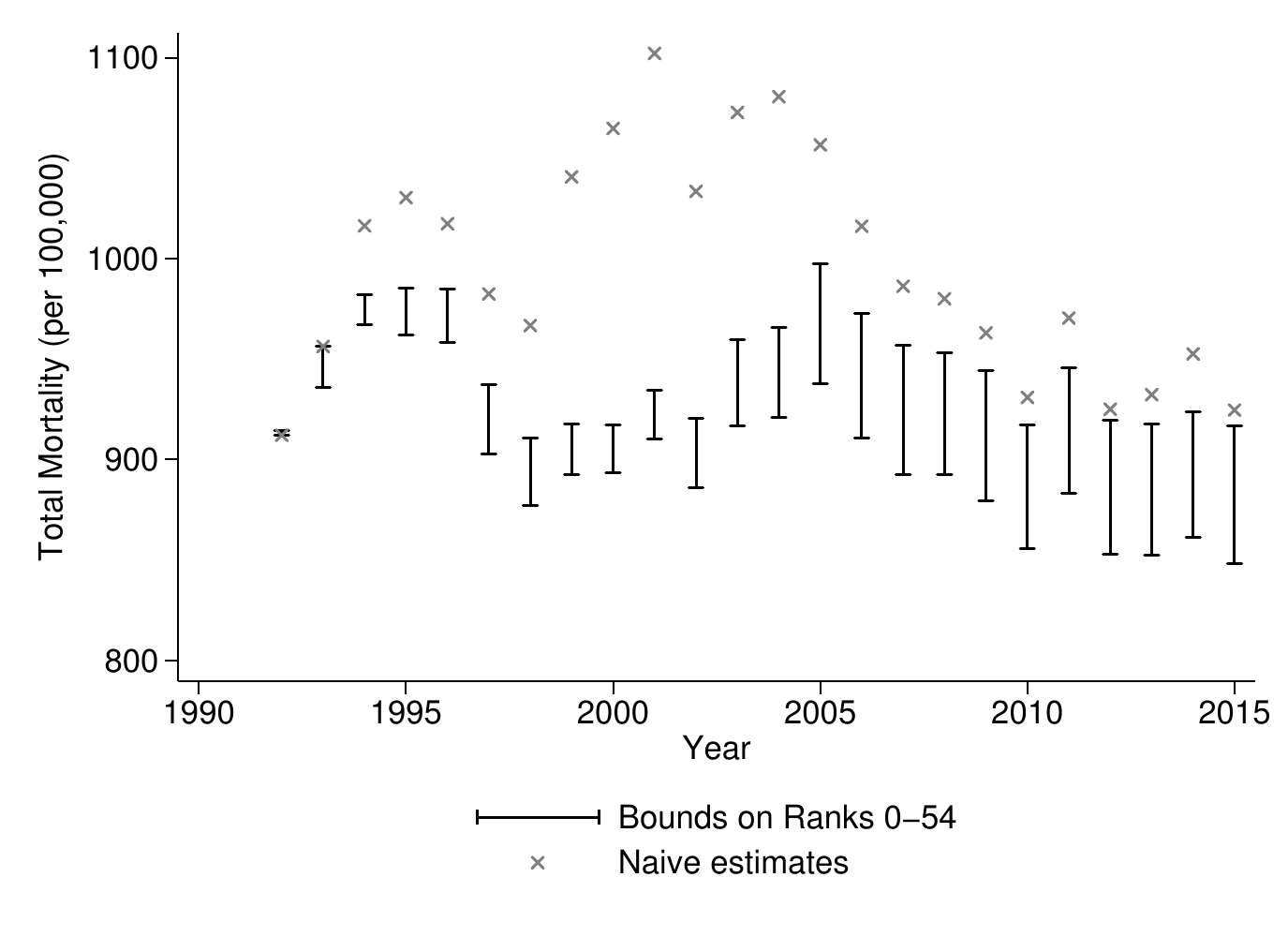}

    \\

    Panel C: Women (Deaths of Despair) & Panel D: Men (Deaths of Despair) \\

    \includegraphics[scale=.54]{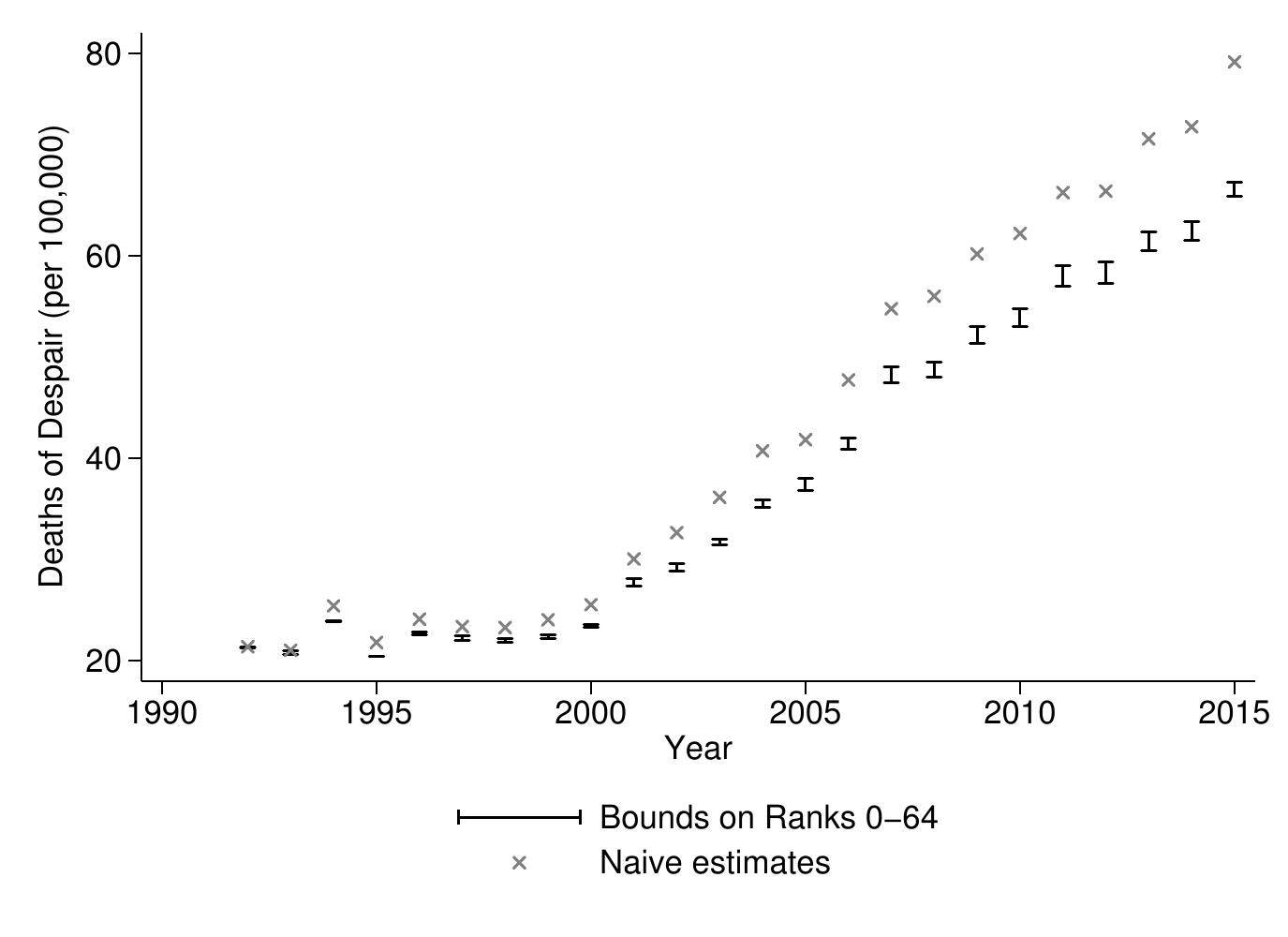}
    &
    \includegraphics[scale=.54]{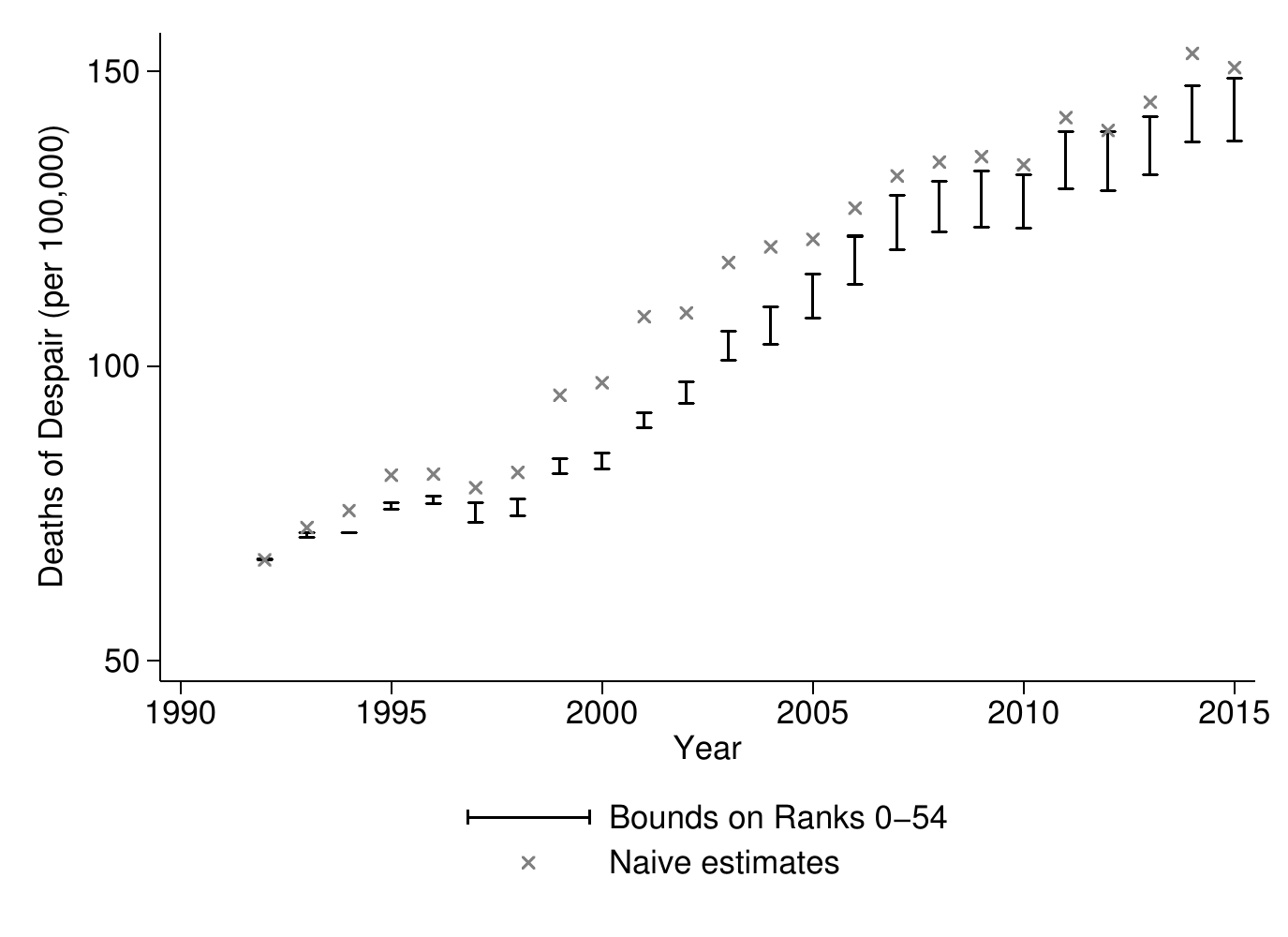}

    \\
    
    \hline

  \end{tabular}

\end{center}
\noindent
\footnotesize{Figure \ref{fig:mort_change} shows bounds on mortality
  change for less educated men and women aged 50--54 over time, as
  well unadjusted estimates. The points show the total mortality of
  men and women with high school education or less (LEHS), from
  1992--2015. The vertical lines show the bounds on mortality for the
  group of men or women who occupy a constant set of ranks
  corresponding to the ranks of men and women with LEHS in 1992. For
  women, these are ranks 0-64, and for men they are ranks 0-54. Panel
  A shows estimates for women age 50-54, and Panel B for men. Panels C
  and D show analogous plots for mortality deaths of despair for both
  groups, defined as deaths from suicide, poisoning or chronic liver
  disease. All bounds are calculated analytically under the
  assumptions of monotonicity and unconstrained curvature.}
\end{figure}

\begin{figure}[H]
  \caption{Changes in Intergenerational Educational Mobility in India \cnewline 
    from 1950s to 1980s Birth Cohorts}
  \label{fig:mob_changes}
  \begin{center}

    \begin{tabular}{c}
      Panel A: Rank Bin Midpoints
      \\
      \includegraphics[scale=0.45]{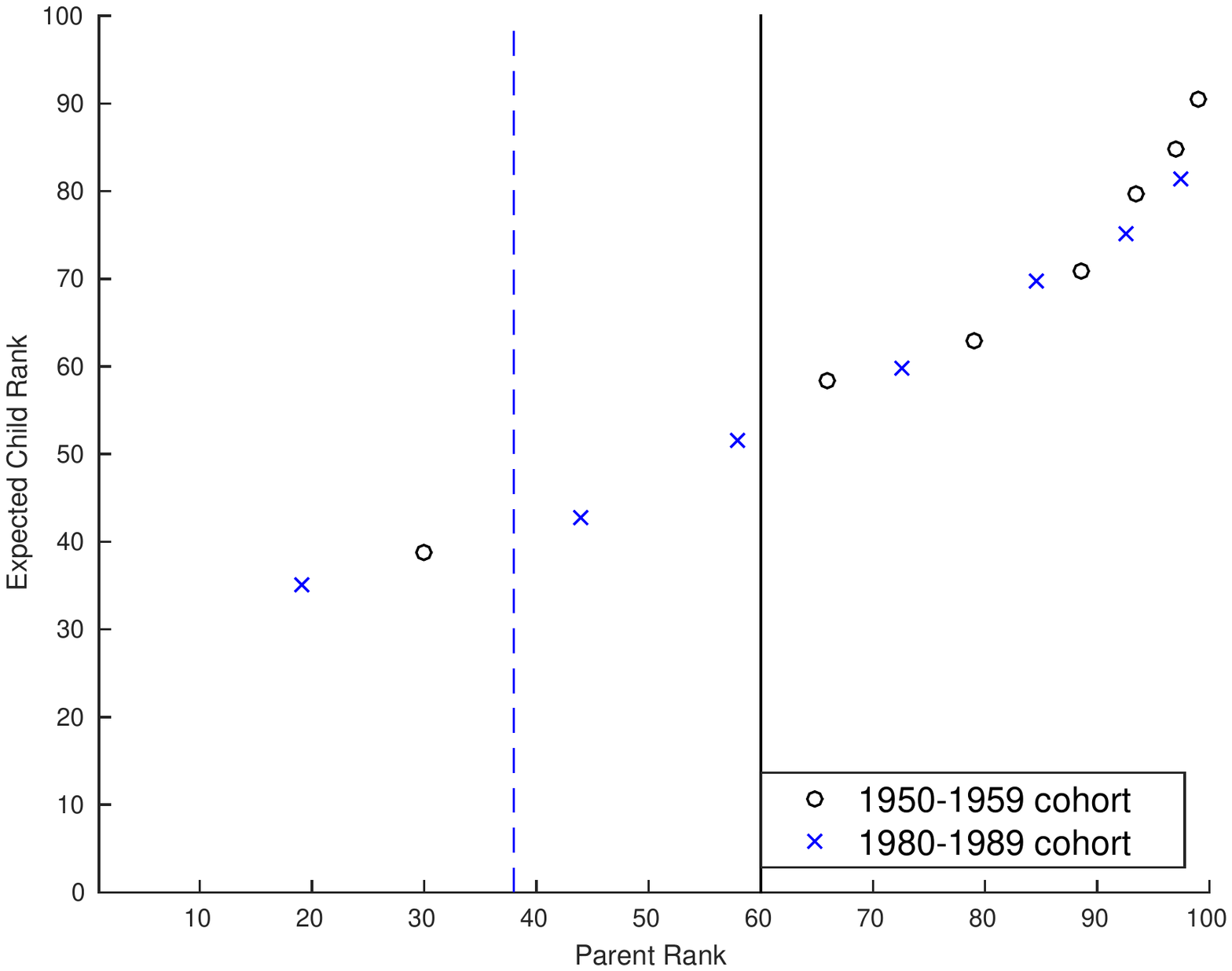} \\
      Panel B: CEF Bounds   \\
      \includegraphics[scale=0.45]{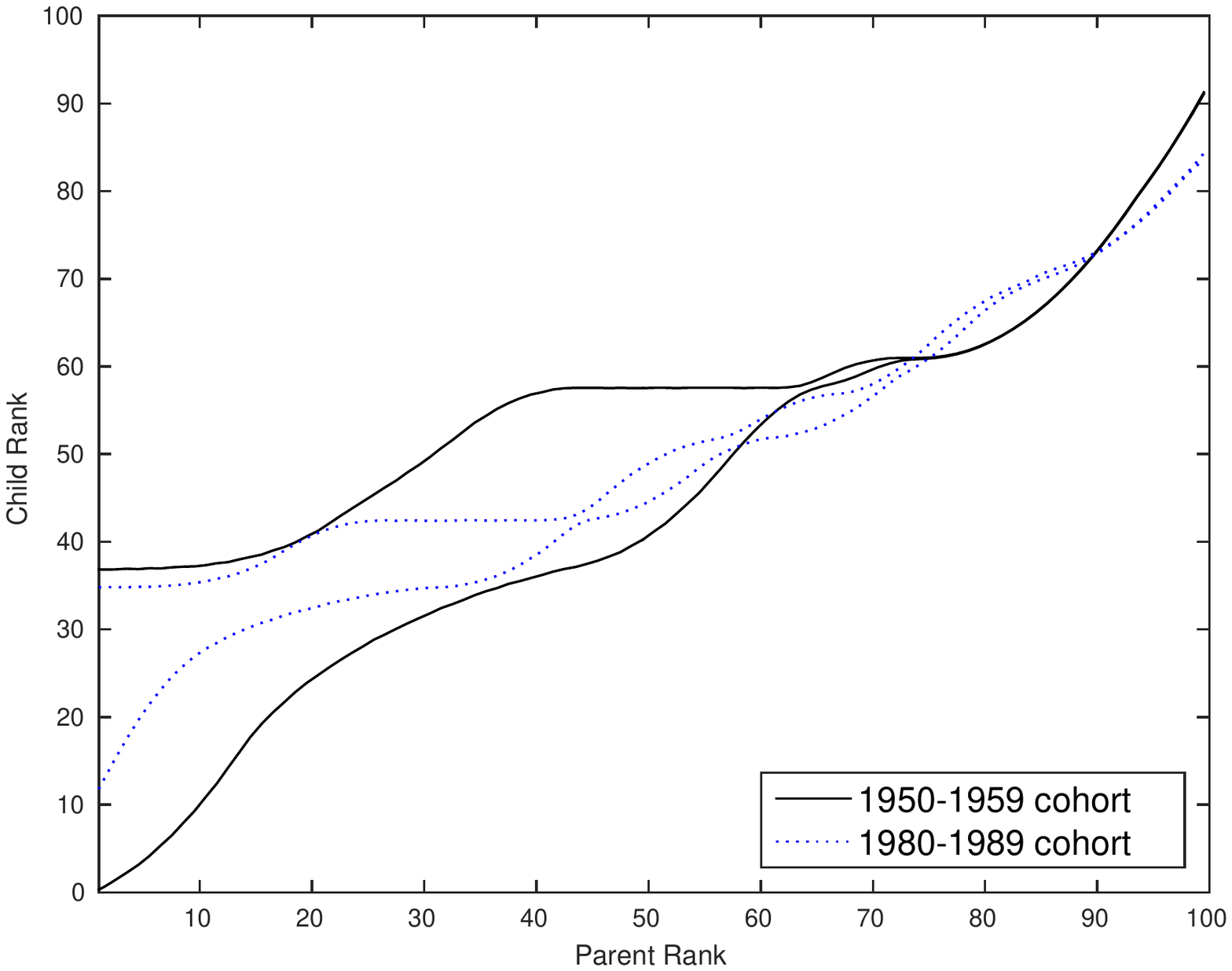}  \\

      \hline

    \end{tabular}

  \end{center}
  \footnotesize{Figure \ref{fig:mob_changes} presents the change over time in the
  rank-rank relationship between Indian fathers and sons born in the
  1950s and the 1980s. Panel A presents the
  raw bin means in the data. The vertical lines indicate the size of the
  lowest parent education rank bin, representing fathers with less
  than two years
  of education; the solid line shows this value for the 1950s cohort,
  and the dashed line for the 1980s cohort.  Panels B presents the
  bounds on the CEF of child rank at each
  parent rank, under the curvature constraint
  $\overline{C}=0.10$.}

\end{figure}

\begin{figure}[H]
  \caption{Mobility Bounds for 1950s to 1980s Birth Cohorts}
  \label{fig:mob_time_stats}
  \begin{center}
    \begin{tabular}{c}

      Panel A: Rank-Rank Gradient
      \\
      \includegraphics[scale=0.80]{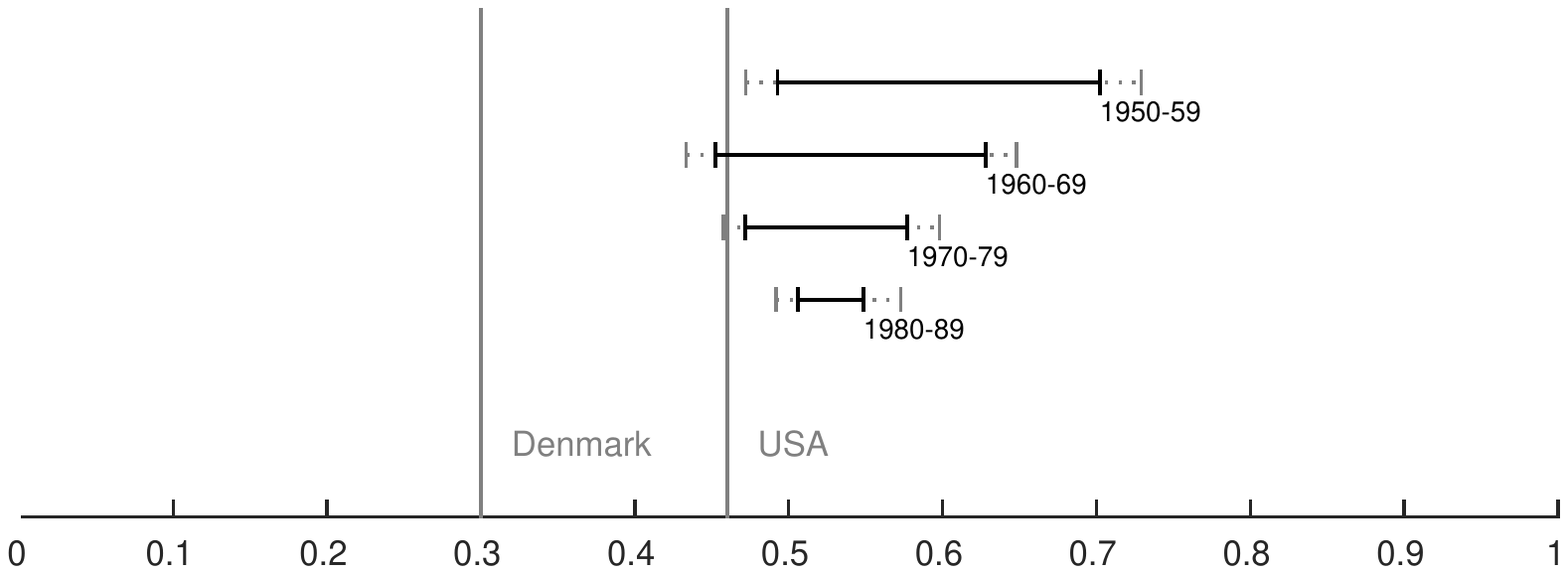} \\
      Panel B: Absolute and Interval Mobility: $p_{25}$ and $\mu_0^{50}$ 
      \\
      \includegraphics[scale=0.80]{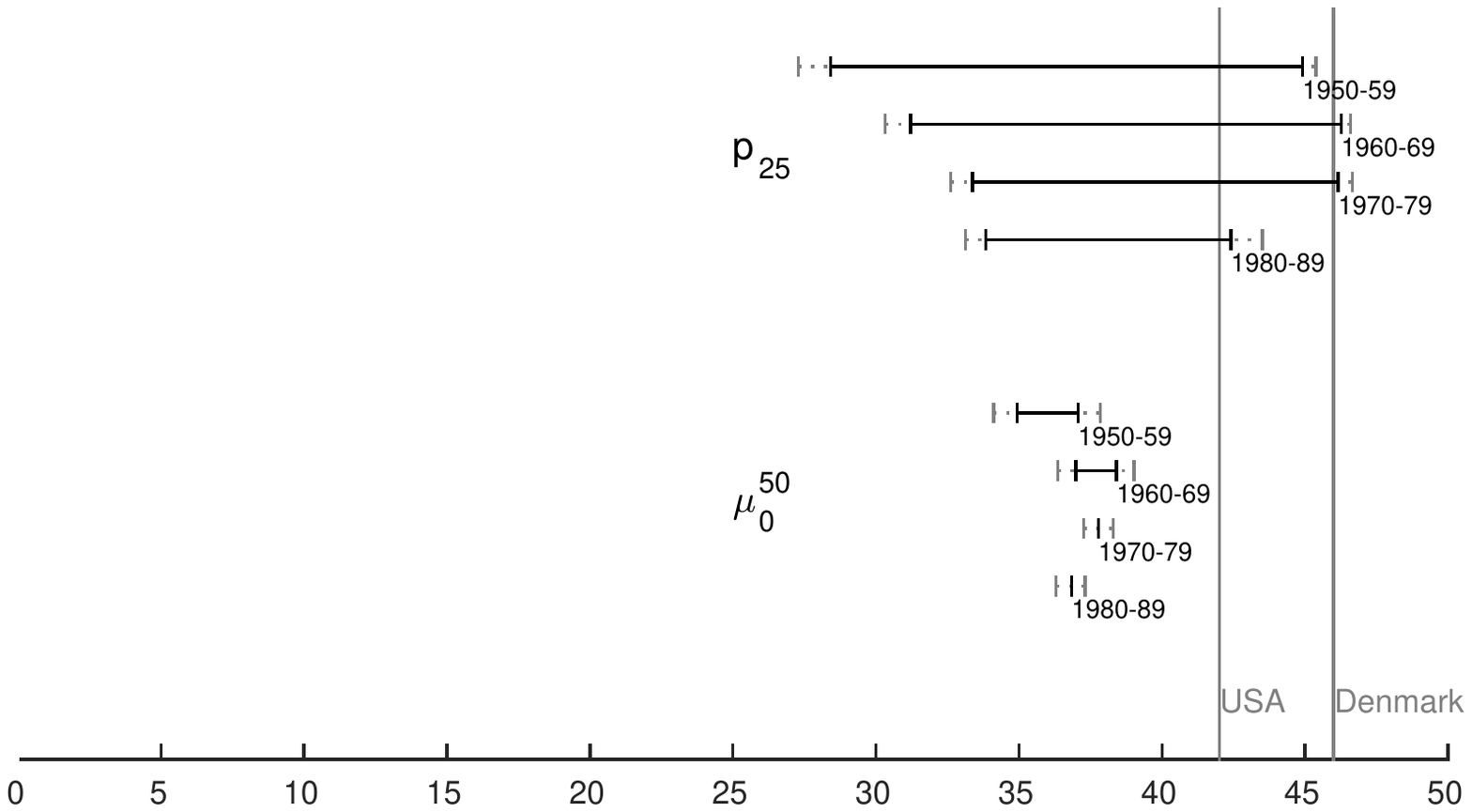} \\

    \end{tabular}
  \end{center}

\footnotesize{  Figure~\ref{fig:mob_time_stats} shows bounds on three mobility
  statistics, estimated on four decades of matched Indian father-son
  pairs. The solid lines show the estimated bounds on each statistic and
  the gray dashed lines show the 95\% bootstrap confidence sets, based on
  1000 bootstrap samples. Each of
  these statistics was calculated using monotonicity and the curvature
  constraint $\overline{C}=0.10$. For reference, we display the
  rank-rank education gradient for USA and Denmark (from
  \citeasnoun{Hertz2008}), and $p_{25}$ for USA and Denmark (from
  \citeasnoun{Chetty2014c}). The rank-rank
  gradient is the slope coefficient from a regression of son education
  rank on father education rank. $p_{25}$ is absolute upward mobility,
  which is the expected rank of a son born to a family at the 25th
  percentile. $\mu_{0}^{50}$ is  upward interval mobility, which is
  the expected rank of a son born below to a family below the 50th
  percentile. }

\end{figure}

\begin{landscape}
\begin{table}[H]
  \caption{Sample Statistics for Mortality of Less Educated Women Ages 50--54}
  \label{tab:ests_mortality}                  

\centering

\begin{center}
\textbf{Panel A: 1992} 
\end{center}

\small{\begin{tabular}{lcccc}
\hline
Statistic                    &  Monotonicity Only        & Curvature Only     & Monotonicity and           & Linear Fit          \\
                             &  $(\overline{C}=\infty)$  & $(\overline{C}=3)$ & Curvature $\overline{C}=3$ & $(\overline{C}=0)$  \\
\hline
$p_{10}$: First Quintile Median              & [314.0, 1236.1]  & [223.8, 1008.0]  & [453.9, 813.9]  & 526.4 \\
$p_{25}$: Bottom Half Median                 & [314.0, 683.1]  & [226.1, 738.7]  & [384.9, 585.5]  & 479.3 \\
$p_{32}$: Median $\leq$ High School (1992)   & [314.0, 602.4]  & [176.9, 694.9]  & [346.1, 536.3]  & 457.3 \\
$p_{19}$: Median $\leq$ High School (2015)   & [314.0, 799.6]  & [253.4, 750.3]  & [421.2, 642.8]  & 498.1 \\
\rule{0pt}{2ex}  & & & & \\
$\mu_0^{20}$: First Quintile Mean            & [459.0, 775.5] & [11.2, 1287.5] & [467.6, 749.0] & 526.4 \\
$\mu_0^{50}$: Bottom Half Mean               & [459.0, 498.6] & [423.1, 565.1] & [460.7, 496.2] & 479.3 \\
$\mu_0^{64}$: Mean $\leq$ High School (1992) & [458.2, 459.0] & [459.0, 459.0] & [459.0, 459.0] & 457.3 \\
$\mu_0^{39}$: Mean $\leq$ High School (2015) & [459.0, 550.7] & [316.5, 608.4] & [462.1, 544.9] & 496.5 \\
\hline
\end{tabular}

}

\begin{center}
\textbf{Panel B: 2015} 
\end{center}

\small{\begin{tabular}{lcccc}
\hline
Statistic                    &  Monotonicity Only        & Curvature Only     & Monotonicity and           & Linear Fit          \\
                             &  $(\overline{C}=\infty)$  & $(\overline{C}=3)$ & Curvature $\overline{C}=3$ & $(\overline{C}=0)$  \\
\hline
$p_{10}$: First Quintile Median               & [335.1, 1313.5]  & [473.4, 899.1]  & [577.2, 846.8]  & 640.8      \\
$p_{25}$: Bottom Half Median                  & [335.1, 726.7]  & [345.3, 718.2]  & [376.9, 628.8]  & 542.6      \\
$p_{32}$: Median $\leq$ High School (1992)    & [335.1, 641.1]  & [260.5, 721.5]  & [345.3, 593.8]  & 496.7      \\
$p_{19}$: Median $\leq$ High School (2015)    & [335.1, 850.3]  & [434.0, 746.4]  & [464.5, 672.8]  & 581.9      \\
\rule{0pt}{2ex}  & & & & \\
$\mu_0^{20}$: First Quintile Mean              & [587.7, 824.8] & [326.5, 956.9] & [590.0, 806.9] & 640.8 \\
$\mu_0^{50}$: Bottom Half Mean                 & [531.0, 587.7] & [523.1, 576.9] & [534.1, 567.5] & 542.6 \\
$\mu_0^{64}$: Mean $\leq$ High School (1992)   & [488.2, 497.3] & [488.4, 498.4] & [490.2, 498.0] & 496.7 \\
$\mu_0^{39}$: Mean $\leq$ High School (2015)   & [586.3, 587.7] & [587.5, 587.5] & [587.5, 587.5] & 578.6 \\
\hline
\end{tabular}

}

\end{table}
\noindent \footnotesize{Table \ref{tab:ests_mortality} presents bounds
  on various mortality statistics under different constraints.  The last column
  in each panel presents point estimates obtained from the best linear
  approximation to the mean mortality observed
  in each bin. $p_x$ is the value of the CEF at $x$; $\mu_a^b$ is the
  average value of the CEF between points $a$ and $b$. Panel A
  presents statistics for women in 1992, and Panel B for 2015.}
\end{landscape}

  \begin{table}[H]
    \caption{Simulation Results: \cnewline Mortality at Different Income Ranks}
    \label{tab:sims_stats}
\centering 
\begin{tabular}[t]{lcccc}
 & Value from &  &  \\ 
Statistic & Linear Fit & True Value & $\overline{C} = \infty$ \\ 
$p_{10}$: First Quintile Median           & 393.4 & 444.9 & $[213.7, 797.3]$  \\ 
$p_{25}$: Bottom Half Median              & 337.1 & 272.3 & $[213.7, 447.1]$  \\ 
$p_{32}$: Median $\leq$ High School (1992) & 310.8 & 251.0 & $[213.7, 396.1]$  \\ 
$p_{19}$: Median $\leq$ High School (2015) & 359.6 & 315.4 & $[213.7, 520.9]$  \\ 
$\mu_0^{20}$: First Quintile Mean            & 393.4 & 456.7 & $[361.4, 505.5]$  \\ 
$\mu_0^{50}$: Bottom Half Mean               & 337.1 & 335.3 & $[330.4, 361.4]$  \\ 
$\mu_0^{64}$: Mean $\leq$ High School (1992)  & 310.8 & 307.6 & $[304.9, 312.5]$  \\ 
$\mu_0^{39}$: Mean $\leq$ High School (2015)  & 357.7 & 361.4 & $[361.4, 363.3]$  \\ 
\end{tabular}%
\begin{tabular}[t]{*{1}{c}}
   \\ 
$\overline{C} = 3$ \\ 
$[ 352.6, 519.5 ]$ \\ 
$[ 285.6, 393.7 ]$ \\ 
$[ 236.3, 367.3 ]$ \\ 
$[ 323.8, 429.5 ]$ \\ 
$[ 362.2, 500.1 ]$ \\ 
$[ 332.2, 353.7 ]$ \\ 
$[ 306.4, 311.7 ]$ \\ 
$[ 361.9, 364.9 ]$ \\ 
\hline
\end{tabular}
\end{table}
\footnotesize{Table~\ref{tab:sims_stats} presents bounds on mortality
  statistics computed in a simulation exercise. We begin with
  mortality-income rank data on women aged 52 in 2014 from
  \citeasnoun{Chetty2016b} and compute the best-fit spline to the data
  to obtain a close estimate of the true CEF for this distribution. We
  then simulate interval censoring according to the education bins for
  women aged 50--54 in 2014 used elsewhere in the paper. We then
  compute bounds on mortality statistics obtained data with simulated
  censoring. $p_x$ is the value of the CEF at $x$; $\mu_a^b$ is the
  average value of the CEF between points $a$ and $b$.}


\begin{landscape}
\begin{table}[H]
  \caption{Unadjusted and Constant Rank Estimates of Mortality
    Changes for \cnewline Women 50-54,
    1992 to 2015}
  \label{tab:mort_changes}
  \centering 
  \begin{tabular}{lcccccc} 
    & \multicolumn{2}{c}{\underline{$\leq$ High School}} & \multicolumn{2}{c}{\underline{Some College}} & \multicolumn{2}{c}{\underline{B.A. or Higher}} \\
Age & Unadjusted & Constant Rank & Unadjusted & Constant Rank & Unadjusted & Constant Rank          \\ 
    & Estimate   & Bounds        & Estimate   & Bounds        & Estimate   & Bounds                 \\
\hline
25-29  &  42.8  &   [ 16.1, 24.1 ] &  9.6 &   [ -28.0, 6.3 ] &  -8.8 &   [ -28.6, -8.8 ]  \\ 
30-34  &  46.7  &   [ 17.6, 27.4 ] &  22.2 &   [ -22.6, 21.3 ] &  -9.0 &   [ -42.5, -9.0 ]  \\ 
35-39  &  33.1  &   [ 8.9, 26.3 ] &  24.4 &   [ -39.0, 24.4 ] &  -17.0 &   [ -53.9, -17.0 ]  \\ 
40-44  &  47.6  &   [ 16.4, 44.2 ] &  35.8 &   [ -55.7, 35.8 ] &  -34.5 &   [ -82.9, -34.5 ]  \\ 
45-49  &  74.9  &   [ 22.3, 46.6 ] &  33.7 &   [ -80.4, 33.7 ] &  -56.9 &   [ -127.5, -56.9 ]  \\ 
50-54  &  128.6  &   [ 30.0, 40.1 ] &  21.1 &   [ -145.5, 21.1 ] &  -111.8 &   [ -272.8, -111.8 ]  \\ 
55-59  &  84.6  &   [ -47.4, -37.2 ] &  39.3 &   [ -168.6, 39.3 ] &  -158.4 &   [ -398.9, -158.4 ]  \\ 
60-64  &  5.0  &   [ -177.8, -168.5 ] &  -89.7 &   [ -372.7, -89.7 ] &  -242.1 &   [ -625.3, -242.1 ]  \\ 
65-69  &  -82.7  &   [ -266.8, -293.7 ] &  -135.9 &   [ -475.9, -328.0 ] &  -627.1 &   [ -1093.1, -627.1 ]  \\ 
\hline \\ 
\end{tabular}

\end{table}  
\footnotesize{Table \ref{tab:mort_changes} compares unadjusted
  estimates to bounds on total mortality changes for less educated
  women age 50-54.  The unadjusted estimate is the change in total
  mortality of women with high school education or less (LEHS), from
  1992--2015.  The bounds describe the mortality change for the group
  of women who occupy ranks 0-64 in the education distribution, which
  are the ranks occupied by LEHS women in 1992. The lower bound on
  mortality increase is the lower bound in 2015 minus the upper bound
  in 1992, and vice versa for the upper bound on mortality
  increase. Bounds are computed analytically under the assumptions of
  monotonicity and unconstrained curvature.}
\end{landscape}


\begin{table}[H]                                                    
  \caption{Bounds on Intergenerational Educational Mobility in India} 
  \label{tab:ests_india}
  \begin{center}            

    \small {
      \begin{tabular}{l*{6}{c}}

        & \multicolumn{3}{c}{$\overline{C} = \infty$}  & 
        \multicolumn{3}{c}{$\overline{C} = 0.20$}    \\
        Cohort & 
        Gradient & $p_{25}$ & $\mu_0^{50}$ & 
        Gradient & $p_{25}$ & $\mu_0^{50}$ \\
        \hline
        1950-59  & [0.457,  0.742] & [13.0,  58.3] & [34.8,  38.7]  & [0.474,  0.722] & [25.5,  48.1] & [34.8,  37.9] \\ 
         & (0.447,  0.763) & (10.2,  59.8) & (34.0,  39.0)  & (0.464,  0.745) & (24.2,  48.3) & (34.0,  38.4) \\ 
1960-69  & [0.436,  0.655] & [22.2,  54.5] & [37.0,  39.1]  & [0.444,  0.639] & [29.3,  49.3] & [37.0,  38.8] \\ 
         & (0.421,  0.677) & (19.7,  54.8) & (36.3,  39.5)  & (0.429,  0.661) & (28.0,  49.5) & (36.3,  39.3) \\ 
1970-79  & [0.463,  0.595] & [29.0,  48.6] & [37.8,  37.8]  & [0.468,  0.584] & [32.2,  48.3] & [37.8,  37.8] \\ 
         & (0.455,  0.616) & (26.8,  49.7) & (37.3,  38.0)  & (0.461,  0.603) & (31.9,  49.1) & (37.3,  38.0) \\ 
1980-89  & [0.500,  0.565] & [32.3,  42.3] & [36.8,  36.8]  & [0.505,  0.556] & [33.3,  42.8] & [36.8,  36.8] \\ 
         & (0.488,  0.591) & (30.2,  43.6) & (36.4,  37.3)  & (0.492,  0.582) & (32.8,  43.6) & (36.4,  37.3) \\ 

        \hline
        \\

        & \multicolumn{3}{c}{$\overline{C} = 0.10$}    & 
        \multicolumn{3}{c}{$\overline{C} = 0$}       \\
        Cohort & 
        Gradient & $p_{25}$ & $\mu_0^{50}$ & 
        Gradient & $p_{25}$ & $\mu_0^{50}$ \\
        \hline
        1950-59  & [0.492,  0.702] & [28.4,  44.9] & [34.9,  37.1]  & 0.587 & 35.6 & 35.6 \\ 
         & (0.480,  0.727) & (27.2,  45.3) & (34.0,  37.8)  & (0.577,  0.595) & (35.4,  35.9) & (35.4,  35.9) \\ 
1960-69  & [0.452,  0.629] & [31.2,  46.3] & [37.0,  38.5]  & 0.538 & 36.8 & 36.8 \\ 
         & (0.436,  0.629) & (31.1,  46.5) & (36.9,  38.9)  & (0.530,  0.553) & (36.5,  37.0) & (36.5,  37.0) \\ 
1970-79  & [0.472,  0.577] & [33.4,  46.2] & [37.8,  37.8]  & 0.534 & 36.9 & 36.9 \\ 
         & (0.465,  0.597) & (32.7,  46.5) & (37.3,  38.0)  & (0.524,  0.549) & (36.6,  37.2) & (36.6,  37.2) \\ 
1980-89  & [0.506,  0.548] & [33.8,  42.4] & [36.8,  36.8]  & 0.537 & 36.8 & 36.8 \\ 
         & (0.494,  0.575) & (33.3,  43.1) & (36.3,  37.3)  & (0.523,  0.551) & (36.5,  37.2) & (36.5,  37.2) \\

        \hline\hline
    \end{tabular}}
  \end{center}                                                        
\end{table}                                                         

The table shows estimates of bounds on three scalar mobility
statistics, for different decadal cohorts and under different
restrictions $\overline{C}$ on the curvature of the child rank
conditional expectation function given parent rank. The rank-rank
gradient is the slope coefficient from a regression of son education
rank on father education rank. $p_{25}$ is absolute upward mobility,
which is the expected rank of a son born to a family at the 25th
percentile. $\mu_{0}^{50}$ is  upward interval mobility, which is
the expected rank of a son born below to a family below the 50th
percentile. When $\overline{C}=0$, the bounds shrink to point
estimates.  Bootstrap 95\% confidence sets are displayed in
parentheses below each estimate based on 1000 bootstrap samples.

\newpage
\section{Appendix A: Additional Tables and Figures}

\setcounter{table}{0}
\renewcommand{\thetable}{A\arabic{table}}
\setcounter{figure}{0}
\renewcommand{\thefigure}{A\arabic{figure}}

\begin{figure}[H]
  \caption{Spline Approximations to the Empirical Mortality-Income CEF
  \cnewline 52-Year-Old Women in 2014 }
  \label{fig:mort_splines}

  \begin{center}
      \includegraphics[scale=0.75]{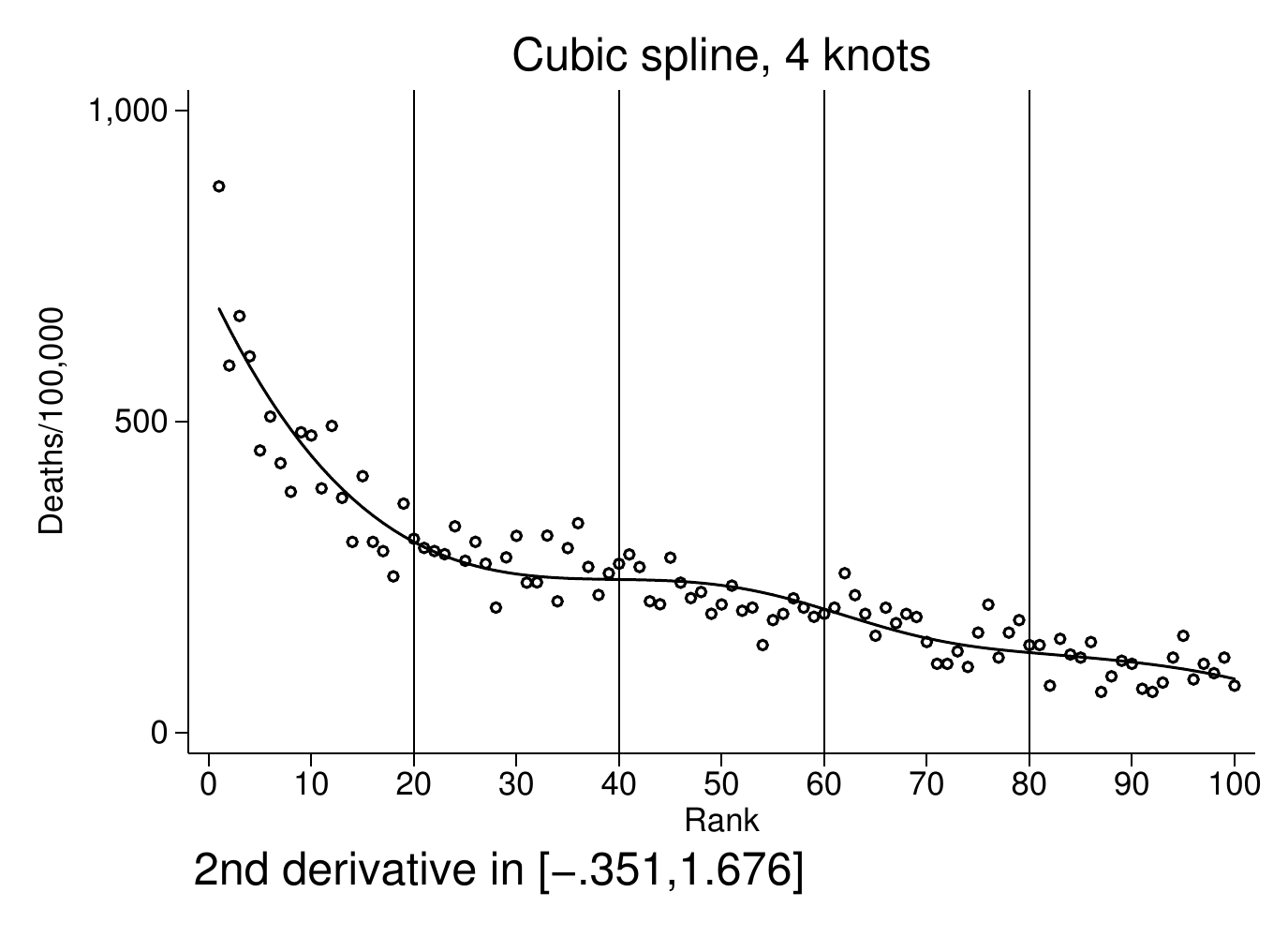}
  \end{center}

  \noindent
  \footnotesize{Figure \ref{fig:mort_splines} presents estimates of the
    conditional expectation function of U.S. mortality given income
    rank, using data from  \citeasnoun{Chetty2016b}. The CEF is
    fitted using a four-knot cubic spline. The function plots the best cubic spline
    fit to the data series, and the circles plot the underlying
    data. The text under the
    graph shows the range of the second derivative across the
    support of the function.}

\end{figure}

\begin{figure}[H]
\caption{Changes in U.S. Mortality, Age 50-54, 1992-2015:
  \cnewline Constant Rank Interval Estimates (High School or Less in
  1992) \cnewline Curvature Constraint Only}
\label{fig:mort_change_nomon}
\begin{center}

  \begin{tabular}{cc}
    Panel A: Women (Total Mortality) & Panel B: Men (Total Mortality) \\

    \includegraphics[scale=.54]{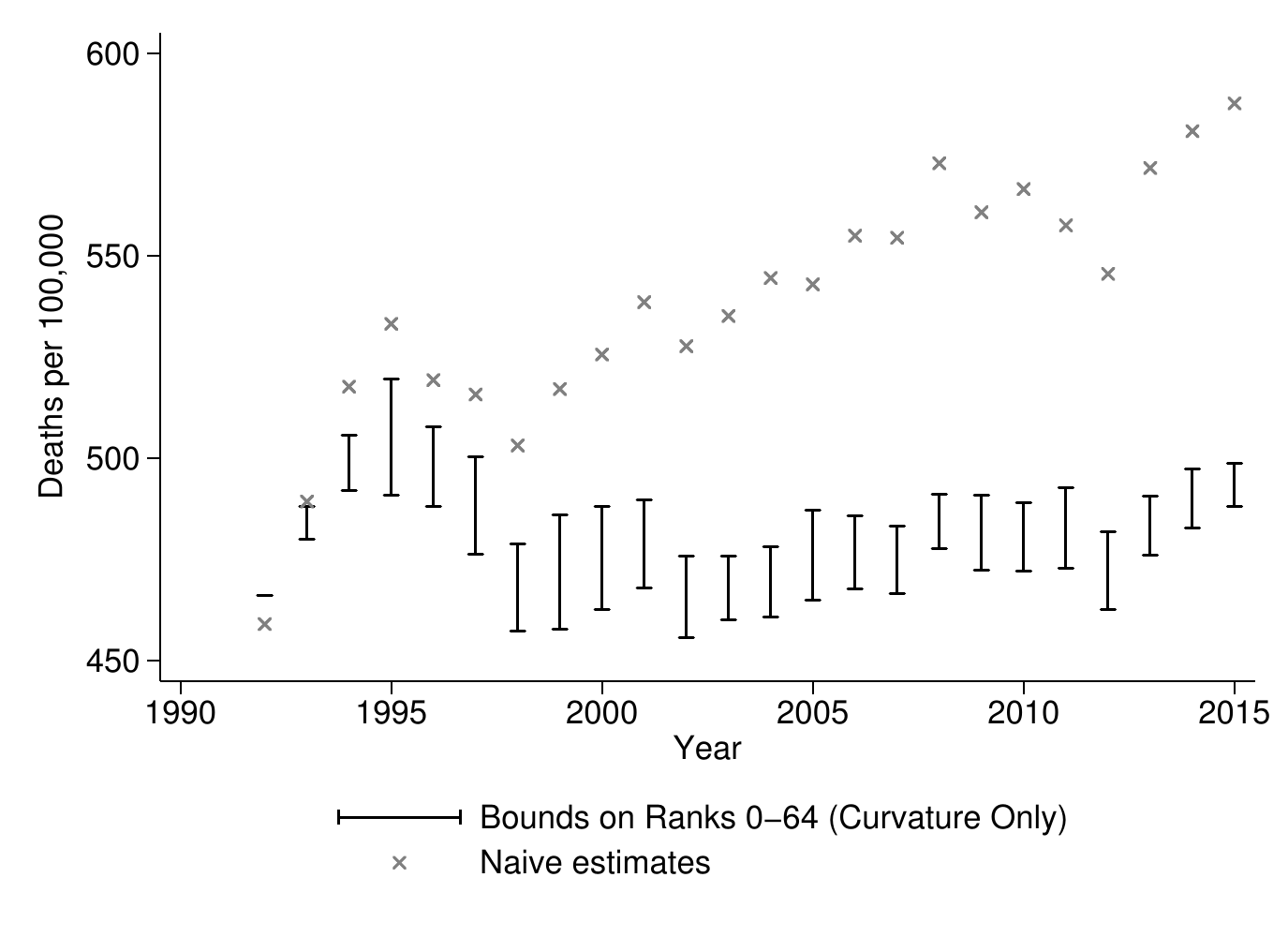}
    &
    \includegraphics[scale=.54]{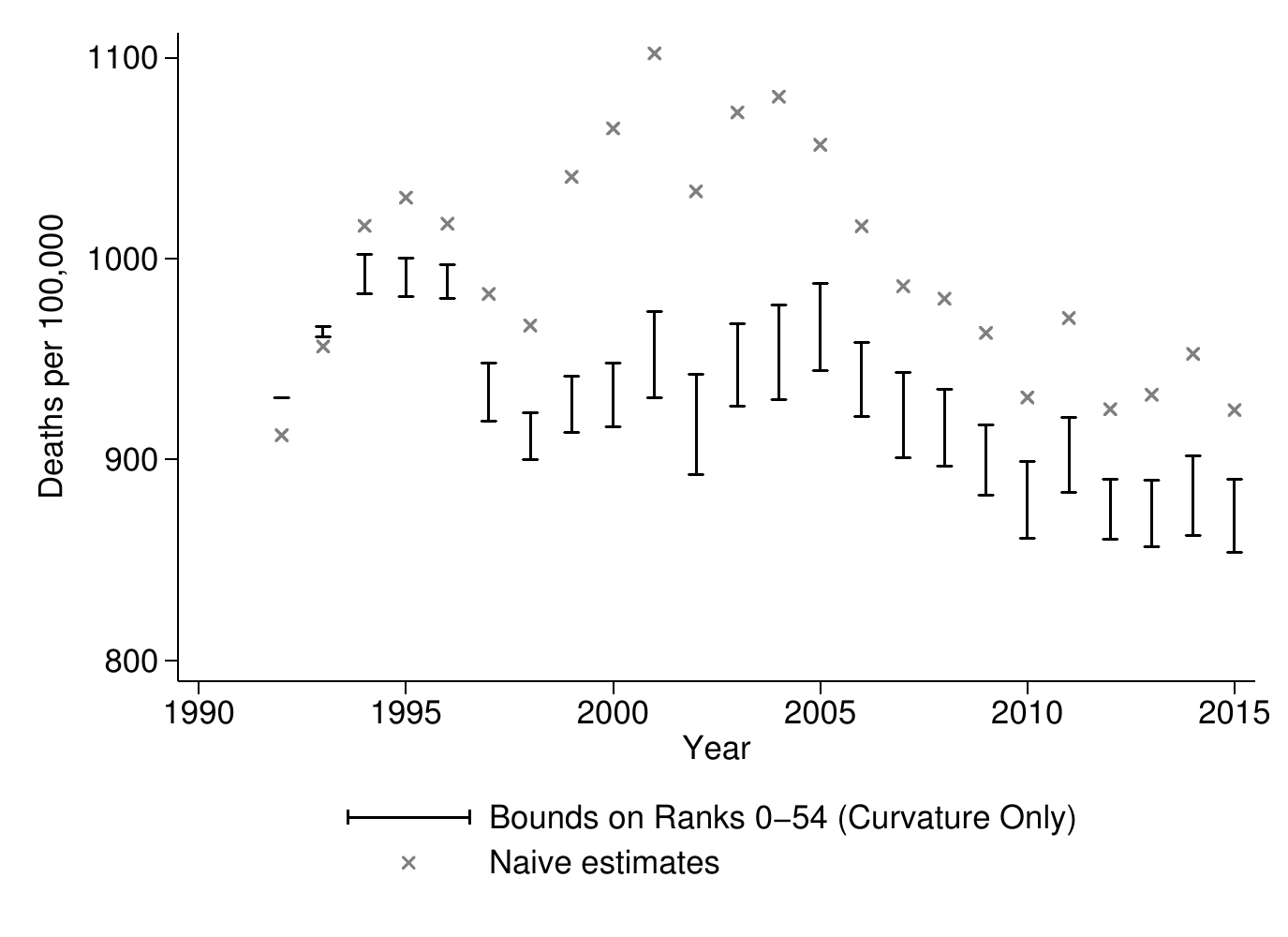}

    \\

    Panel C: Women (Deaths of Despair) & Panel D: Men (Deaths of Despair) \\

    \includegraphics[scale=.54]{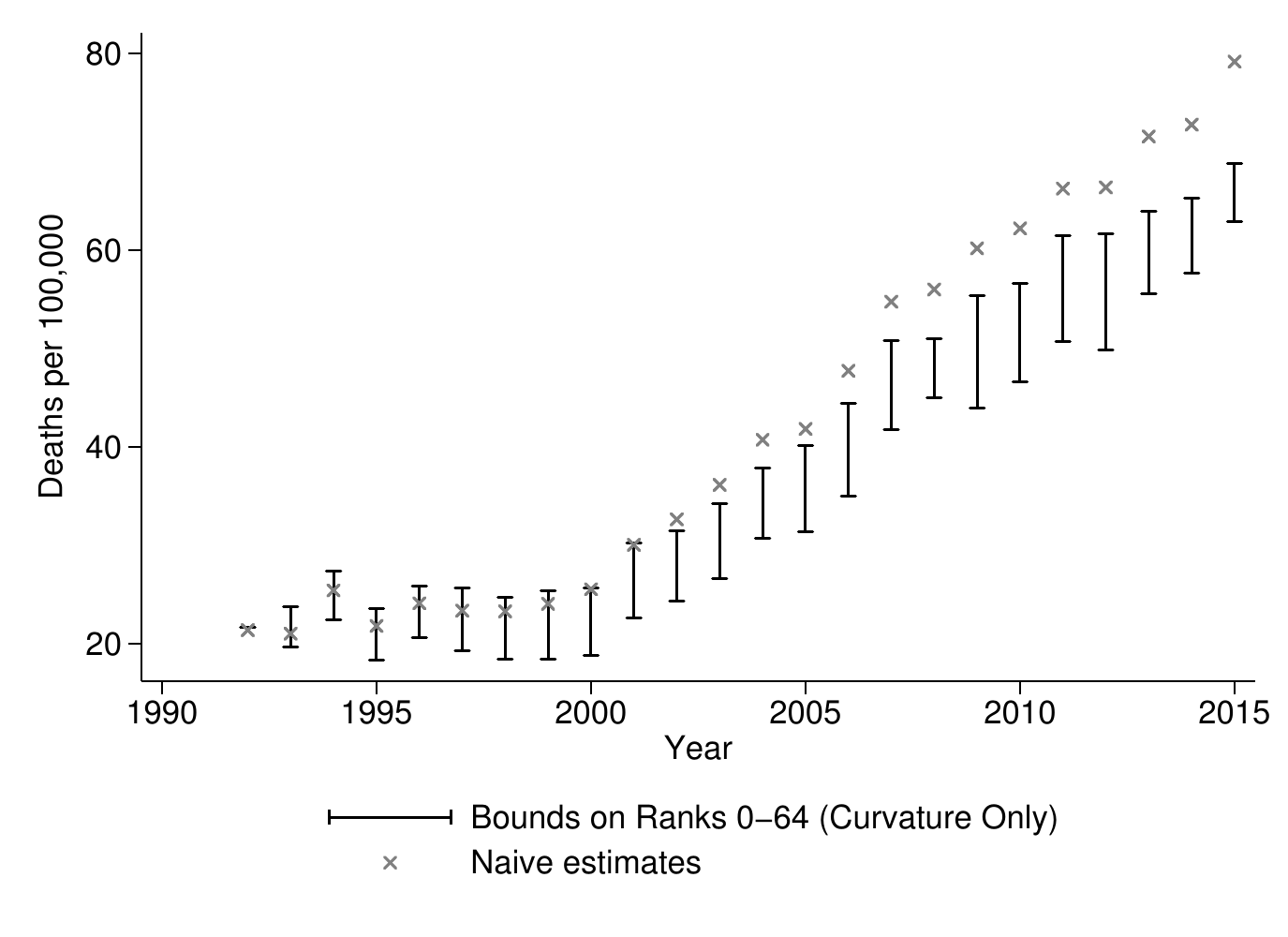}
    &
    \includegraphics[scale=.54]{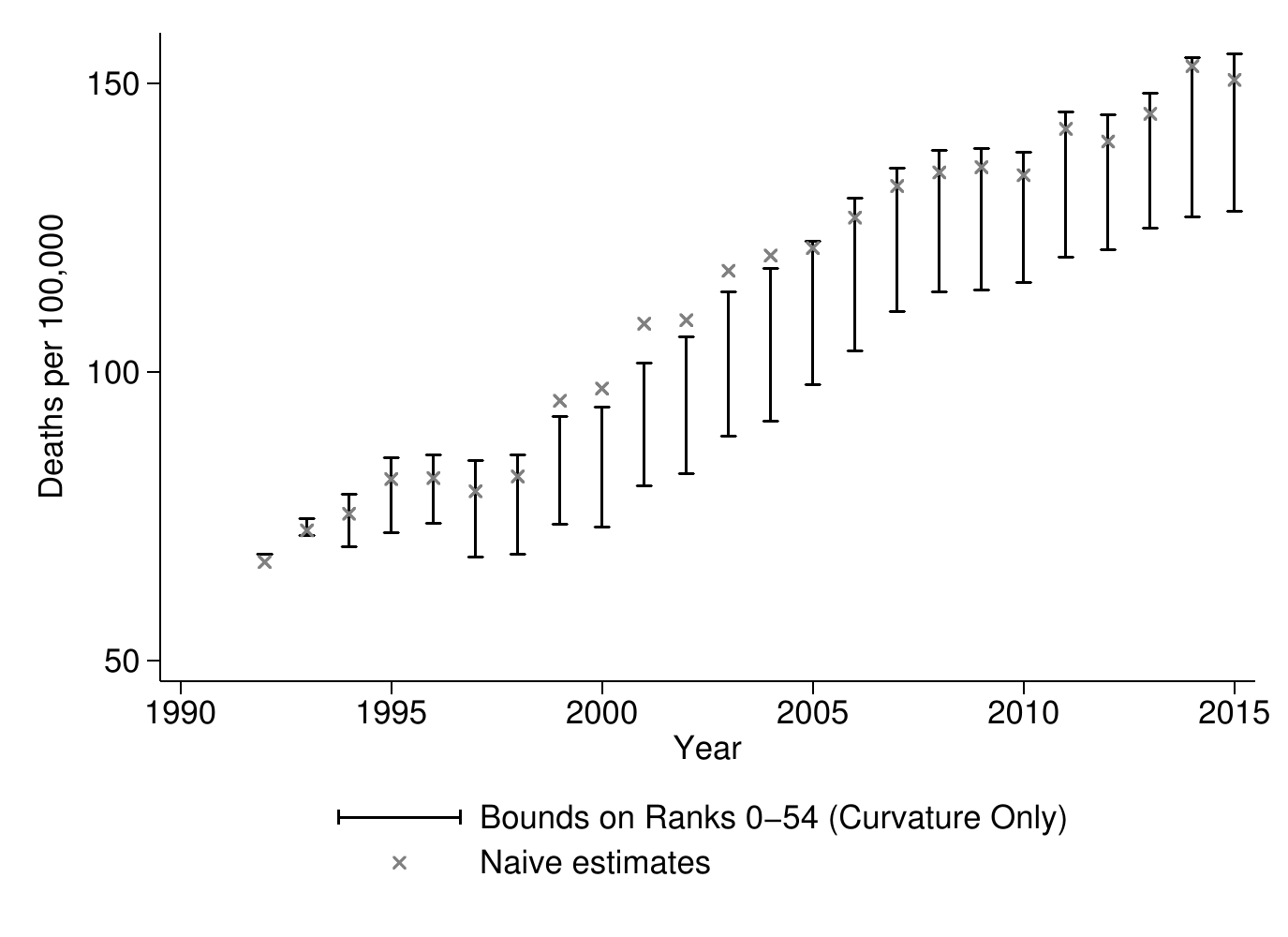}

    \\
    
    \hline

  \end{tabular}

\end{center}
\noindent
\footnotesize{Figure \ref{fig:mort_change_nomon} shows bounds on
  estimates of mortality for men and women aged 50--54 over time. The
  figure is similar to Figure~\ref{fig:mort_change}, but the bounds
  here are generated under the assumption of a curvature constraint
  ($\overline{C}=3$) but without the requirement of monotonicity. In
  contrast, Figure~\ref{fig:mort_change} calculates bounds under the
  assumption of a monotonic CEF with no curvature constraint. The
  sample is defined by the set of latent education ranks corresponding
  to a high school education or less in 1992, or ranks 0-64 for
  women and 0-54 for men. Panel A shows total mortality for women age
  50--54, and Panel B shows total mortality for men age 50--54. Panels
  C and D show mortality from deaths of despair for both groups.}
\end{figure}

\newpage
\begin{figure}[H]
  \caption{Spline Approximations to Empirical Parent-Child Rank Distributions}

  \label{fig:mob_splines}
  \begin{center}

    \begin{tabular}{cc}
      Panel A: U.S.A. & Panel B: Denmark \\
      \includegraphics[scale=0.55]{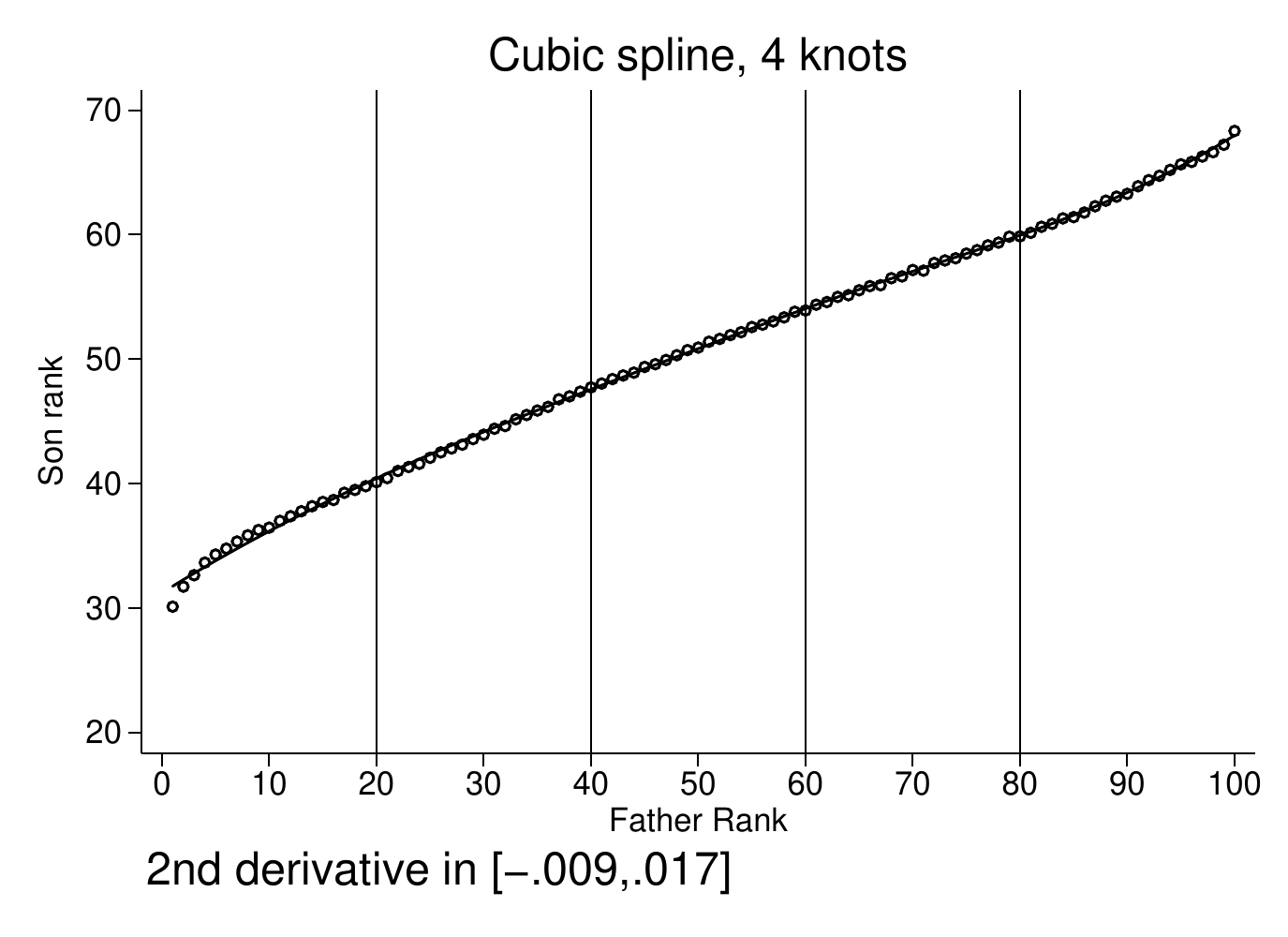}
      &
      \includegraphics[scale=0.55]{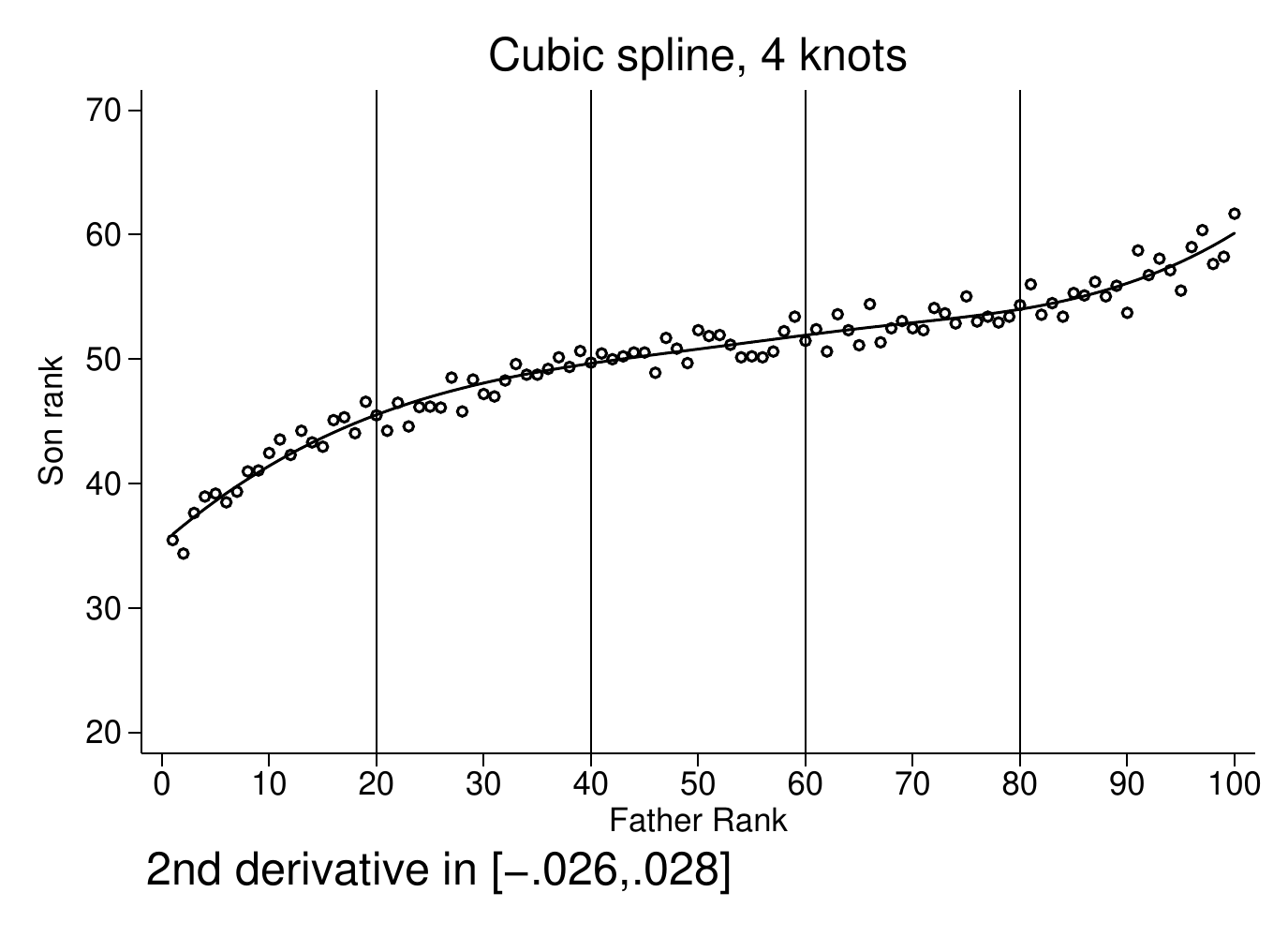}
      \\

      Panel C: Sweden & Panel D: Norway \\
      \includegraphics[scale=0.55]{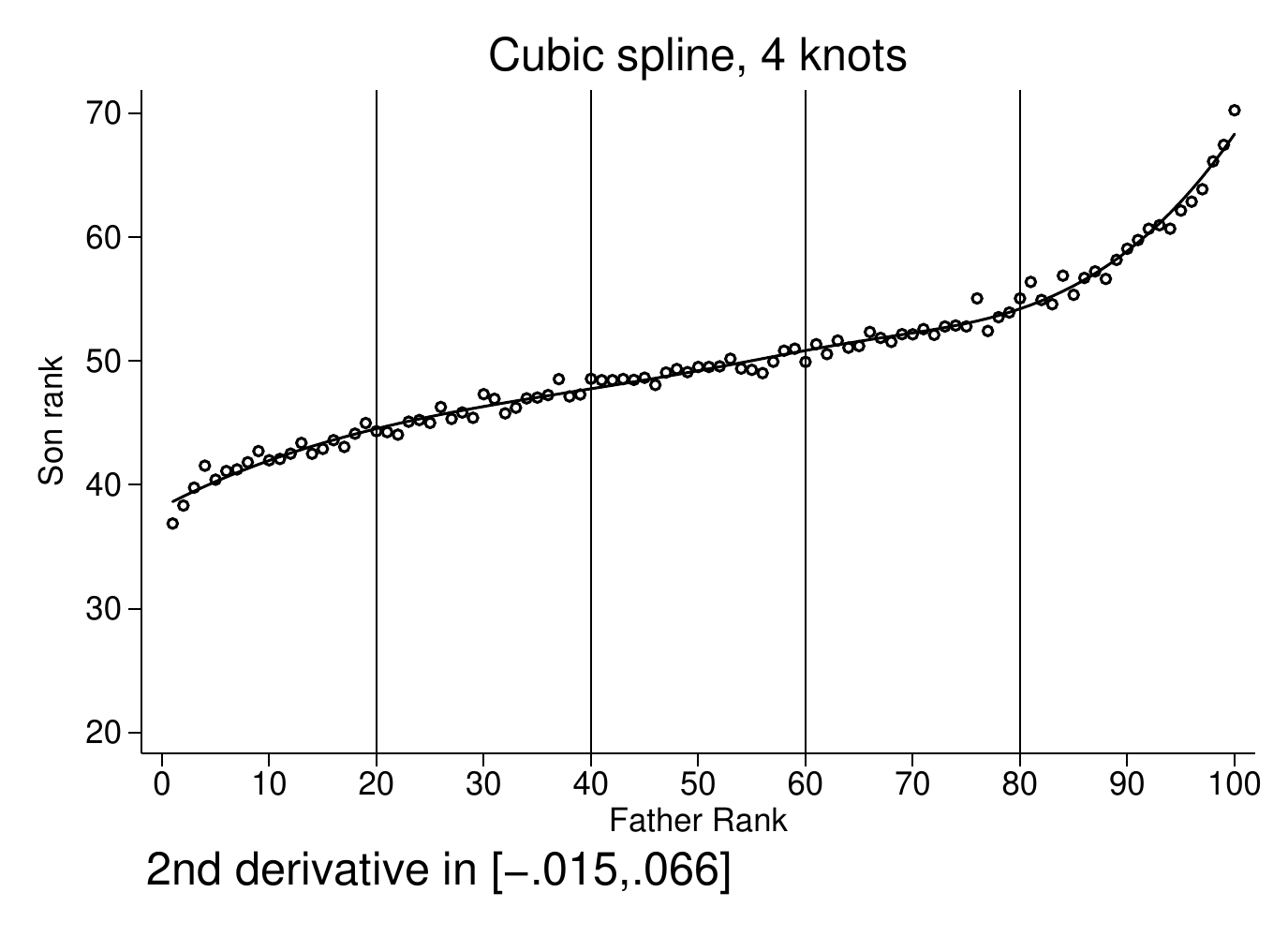}
      &
      \includegraphics[scale=0.55]{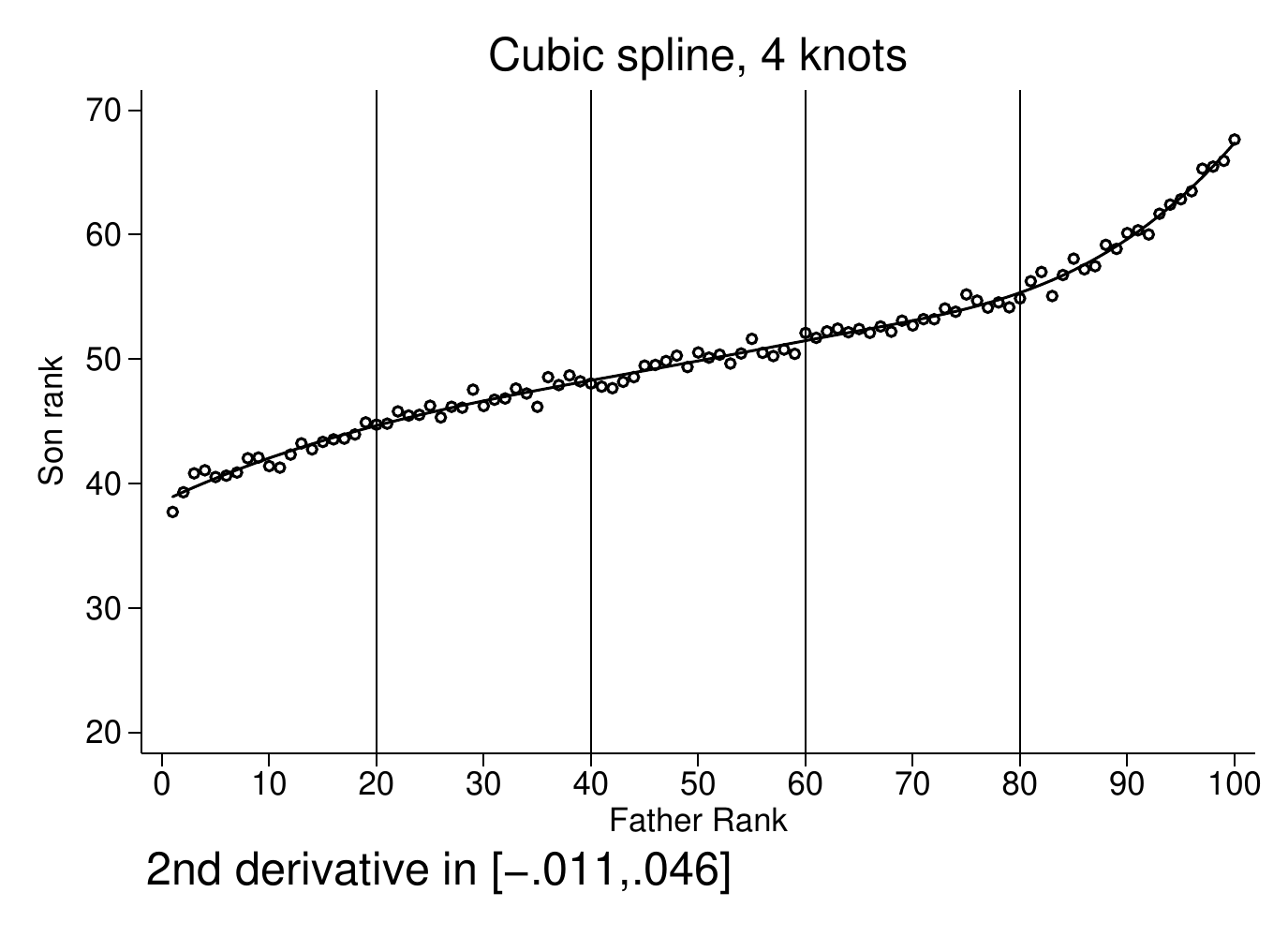}
      \\

      \hline

    \end{tabular}

  \end{center}

  \noindent
  \footnotesize{Figure \ref{fig:mob_splines} presents estimates of the
    conditional expectation functions obtained from fully supported
    parent-child rank-rank income distribution in several developed
    countries. The data for U.S.A. and Denmark come from
    \citeasnoun{Chetty2014c}, who obtained the Denmark data from
    \citeasnoun{Boserup2014}. The data for Sweden and Norway come from
    \citeasnoun{Bratberg2017}. The CEFs were fitted using cubic
    splines, with knots at 20, 40, 60, and 80 (as indicated by the
    vertical lines). The functions plot the best cubic spline fit to
    each series, and the circles plot the underlying data. The text
    under each graph shows the range of the second derivative across
    the support of the function.}

\end{figure}

\begin{table}[H]
  \caption{Bin Sizes in Studies of Intergenerational Mobility}
  \begin{center}
\small{    \begin{tabular}{lcccc}
\hline\hline
Study                    &  Country        & Birth Cohort                     & Number of Parent & Population Share in   \\
                         &                 &   of Son                         & Outcome Bins     &   Largest Bin         \\
\hline
\citeasnoun{Aydemir2016} &  Turkey         & ~1990\protect\footnotemark       & 15               &  39\%          \\
                         &  Turkey         & ~1960\protect\footnotemark       & 15               &  78\%          \\

\citeasnoun{Dunn2007}    &  Brazil         & 1972--1981                       & $>$ 18           & $~$ 20\%\protect\footnotemark   \\

\citeasnoun{Emran2011}   & Nepal, Vietnam  & 1992-1995                        & 2                & 83\%           \\

\citeasnoun{Guell2013}   &  Spain          &  $\sim$ 2001                     & 9                &  27\%\protect\footnotemark      \\

\citeasnoun{Guest1989}   &  USA            &  $\sim$ 1880                     & 7                &  53.2\%        \\


\citeasnoun{Hnatkovska2013} &  India       &  1918-1988                       & 5                &  Not reported  \\

\citeasnoun{Knight2011}  &  China          & 1930--1984                       & 5                & 29\%\protect\footnotemark \\

\citeasnoun{Lindahl2012} & Sweden          & 1865-2005                        & 8                & 34.5\%       \\

\citeasnoun{Long2013}    & Britain         & $\sim$    1850                   &  4               & 57.6\%       \\
                         & Britain         & $\sim$ 1949-55                   &  4               & 54.2\%       \\
                         & USA             & $\sim$ 1850-51                   &  4               & 50.9\%       \\
                         & USA             & $\sim$ 1949-55                   &  4               & 48.3\%       \\

\citeasnoun{Piraino2015} & South Africa    & 1964--1994                       & 6                &  36\%               \\
\hline
\end{tabular}

 }
  \end{center}
  \label{tab:mob_ests}
\end{table}
Table \ref{tab:mob_ests} presents a review of papers analyzing
educational and occupational mobility. The sample is not
representative: we focus on papers where interval censoring may
be a concern. The column indicating number of parent outcome bins
refers to the number of categories for the parent outcome used in the
main specification. The outcome is education in all studies with the
exception of \citeasnoun{Long2013} and \citeasnoun{Guest1989}, where
the outcome is occupation.

\addtocounter{footnote}{-3}
\footnotetext{Includes all people born after about 1990.}
\addtocounter{footnote}{1}
\footnotetext{Includes all people born after about 1960.} 
\addtocounter{footnote}{1}
\footnotetext{This is the proportion of sons in 1976 who had not
  completed one year of education --- an estimate of the proportion of
  fathers in 2002 with no education, which is not reported.}

\addtocounter{footnote}{1}
\footnotetext{Estimate is from the full population rather than just fathers.}

\addtocounter{footnote}{1}
\footnotetext{This reported estimate does not incorporate sampling
  weights; estimates with weights are not reported.}

  \begin{table}[H]
  \thisfloatpagestyle{empty}
  \caption{Transition Matrices for Father and Son Education in India}
  \label{tab:trans_matrices}
        {\scriptsize

          \begin{center}
            \textbf{A: Sons Born 1950-59} 
          \end{center}

          \begin{tabular}{c |c c c c c c c}
            \hline
            \hline
            & \multicolumn{7}{c}{Son highest education attained} \\
             & $<$ 2 yrs. & 2-4 yrs. & Primary & Middle & Sec. & Sr. sec. & Any higher \\Father ed attained & (31\%) &  (11\%) & (17\%) & (13\%) &  (13\%) &  (6\%) &  (8\%) \\ 
            \hline
            $<$2 yrs. (60\%) & 0.47 & 0.12 & 0.17 & 0.11 & 0.09 & 0.03 & 0.03 \\ 
2-4 yrs. (12\%) & 0.10 & 0.18 & 0.22 & 0.19 & 0.16 & 0.09 & 0.06 \\ 
Primary (13\%) & 0.07 & 0.08 & 0.31 & 0.16 & 0.19 & 0.08 & 0.10 \\ 
Middle (6\%) & 0.06 & 0.05 & 0.09 & 0.30 & 0.17 & 0.14 & 0.18 \\ 
Secondary (5\%) & 0.03 & 0.02 & 0.04 & 0.12 & 0.37 & 0.11 & 0.30 \\ 
Sr. secondary (2\%) & 0.02 & 0.00 & 0.03 & 0.11 & 0.11 & 0.35 & 0.38 \\ 
Any higher ed (2\%) & 0.01 & 0.01 & 0.01 & 0.03 & 0.08 & 0.13 & 0.72 \\ 
 
            \hline

          \end{tabular}

          \vspace{.3 cm} 

          \begin{center}
            \textbf{B: Sons Born 1960-69} 

          \end{center}

          \begin{tabular}{c |c c c c c c c}

            \hline
            \hline
            & \multicolumn{7}{c}{Son highest education attained} \\
             & $<$ 2 yrs. & 2-4 yrs. & Primary & Middle & Sec. & Sr. sec. & Any higher \\Father ed attained & (27\%) &  (10\%) & (16\%) & (16\%) &  (14\%) &  (7\%) &  (10\%) \\ 
            \hline
            $<$2 yrs. (57\%) & 0.41 & 0.12 & 0.16 & 0.14 & 0.09 & 0.04 & 0.04 \\ 
2-4 yrs. (13\%) & 0.12 & 0.17 & 0.18 & 0.22 & 0.15 & 0.08 & 0.08 \\ 
Primary (14\%) & 0.09 & 0.05 & 0.26 & 0.18 & 0.20 & 0.09 & 0.13 \\ 
Middle (6\%) & 0.06 & 0.04 & 0.09 & 0.29 & 0.21 & 0.13 & 0.19 \\ 
Secondary (6\%) & 0.03 & 0.02 & 0.08 & 0.12 & 0.35 & 0.16 & 0.25 \\ 
Sr. secondary (2\%) & 0.02 & 0.02 & 0.03 & 0.07 & 0.19 & 0.25 & 0.41 \\ 
Any higher ed (2\%) & 0.01 & 0.01 & 0.02 & 0.03 & 0.09 & 0.11 & 0.73 \\ 
 
            \hline 
          \end{tabular}

          \vspace{.3 cm} 

          \begin{center}
            \textbf{C: Sons Born 1970-79} 
          \end{center}

          \begin{tabular}{c |c c c c c c c}

            \hline
            \hline
            & \multicolumn{7}{c}{Son highest education attained} \\
             & $<$ 2 yrs. & 2-4 yrs. & Primary & Middle & Sec. & Sr. sec. & Any higher \\Father ed attained & (20\%) &  (8\%) & (17\%) & (18\%) &  (16\%) &  (10\%) &  (12\%) \\ 
            \hline
            $<$2 yrs. (50\%) & 0.33 & 0.10 & 0.19 & 0.17 & 0.12 & 0.05 & 0.04 \\ 
2-4 yrs. (11\%) & 0.11 & 0.16 & 0.20 & 0.22 & 0.15 & 0.08 & 0.08 \\ 
Primary (15\%) & 0.08 & 0.06 & 0.24 & 0.23 & 0.18 & 0.11 & 0.11 \\ 
Middle (8\%) & 0.05 & 0.03 & 0.09 & 0.29 & 0.21 & 0.17 & 0.16 \\ 
Secondary (9\%) & 0.03 & 0.02 & 0.06 & 0.12 & 0.31 & 0.19 & 0.27 \\ 
Sr. secondary (3\%) & 0.01 & 0.01 & 0.02 & 0.08 & 0.17 & 0.29 & 0.42 \\ 
Any higher ed (4\%) & 0.00 & 0.00 & 0.02 & 0.05 & 0.10 & 0.17 & 0.66 \\ 
 
            \hline 
          \end{tabular}

          \vspace{.3 cm} 

          \begin{center}
            \textbf{D: Sons Born 1980-89} 
          \end{center}

          \begin{tabular}{c | c c c c c c c}
            \hline
            \hline
            & \multicolumn{7}{c}{Son highest education attained} \\
             & $<$ 2 yrs. & 2-4 yrs. & Primary & Middle & Sec. & Sr. sec. & Any higher \\Father ed attained & (12\%) &  (7\%) & (16\%) & (20\%) &  (16\%) &  (12\%) &  (17\%) \\ 
            \hline
            $<$2 yrs. (38\%) & 0.26 & 0.10 & 0.21 & 0.20 & 0.12 & 0.06 & 0.05 \\ 
2-4 yrs. (11\%) & 0.08 & 0.17 & 0.19 & 0.24 & 0.15 & 0.09 & 0.08 \\ 
Primary (17\%) & 0.05 & 0.04 & 0.22 & 0.23 & 0.20 & 0.13 & 0.13 \\ 
Middle (12\%) & 0.03 & 0.02 & 0.10 & 0.28 & 0.20 & 0.17 & 0.20 \\ 
Secondary (11\%) & 0.02 & 0.01 & 0.05 & 0.13 & 0.23 & 0.24 & 0.32 \\ 
Sr. secondary (5\%) & 0.02 & 0.01 & 0.04 & 0.09 & 0.15 & 0.24 & 0.46 \\ 
Any higher ed (5\%) & 0.01 & 0.01 & 0.02 & 0.05 & 0.10 & 0.16 & 0.65 \\ 
 
            \hline 
          \end{tabular}
        }
        \vspace{.3 cm} 

        Table \ref{tab:trans_matrices} shows transition matrices by decadal birth
        cohort for Indian fathers and sons in the study.

\end{table}

\section{Appendix B: Proofs} 
\label{app:proofs}
\textbf{Proof of Proposition~\ref{eq:cef_bound}.} 

Let the function $Y(x) = E(y|x)$ be defined on a known interval;
without loss of generality, define this interval as $x \in
[0,100]$. Assume $Y(x)$ is integrable. We want to bound $E(y|x)$ when
$x$ is known to lie in the interval $[x_k,x_{k+1}]$; there are $K$
such intervals. Define the expected value of $y$ in bin $k$ as
$$r_k = \int_{x_k}^{x_{k+1}} Y(x) f_k(x)
dx.$$ 
Note that $$r_k = E(y | x \in [x_k, x_{k+1}])$$ via the law of
iterated expectations. Define $r_0 =0$ and $r_{K+1} = 100$. 

Restate the following assumptions from \citeasnoun{Manski2002}: 
\begin{align*}
 \tag{Assumption I} 
  &P(x \in [x_{k}, x_{k+1}]) = 1. \\
  \tag{Assumption M}  
  &E(y|x) \text{ must be weakly increasing in } x. \\ 
  \tag{Assumption MI} 
  &E( y \vert x \text{ is interval censored}) = E(y
  \vert x). \\
\end{align*}

From \citeasnoun{Manski2002}, we have: 
\begin{equation}                                                                                                                                                                                                                                                                                                                                                                                                                               
  r_{k-1} \leq E(y | x) \leq r_{k+1}                                                                                                                                                                                                                                                                                                                                                                                              
  \tag{Manski-Tamer bounds} 
\end{equation}                                                                                                                                                                                                                                                                                                                                                                                                                                 

Suppose also that 
\begin{equation}
x \sim U(0,100). 
  \tag{Assumption U} 
\end{equation}
In that case, $$ r_k = \frac{1}{x_{k+1} - x_k} \int_{x_k}^{x_{k+1}}
Y(x)dx,$$ substituting the probability distribution function for the
uniform distribution within bin $k$. Then we derive the following
proposition. 

\begin{nono-prop} 
\nonumber
  Let $x$ be in bin $k$. Under assumptions IMMI \cite{Manski2002} and
  U, and without additional information, the
  following bounds on $E(y \vert x)$ are sharp:
  $$
  \begin{cases}                                                                                                                          %
    r_{k-1} \leq E(y \vert x) \leq \frac{1}{x_{k+1} - x} \left(
    \left(x_{k+1} - x_k\right) r_k - \left(x - x_k\right) r_{k-1} \right), & x < x_k^* \\                                                                %
    \frac{1}{x - x_k} \left( \left(x_{k+1} - x_k\right) r_k -
    \left(x_{k+1} - x\right) r_{k+1} \right) \leq E(y \vert x) \leq r_{k+1}, & x \geq x_k^*                                                                                                                 %
  \end{cases}
  $$
  where $$x_k^* = \frac{x_{k+1} r_{k+1}
    - \left(x_{k+1} - x_k\right) r_k -
    x_k r_{k-1}  }{r_{k+1} - r_{k-1} }.$$  
\end{nono-prop} 

The intuition behind the proof is as follows. First, find the function $z$ which meets the bin mean and is defined as
$r_{k-1}$ up to some point $j$. Because $z$ is a valid CEF, the lower
bound on $E(y | x)$ is no larger than $z$ up to $j$; we then show that
$j$ is precisely $x_k^*$ from the statement. For points $x > x_k^*$,
we show that the CEF which minimizes the value at point $x$ must be a
horizontal line up to $x$ and a horizontal line at $r_{k+1}$ for
points larger than $x$. But there is only one such CEF, given
that the CEF must also meet the bin mean, and we can solve
analytically for the minimum value the CEF can attain at point $x$. We focus on lower bounds
for brevity, but the proof for upper bounds follows a symmetric
structure. 

\textit{Part 1: Find} $x_k^*$. First define $\mathcal{V}_k$ as the
set of weakly increasing CEFs which meet the bin mean. Put
 otherwise, let $\mathcal{V}_k$ be the set of $v: [x_k,x_{k+1}] \to \mathbb{R}$
satisfying $$r_k = 
\frac{1}{x_{k+1} - x_k} \int_{x_k}^{x_{k+1}} v(x) dx.$$ 
Now choose $z \in \mathcal{V}_k$ such that 
$$
z(x) = \begin{cases} 
r_{k-1}, & x_k \leq x < j \\
r_{k+1}, & j \leq x \leq x_{k+1}.
\end{cases} 
$$

Note that $z$ and $j$
both exist and are unique (it suffices to show that just $j$ exists
and is unique, as then $z$ must be also). We can solve for $j$ by
noting that $z$ lies in $\mathcal{V}_k$, so it must meet the bin mean. Hence, by
evaluating the integrals, $j$ must satisfy: 
\begin{align*}
r_k &= \frac{1}{x_{k+1} - x_k} \int_{x_k}^{x_{k+1}} z(x) dx \\
&= \frac{1}{x_{k+1} - x_k} \left(\int_{x_k}^{j} r_{k-1} dx
+  \int_{j}^{x_{k+1}} r_{k+1} dx \right) \\ 
&= \frac{1}{x_{k+1} - x_k} \left( \left(j - x_k\right)
r_{k-1} + \left(x_{k+1} - j\right) r_{k+1} \right). 
\end{align*}
Note that these expressions invoke assumption U, as the integration
of $z(x)$ does not require any adjustment for the density on the $x$ axis. For a
more general proof with an arbitrary distribution of $x$, see section \ref{sec:gen_distrib}. 

With some algebraic manipulations, we obtain that $j = x_k^*$. 

\textit{Part 2: Prove the bounds.} 
In the next step, we show that $x_k^*$ is the smallest point at which no
$v \in \mathcal{V}_k$ can be $r_{k-1}$, which means that there must be some
larger lower bound on $E(y | x)$ for $x \geq x_k^*$. In other words, we prove
that $$x_k^* = \sup \Big\{x \vert \text{ there exists } v \in \mathcal{V}_k \text{
such that
} v(x) = r_{k-1}. \Big\}.$$ We must show that $x_k^*$ is an upper bound
and that it is the least upper bound. 

First, $x_k^*$ is an upper bound. Suppose that there exists $j' > x_k^*$ such
that for some $w \in \mathcal{V}_k$, $w(j') = r_{k-1}$. Observe that by
monotonicity and the bounds from \citeasnoun{Manski2002}, $w(x) =
r_{k+1}$ for $x \leq j'$; in other words, if $w(j')$ is the mean of
the mean of the prior bin, it can be no lower or higher than the mean
of the prior bin up to point $j'$. But since $j' > j$, this means that 
$$ \int_{x_k}^{j'} w(x)dx < \int_{x_k}^{j'} z(x) dx,$$ since $z(x) >
w(x)$ for all $h \in (j,j')$. But recall that both $z$ and $w$ lie in
$\mathcal{V}_k$ and must therefore meet the bin mean; i.e., 
$$ \int_{x_k}^{x_{k+1}} w(x)dx = \int_{x_k}^{x_{k+1}}
z(x)dx.$$ 
But then $$\int_{j'}^{x_{k+1}} w(x)dx > \int_{j'}^{x_{k+1}} z(x)
dx.$$ That is impossible by the bounds
from \citeasnoun{Manski2002}, since $w(x)$ cannot exceed
$r_{k+1}$, which is precisely the value of $z(x)$ for $x \geq j$. 

Second, $j$ is the least upper bound. Fix $j' < j$. From the definition of $z$, we
have shown that for some $h \in (j',j)$, $z(h) = r_{k-1}$ (and $z \in
\mathcal{V}_k$). So any point $j'$ less than $j$ would not be a lower bound on the
set --- there is a point $h$ larger than $j'$ such that $z(h) =
r_{k-1}$. 

Hence, for all $x < x_k^*$, there exists a function $v \in \mathcal{V}_k$ such
that $v(x) = r_{k-1}$; the lower bound on $E(y | x)$ for $x < x_k^*$
is no greater than $r_{k-1}$. By choosing $z'$ with 
$$ 
z'(x) = \begin{cases} 
r_{k-1}, & x_k \leq x \leq j \\
r_{k+1}, & j < x \leq x_{k+1}, 
\end{cases} 
$$ 
it is also clear that at $x_k^*$, the lower bound is no larger 
than $r_{k-1}$ (and this holds in the proposition itself, substituting in
$x_k^*$ into the lower bound in the second equation).

Now, fix $x' \in (x_k^*, x_{k+1}]$. Since $x_k^*$ is the supremum, there
is no function $v \in \mathcal{V}_k$ such that $v(x') = r_{k-1}$. Thus for
$x' > x_k^*$, we seek a sharp lower bound larger than $r_{k-1}$. Write this lower bound as
$$Y_{x'}^{min} = \min \Big\{ v(x') \text{ for all  } v \in \mathcal{V}_k \Big\},$$
where $Y_{x'}^{min}$ is the smallest value attained by any function $v \in \mathcal{V}_k$ at the point
$x'$. 

We find this $Y_{x'}^{min}$ by choosing the function which maximizes
every point after $x'$, by attaining the value of the
subsequent bin. The function which minimizes $v(x')$ must be a
horizontal line up to this point. 

Pick $\tilde{z} \in \mathcal{V}_k$ such that 
$$ 
\tilde{z}(x) = \begin{cases}
\underline{Y}, &x_k \leq x' \\
r_{k+1}, &x' < x_{k+1} \leq x_{k+1}
\end{cases}. 
$$
By integrating $\tilde{z}(x)$, we claim that $\underline{Y}$ satisfies the following: 
$$ \frac{1}{x_{k+1} - x_k} \left( \left(x' - x_k\right) \underline{Y} + \left(x_{k+1} -
x'\right) r_{k+1}\right) = r_k .$$ As a result, $\underline{Y}$ from this expression exists and is unique,
because we can solve the equation. Note that this integration step
also requires that the distribution of $x$ be uniform, and we
generalize this argument in \ref{sec:gen_distrib}. 

By similar reasoning as above, there
is no $Y' < \underline{Y}$ such that there exists $w \in \mathcal{V}_k$ with 
$w(x') = Y'$. Otherwise there must be some point $x
> x'$ such that $w(x') > r_{k+1}$ in order that $w$ matches the bin
means and lies in $\mathcal{V}_k$; the expression
for $\underline{Y}$ above maximizes every point after $x'$, leaving no
additional room to further depress $\underline{Y}$. 

Formally, suppose there exists $w \in \mathcal{V}_k$ such that $w(x') =
Y' < \underline{Y}$. Then $w(x') < \tilde{z}(x')$ for all $x < x'$, since $w$ is
monotonic. As a result, $$\int_{x_k}^{x'}\tilde{z}(x)dx
>  \int_{x_k}^{x'}w(x)dx.$$ But recall that 
$$\int_{x_k}^{x_{k+1}} w(x) dx = \int_{x_k}^{x_{k+1}} \tilde{z}(x) dx,$$ so 
$$\int_{x'}^{x_{k+1}} w(x) dx > \int_{x'}^{x_{k+1}} \tilde{z}(x) dx. $$ This is
impossible, since $\tilde{z}(x) = r_{k+1}$ for all $x > x'$, and
by \citeasnoun{Manski2002}, $w(x) \leq r_{k+1}$ for all $w \in \mathcal{V}_k$. Hence there
is no such $w \in \mathcal{V}_k$, and therefore $\underline{Y}$ is smallest possible
value at $x'$, i.e. $\underline{Y} = Y_{x'}^{min}$.

By algebraic manipulations, the expression for $\underline{Y} = Y_{x}^{min}$ reduces
to $$Y_{x}^{min} = \frac{ \left(x_{k+1} - x_k\right) r_k  - (x_{k+1}
- x) r_{k+1}}{x - x_k}, \ x \geq x_k^*.$$  

The proof for the upper bounds uses the same structure as the proof of the lower bounds.

Finally, the body of this proof gives sharpness of the bounds. For we
have introduced a CEF $v \in \mathcal{V}_k$ that obtains the value of the upper and lower
bound for any point $x \in [x_k,x_{k+1}]$. For any value $y$ within the
bounds, one can generate a CEF $v \in \mathcal{V}_k$ such that $v(x) =
y$. \qed

\vspace{1em}


\vspace{2em}

\label{sec:gen_distrib} 
\textbf{Proof of Proposition~\ref{eq:cef_distrib}.} Suppose we relax
assumption $U$ and merely characterize $x$ by some known probability
density function. Then we can derive the following bounds. 

\begin{nono-prop} 
\label{eq:bound_arb_distrib}
Let $x$ be in bin $k$. Let $f_k(x)$ be the probability density function
of $x$ in bin $k$. Under assumptions IMMI \cite{Manski2002}, and
without additional information, the
following bounds on $E(y \vert x)$ are sharp:
$$
\begin{cases}                                                                                                                          
r_{k-1} \leq E(y \vert x) \leq \frac{r_k - r_{k-1} \int_{x_{k}}^x
  f_k(s)ds}{\int_x^{x_{k+1}}f_k(s)ds}, & x < x_k^*      \\           
\frac{r_k - r_{k+1}\int_x^{x_{k+1}} f_k(s)ds }{\int_{x_k}^x f_k(s)ds}  \leq E(y \vert x)  \leq                                                                         
r_{k+1} , & x \geq x_k^*                                                                                                             
\end{cases}
$$
where $x_k^*$ satisfies: 
$$r_k = r_{k-1} \int_{x_k}^{x_k^*} f_k(s) ds + r_{k+1}
\int_{x_k^*}^{x_{k+1}} f_k(s) ds.$$ 
\end{nono-prop} 

The proof follows the same argument as in proposition
\ref{eq:cef_bound}. With an arbitrary distribution, $\mathcal{V}_k$ now
constitutes the functions $v: [x_k,x_{k+1}] \to \mathbb{R}$ which satisfy: 
$$ \int_{x_k}^{x_{k+1}} v(s)f_k(s) ds = r_k.$$ 

As before, choose $z \in \mathcal{V}_k$ such that 
$$
z(x) = \begin{cases} 
r_{k-1}, & x_k \leq x < j \\
r_{k+1}, & j \leq x \leq x_{k+1}. 
\end{cases}
$$

Because the distribution of $x$ is no longer uniform, $j$ must now satisfy 
\begin{align*}
r_k &= \int_{x_k}^{x_{k+1}} z(s)f_k(s) ds \\ 
&= r_{k-1} \int_{x_k}^j f_k(s)ds + r_{k+1} \int_{j}^{x_{k+1}}
f_k(s)ds. 
\end{align*} 
This implies that $j = x_k^*$, precisely. 

The rest of the arguments follow identically, except we now claim that
for $x > x_k^*$, 
$\underline{Y} = Y_{x}^{min}$ satisfies the following: 
$$ r_k = \int_{x_k}^{x} Y_{x}^{min} f_k(s)ds + \int_{x}^{x_{k+1}}
r_{k+1} f_k(s)ds.$$ 

By algebraic manipulations, we obtain: 
$$ Y_{x}^{min}= \frac{r_k - r_{k+1} \int_{x}^{x_{k+1}}
f_k(s)ds}{\int_{x_k}^{x}  f_k(s)ds}$$ and the proof of the lower
bounds is complete. As before, the proof for upper bounds follows from identical logic. \qed

\textbf{Proof of Proposition~\ref{eq:bound_mu}}. 
Define $$  \mu_a^{b} = \frac{1}{b - a} \int_a^{b} E(y | x) di. $$ Let
$Y_x^{min}$ and $Y_x^{max}$ be the lower and upper bounds respectively
on $E(y | x)$ given by Proposition \ref{eq:cef_bound}. 
We seek to bound $\mu_a^b$ when $x$ is observed only in discrete intervals. 

\begin{nono-prop} 
  Let $b \in [x_k, x_{k+1}]$ and $a \in [x_h, x_{h+1}]$ with $a<b$. Let
  assumptions IMMI \cite{Manski2002} and U hold. Then, if there is no
  additional information available, the
  following bounds are sharp: 
  \label{eq:bound_mu} 
$$ 
  \begin{cases} 
     Y_b^{min} \leq \mu_a^b \leq Y_a^{max}, & h = k \\
    \frac{r_h (x_k - a) + Y_b^{min}(b - x_k)}{b-a} \leq
    \mu_a^b \leq \frac{Y_a^{max} (x_k - a) + r_k
      (b-x_k)}{b-a}, & h +
    1 = k \\
    \frac{r_h (x_{h+1} - a) + \sum_{\lambda = h+1}^{k-1} r_{\lambda}
      (x_{\lambda+1} - x_{\lambda}) + Y_b^{min}(b - x_k)}{b-a} \leq
    \mu_a^b 
    \leq \frac{Y_a^{max} (x_{h+1} - a) + \sum_{\lambda = h+1}^{k-1} r_{\lambda}
      (x_{\lambda+1} - x_{\lambda}) + r_k (b-x_k)}{b-a}, & h +
    1 < k. 
  \end{cases} 
$$ 
\end{nono-prop} 

The order of the proof is as follows. If $a$ and $b$ lie in the same
bin, then $\mu_a^b$ is maximized only if the CEF is minimized prior to
$a$. As in the proof of proposition \ref{eq:cef_bound}, that occurs when
the CEF is a horizontal line at $Y_x^{min}$ up to $a$, and a
horizontal line $Y_x^{max}$ at and after $a$. If $a$ and $b$ lie in separate bins, the
value of the integral in bins that are contained between $a$ and $b$ is determined by
the observed bin means. The portions of the integral that are not
determined are maximized by a similar logic, since they both lie
within bins. We prove the bounds for
maximizing $\mu_a^b$, but the proof is symmetric
for minimizing $\mu_a^b$. 

\textit{Part 1: Prove the bounds if $a$ and $b$ lie in the same bin.} We
seek to maximize $\mu_a^b$ when $a, b \in [x_k,x_{k+1}]$. This
requires finding a candidate CEF $v \in \mathcal{V}_k$ which maximizes $\int_a^b
v(x) dx$. Observe that the function
$v(x)$ defined as $$ v(x) = \begin{cases}
Y_a^{min}, &x_k \leq x < a \\ 
Y_a^{max}, &a \leq x \leq x_{k+1} 
\end{cases}
$$ 
has the property that $v \in \mathcal{V}_k$. For if $a \geq x_k^*$, $v = \tilde{z}$
from the second part of the proof of
proposition \ref{eq:cef_bound}. If $a < x_k^*$, the CEF in $\mathcal{V}_k$ which
yields $Y_a^{max}$ is precisely $v$ (by a similar argument which delivers the upper bounds in
proposition \ref{eq:cef_bound}). 

This CEF maximizes $\mu_a^b$, because
there is no $w \in \mathcal{V}_k$ such that $$\frac{1}{b-a} \int_{a}^{b}
w(x) dx > \frac{1}{b-a}
\int_{a}^{b} v(x)dx.$$ Note that for any $w \in \mathcal{V}_k$, $\frac{1}{x_{k+1} - x_k} \int_{x_k}^{x_{k+1}} w(x)dx =
\frac{1}{x_{k+1} - x_k} \int_{x_k}^{x_{k+1}} v(x)dx = r_k$. Hence in
order that $\int_{a}^{b} w(x)dx > \int_{a}^{b} v(x)dx$, there are two
options. The first option is that $$ \int_{x_k}^{a}
w(x)dx < \int_{x_k}^{a} v(x)dx.$$ That is impossible, since there is
no room to depress $w$ given the value of $v$ after $a$. If $a < x_k^*$, then it is
clear that there is no $w$ giving a larger $\mu_a^b$, since
$r_{k-1} \leq w(x)$ for $x_{k-1} \leq x \leq a$, so $w$ is bounded below by $v$. If $a \geq x_k^*$, then $v(x) = r_{k+1}$ for all $a \leq x \leq
x_{k+1}$. That would leave no room to depress $w$ further; if $ \int_{x_k}^{a}
w(x)dx < \int_{x_k}^{a} v(x)dx$, then $\int_{a}^{x_{k+1}} w(x) dx
> \int_{a}^{x_{k+1}} v(x) dx $, which cannot be the case if $v =
r_{k+1}$, by the bounds given in \citeasnoun{Manski2002}. 

The second option is that $$ \int_{b}^{x_k}
w(x)dx < \int_{b}^{x_k}
v(x)dx .$$ This is impossible due to monotonicity. For if $ \int_{a}^{b}
w(x)dx > \int_{a}^{b}
v(x)dx$, then there must be some point $x' \in [a,b)$ such that $w(x')
> v(x')$. By monotonicity, $w(x) > v(x)$ for all $x \in [x',x_{k+1}]$
since $v(x) =Y_a^{max}$ in that interval. As a result,  $$ \int_{b}^{x_k}
w(x)dx > \int_{b}^{x_k}
v(x)dx,$$ since $b \in (x',x_{k+1})$. (If $b = x_{k+1}$, then only the
first option would allow $w$ to maximize the desired $\mu_a^b$.) 

Therefore, there is no such $w$, and $v$ indeed maximizes the desired integral. Integrating $v$ from
$a$ to $b$, we obtain that the upper bound on $\mu_a^b$ is
$\frac{1}{b-a} \int_a^b Y_a^{max} dx = Y_a^{max}$. Note that there may
be many functions which maximize the integral; we only needed to show
that $v$ is one of them. 

To prove the lower bound, use an analogous argument. 

\textit{Part 2: Prove the bounds if $a$ and $b$ do not lie in the same
bin.} We now generalize the set up and permit $a,b \in [0,100]$. Let
  $\mathcal{V}$ be the set of weakly increasing functions such that $\frac{1}{x_{k+1} -
  x_k} \int_{x_k}^{x_{k+1}}
v(x) dx = r_k$ for all $k \leq K$. In other words, $\mathcal{V}$ is the set of
  functions which match the means of every bin. Now observe that for all $v \in \mathcal{V}$, 
\begin{align*}
\mu_a^b &= \frac{1}{b-a}\int_a^b v(x) dx \\ 
&= \frac{1}{b-a} \left(
\int_a^{x_{h+1}} v(x)dx + \int_{x_{h+1}}^{x_k} v(x)dx +
\int_{x_k}^{b} v(x) dx \right), 
\end{align*} 
by a simple expansion of the integral. 

But for all $v \in \mathcal{V}$, $$\int_{x_{h+1}}^{x_k} v(x)dx = \sum_{\lambda =
  h+1}^{k-1} r_{\lambda}
    (x_{\lambda+1} - x_{\lambda})$$ if $h + 1 <k$
and $$\int_{x_{h+1}}^{x_k} v(x)dx = 0$$ if $h + 1 = k$. For in
  bins completely contained inside $[a,b]$, there is no room for any
  function in $\mathcal{V}$ to vary; they all must meet the bin means. 

We proceed to prove the upper bound. We split this into two portions:
we wish to maximize $\int_a^{x_{h+1}}v(x)dx $ and we also wish to maximize $\int_{x_k}^b v(x)dx $. The values of
these objects are not codependent. But observe that the CEFs $v \in
\mathcal{V}_k$ which yield upper bounds on these integrals are the very same functions which yield upper bounds on
$\mu_a^{x_{h+1}} $ and $\mu_{x_k}^b$, since $\mu_s^t
= \frac{1}{t-s} \int_s^t v(x) dx$ for any $s$ and $t$. Also notice
that $a$ and $x_{h+1}$ both lie in bin $h$, while $b$ and $x_k$ both
lie in bin $k$, so we can make use of 
the first portion of this proof. 

In part 1, we showed that the function $v \in \mathcal{V}$, 
$v:[x_h,x_{h+1}] \to \mathbb{R}$, which maximizes $\mu_a^{x_{h+1}}
$ is
$$ v(x) = \begin{cases}
Y_a^{min}, & x_{h} \leq x < a \\
Y_a^{max}, & a \leq x \leq x_{h+1}. 
\end{cases}
$$ 
As a result $$\underset{v \in \mathcal{V}}{\max}\Bigg\{ \int_a^{x_{h+1}} v(x) dx \Bigg\} = \int_a^{x_{h+1}} Y_a^{max} dx = Y_a^{max} (x_{h+1} -a).$$

Similarly, observe that $x_k$ and $b$ lie in the same bin, so the function
$v:[x_k,x_{k+1}] \to \mathbb{R}$, with $v \in \mathcal{V}$  which maximizes $\int_{x_k}^b v(x)dx$ must be of the form 
$$ v(x) = \begin{cases}
Y_{x_k}^{min}, & x_{k} \leq x < a \\
Y_{x_k}^{max}, & b \leq x \leq x_{k+1}. 
\end{cases}
$$ 

With identical logic, 
$$\underset{v \in \mathcal{V}}{\max}\Bigg\{ \int_{x_k}^{b} v(x) dx \Bigg\}
= \int_{x_k}^{b} Y_{x_k}^{max} dx = Y_{x_k}^{max} (b -x_k).$$
And by proposition \ref{eq:cef_bound}, $x_k \leq x_k^*$ so $Y_{x_k}^{max} =
r_{k}$. (Note that if $x_k = x_k^*$, substituting $x_k^*$ into the second
expression of proposition \ref{eq:cef_bound} still yields that
$Y_{x_k}^{max} = r_k$.) 

Now we put all these portions together. First let $h + 1 = k$. Then
$\int_{x_{h+1}}^{x_k} v(x) dx = 0$, so 
we maximize $\mu_a^b$ by
$$\frac{1}{b-a} \left( Y_a^{max} (x_{h+1} -a) + r_k (b
-x_k) \right). $$ Similarly, if $h +1 < k$ and there are entire bins completely
contained in $[a,b]$, then we maximize $\mu_a^b$ by 
$$\frac{1}{b-a} \left( Y_a^{max} (x_{h+1} -a) + \sum_{\lambda =
  h+1}^{k-1} r_{\lambda}
    (x_{\lambda+1} - x_{\lambda}) + r_k  (b -x_k) 
\right). $$ 

The lower bound is proved analogously. Sharpness is immediate, since
we have shown that the CEF which delivers the endpoints of the
bounds lies in $\mathcal{V}$. As a result, there is a function delivering any intermediate
value for the bounds. \qed 

\vspace{1em}

\noindent 
\textbf{Extension of Proposition~\ref{eq:bound_mu} to an arbitrary
  known distribution}

\begin{proposition}
\label{prop:mu_arb_distrib}
  Let $a \in [x_h, x_{h+1}]$ and $b \in [x_k, x_{k+1}]$. Let
  assumptions IMMI hold. Let
$Y_x^{min}$ and $Y_x^{max}$ be the lower and upper bounds respectively
on $E(y | x)$ given by proposition \ref{eq:bound_arb_distrib}.  Let the probability distribution of $x$ be $f(x)$. Then, if no
  additional information is available, the
  following bounds are sharp: 
  \label{eq:bound_mu_arb_distrib} 
$$ 
  \begin{cases} 
    Y_b^{min} \leq \mu_a^b \leq Y_a^{max} & h = k \\
    \frac{r_h \int_a^{x_k} f(x)dx + Y_b^{min}\int_{x_k}^b f(x)dx }{\int_a^b
      f(x)dx} \leq
    \mu_a^b \leq \frac{Y_a^{max} \int_a^{x_k} f(x)dx + r_k
      \int_{x_k}^b f(x)dx}{\int_a^b f(x)dx } & h +
    1 = k \\

\frac{r_h \int_a^{x_{h+1}} f(x)dx + \sum_{\lambda = h+1}^{k-1} r_{\lambda}
      \int_{x_\lambda}^{x_{\lambda+1}}f(x)dx + Y_b^{min} \int_{x_k}^b
      f(x)dx }{\int_a^b f(x)dx}  \leq
    \mu_a^b \\
    \hspace{7cm} 

    \leq \frac{Y_a^{max} \int_a^{x_{h+1}} f(x)dx + \sum_{\lambda = h+1}^{k-1} r_{\lambda}
      \int_{x_\lambda}^{x_{\lambda+1}} f(x)dx + r_k \int_{x_k}^b
      f(x)dx }{\int_a^b f(x)dx}  & h +
    1 < k. 
  \end{cases} 
$$ 
\end{proposition} 

Proposition \ref{eq:bound_mu_arb_distrib} generalizes
proposition \ref{eq:bound_mu} to an arbitrary distribution, but its
proof is identical. The only difference is in the weight given to
components of $\mu_a^b$ that lie in different bins; these weights are
given by integrating the maximizing function $v \in \mathcal{V}$,
while accounting for the probability distribution $f(x)$. 

We consider only maximizing $\mu_a^b$. To prove the first part of proposition \ref{eq:bound_mu_arb_distrib}, we obtain $v
\in \mathcal{V}$ defined as 
$$ v(x) = \begin{cases}
Y_a^{min}, &x_k \leq x < a \\ 
Y_a^{max}, &a \leq x \leq x_{k+1}. 
\end{cases}
$$ 

As before, $\mu_a^b$ given by $\frac{\int_a^b v(x) f(x) dx}{\int_a^b
  f(x) dx}$ will maximize $\mu_a^b$. 

If the first part of the proposition holds, then the rest follows. For
$h \neq k$, then 

\begin{align*}
\mu_a^b &= \frac{1}{\int_a^b f(x)dx}\int_a^b v(x) f(x) dx \\ 
&=   \frac{1}{\int_a^b f(x)dx}   \left(
\int_a^{x_{h+1}} v(x) f(x)dx + \int_{x_{h+1}}^{x_k} v(x)f(x)dx +
\int_{x_k}^{b} v(x) f(x) dx \right), 
\end{align*} 

As before, $\int_{x_{h+1}}^{x_k} v(x)f(x)dx  = 0 $ if $h + 1 = k$. If $h
+ 1 < k$, then 

$$ \int_{x_{h+1}}^{x_k} v(x)f(x) dx = \sum_{\lambda = h+1}^{k-1}
r_\lambda \int_{x_\lambda}^{x_{\lambda+1}} f(x) dx.$$ 

We maximize the objects $\int_a^{x_{h+1}} v(x) f(x)dx$ and
$\int_{x_k}^{b} v(x) f(x)dx$ by using the expression from the first
part of this proof. We therefore have that the maximum of 
$\int_a^{x_{h+1}} v(x) f(x)dx$ over $v \in \mathcal{V}$ is obtained by $Y_a^{max} \int_a^{x_{h+1}}
f(x)dx$. By the same argument, we maximize 
$\int_{x_k}^b v(x)f(x)dx$ with $r_k \int_{x_k}^b f(x)dx$. Putting
these expressions together, the proof is complete. \qed

\newpage
\doublespacing
\section{Appendix C: CEF Bounds When $x$ and $y$ are Interval Censored}
\label{app:impute}
\setcounter{table}{0}
\renewcommand{\thetable}{C\arabic{table}}
\setcounter{figure}{0}
\renewcommand{\thefigure}{C\arabic{figure}}
In the main part of the paper, we focus on bounding a function $Y(x) =
E(y|x)$ when $y$ is observed without error, but $x$ is observed with
interval censoring. In this section, we modify the setup to consider
simultaneous interval censoring in the conditioning variable $x$ and
in observed outcomes $y$. This arises, for example, in the study of
educational mobility, where latent education ranks of both parents and
children are observed with interval censoring.

We first present a setup that takes a similar approach to the bounding
method presented in Section~\ref{sec:method}. We can define bounds on
the CEF $E(y|x)$ when both $y$ and $x$ are interval-censored as a
solution to a constrained optimization problem. The 
number of parameters is an order of magnitude higher than the problem in
Section~\ref{sec:method}, and proved too computationally intensive to
solve in the Indian test case (where interval censoring is severe). We
therefore present a sequential approach that yields theoretical bounds
on the double-censored CEF for the case of intergenerational mobility.

Specifically, we define the theoretical best- and worst-case latent
distributions of $y$ variables for a given intergenerational mobility
statistic. The best- and worst-case assumptions each generate a bound
on the feasible value of $y$ for each $x$ bin. We then use the method in
Section~\ref{sec:method} to calculate bounds on the mobility statistic
under each case. The union of these bounds is a conservative bound on
the mobility statistic given censoring in both the $y$ and $x$
variables.

Finally, we can shed light on the distribution of the true value of
$y$ in each $x$ bin if other data is available. In the context of
intergenerational mobility, and in our specific empirical context, it
is frequently the case that more information is available about
children than about their parents.  We use data on child wages to
predict whether the true latent child rank distribution ($y$) is
better represented by the best- or worst-case mobility scenario. The
joint wage distribution suggests that the true latent distribution of
$y$ in each bin is very close to the best case distribution, which we used in
Section~\ref{sec:mob}, because there is little effect of parent
education on child wages after conditioning on child education.

\subsection{Solution Definition for CEF Bounds with Double Censoring}

We are interested in bounding a function $E(y|x)$, where $y$ is known
only to lie in one of $H$ bins defined by intervals of the form $[y_h,
  y_{h+1}]$, and $x$ is known only to lie in one of $K$ bins defined
by intervals of the form $[x_k, x_{k+1}]$. For simplicity, we focus
on the case where both $y$ and $x$ are
uniformly distributed on the interval $[0, 100]$.\footnote{Taking a different known
  distribution into account would require imposing different weights
  on the mean-squared error function and budget constraint below, but
  would otherwise not be substantively different.}

Where Section~\ref{sec:method} focused on bounding the
\textit{cumulative expectation function} (CEF) of $y$ given $x$, we focus
here on bounding a separate \textit{conditional distribution function}
(CDF) for $y$, given each value of $x$. Each value of $x$ implies a different CDF for
$y$, as follows:

\begin{equation}
  \label{eq:cdf}
  F(r, x) = P(y \leq r | x=X) 
\end{equation}

\noindent This CDF is related to the CEF $E(y|x)$ as follows:

\begin{equation}
  \label{eq:son_cef}
  E(y|x) = \int_0^{100} rf(x,r) dr 
\end{equation}

\noindent where $f(x,r)$ is the probability density function
corresponding to the CDF in
Equation~\ref{eq:cdf}, when the conditioning variable takes the value
$x$. Note that $r$ in this case represents a child rank. This
expression simply denotes that $E(y|x)$ is the average value from $0$
to $100$ on the $y$-axis, holding $x$ fixed.

We do not observe the sample analog of $F(x,r)$ directly. Rather, we
observe the sample analog of the following expression for each of $H*K$ bin
combinations:
\begin{equation}
  \label{eq:analog}
  P(y \leq y_{h+1} \Bigm\lvert x \in [x_k, x_{k+1}]) =
  \frac{1}{x_{k+1} - x_k}  \int_{q={x_k}}^{x_{k+1}}F(q,y_{h+1})dq
\end{equation}

\noindent 
We denote this sample analog $\hat{P}( y \leq y_{h+1}\Bigm\lvert x \in [x_k, x_{k+1}])$ as $\hat{R}(k,
h)$. Equation \ref{eq:analog} states that the probability that $y$ is
less than $y_{h+1}$ is the average value of the
CDF in that bin. Since $x$ is uniform, we can write its probability
distribution function within the bin as $\frac{1}{x_{k+1} - x_k}$. 

We parameterize each CDF as $F(x,r) = S(x, r, \gamma_x)$, where $r$ is
the outcome variable, $x$ is the conditioning variable, and $\gamma_x$
is a parameter vector in some parameter space $G_x$. Similarly let $f(x,r)
= s(x, r, \gamma_x)$. In our numerical calculation, we define $G_x$ as
$[0, 1]^{100}$, a vector which gives the value of the cumulative
distribution function at each of
100 conditioning variable percentiles on the $y$-axis, for a given
value of $x$. Put otherwise, holding $x$
fixed, we seek the 100-valued column vector $\gamma_x$ which contains the
value of the CDF at each of the 100 possible $y$
values: $y=1, y=2, \dots, y=100$. As a result, $\gamma_x$ must lie within
$[0,1]^{100}$. Note that there are as many vectors $\gamma_x$ as there
are possible values for the conditioning variable $x$. If
we discretize also $x$ as $1, 2, \dots, 100$, then  
we define the matrix of 100 CDFs,
indexed by $x$, as $\boldsymbol{\gamma^{100}} = [ \gamma_1 \ \gamma_2 \ \dots
  \gamma_{100} ]$. To be explicit, $\boldsymbol{\gamma^{100}}$ is a $100
\times 100$ matrix constructed by setting its $x\superscript{th}$ column as
$\gamma_x$. We write that $\boldsymbol{\gamma^{100}} \in G^{100}$.

We also introduce a new monotonicity condition for this context.  In
this set up, monotonicity implies that the outcome distribution for
any value of $x$ first-order stochastically dominates the outcome
distribution at any lower value of $x$. Put otherwise,
\begin{equation}
  s(x, r, g_x) \text{ is weakly decreasing
    in } x
  \tag{Monotonicity} 
\end{equation}

\noindent In the mobility context, this statement implies that the
child rank distribution of a higher-ranked parent stochastically
dominates the child rank distribution of a lower-ranked
parent.\footnote{\citeasnoun{Dardanoni2012} find that a similar
  conditional monotonicity holds in almost all mobility tables in 35
  countries.}\superscript{,}\footnote{A stronger monotonicity
  assumption would require that the hazard function is decreasing in
  $x$. This is equivalent to stating that the CDF must be weakly
  decreasing in $x$ conditional upon $x$ being above some value. In
  the mobility case, for example, the stronger assumption would imply
  that conditional on being in high school, a child of a better off
  parent must have a higher latent rank than the child of a worse off
  parent.}


The following minimization problem defines the set of feasible values
of $\gamma_x$ for each value $x$:

\begin{align}
  \label{eq:opt_app}
  \Gamma = \underset{\boldsymbol{g} \in \boldsymbol{G^{100}}} {\text{argmin}} \Bigg\{ \sum_{k=1}^K \sum_{h=1}^H 
  \left( \int_{q=x_k}^{x_{k+1}} 
  S(q, h, g_q)dq - \hat{R}(k, h) \right)^2 \Bigg\}
  \\
  \nonumber
  \text{such that}
  \\
  \tag{Monotonicity} 
  s(x, r, g_x) \text{ is weakly decreasing
    in } x
  \\
  \tag{Budget Constraint}
  \frac{1}{100} \sum_{x=1}^{100} S(x, r, g_x) = r
  \\
  \tag{End Points}
  S(x, 0, g_x) = 0
  \\
  \nonumber
  S(x, 100, g_x) = 1
  \text{.}
\end{align}
\noindent
In the above minimization problem, $\boldsymbol{g}$ is a candidate
vector satisfying the conditions; each $g_x$ describes the candidate
CDF holding $x$ fixed. A valid set of cumulative
distribution functions is one that minimizes error with respect to all of
the observed data points and obeys the monotonicity condition. The budget constraint requires that the weighted sums of
CDFs across all conditioning groups must add up to the population
CDF. For example, $J$\% of children must on average attain less than
or equal to the $J^{th}$ percentile. The constraints on the end points
of the CDF are redundant given the other constraints, but are included
to highlight how the end points constrain the set of possible
outcomes.  For simplicity, we have not included a curvature
constraint, but such a constraint would be a sensible further
restriction on the feasible parameter space in many contexts.

Once a set of candidate CDFs have been identified, they have a
one-to-one correspondence with the CEF given an interval censored
conditioning variable, 
(described by Equation~\ref{eq:son_cef}), and thus with any function
of the CEF. These statistics can be numerically bounded as in
Section~\ref{sec:method}.

This problem is computationally more challenging than the problem of
censoring only in the conditioning variable dealt with in
Section~\ref{sec:method}. In the case of the rank distribution, if we
discretize both outcome and conditioning variables into 100 separate
percentile bins, then the problem has 10,000 parameters and 10,000
constraints, and an additional 9800 curvature constraint inequalities
if desired. This problem proved computationally too difficult to
resolve. Restricting the set of discrete bins (e.g. to deciles) is
unsatisfying because it requires significant rounding of the raw data
which could substantively affect results. We proceed instead by taking
advantage of characteristics that are specific to the problem of
intergenerational mobility.

\subsection{Best and Worst Case Mobility Distributions}
\label{sec:best_worst}

Our goal is to bound the parent-child rank CEF given interval censored
data on both parent and child ranks. In this section, we take a
sequential approach to the double-censoring problem. We use additional
information about the structure of the mobility problem to obtain
worst- and best-case parent CDFs for intergenerational mobility. From
these cumulative distribution functions, we can obtain worst- and
best-case CEFs using Equation \ref{eq:son_cef}. First, we calculate
bounds on the average value of the child rank in each child rank *
parent rank cell. We then apply the methods from
Section~\ref{sec:method} on the best and worst case bounds; the union
of resulting bounds describes the bounds on the mobility statistic of
interest. We focus on the rank-rank gradient and on $\mu_0^{50}$.

Given data where child rank is known only to lie in one of $h$ bins,
there are two hypothetical scenarios that describe the best and worst
cases of intergenerational mobility. Mobility will be lowest if child
outcomes are sorted perfectly according to parent outcomes
\textit{within} each child bin, and highest if there is no additional
sorting within bins.\footnote{Specifically, these scenarios
  respectively minimize and maximize both the rank-rank gradient and
  $\mu_0^{x}$ for any value of $x$. To minimize and maximize $p_x$, a different
  within-bin arrangement is required for every $x$. We leave this out
  for the sake of brevity, and because bounds on $p_x$ are minimally
  informative even with uncensored $y$.}

Consider a simple 2x2 case. In the 1960s birth cohort in India, 27\%
of boys attained less than two years of education, the lowest recorded
category. 55\% of these had fathers with less than two years of education, and
45\% had fathers with two or more years of education. We do not
observe how the children of each parent group are distributed
\textit{within} the bottom 27\%. For this case, mobility will be
lowest if children of the least educated parents occupy the bottom
ranks of this bottom bin, or ranks 0 through 15, and children of more
educated parents occupy ranks 16 through 27. Mobility will be
highest if parental education has no relationship with rank,
conditional on the child rank bin. We do not consider the case of
perfectly reversed sorting, where the children of the least educated
parents occupy the highest ranks within each child rank bin, as it
would violate the stochastic dominance condition (and is
implausible).

Appendix Figure~\ref{fig:son_solution} shows two set of CDFs that
correspond to these two scenarios for the 1960--69 birth cohort.  In Panel A, children's ranks are
perfectly sorted according to parent education within bins. Each line
shows the CDF of child rank, given some father education. The points
on the graph correspond to the observations in the data---the value of
each CDF is known at each of this points.  Children below the 27th
percentile are in the lowest observed education bin. Within this bin,
the CDF for children with the least educated parents is concave, and
the CDF for children with the most educated parents is
convex---indicating that children from the best off families have the
highest ranks within this bin. This pattern is repeated within each
child bin. Panel B presents the high mobility scenario, where
children's outcomes are uniformly distributed within child education
bins, and are independent of parent education within child bin.

According to Equation~\ref{eq:son_cef}, each of these CDFs corresponds
to a single mean child outcome in a given parent bin, or $Y=E(y|x \in
[x_k, x_{k+1}])$.  From these expected values, we can then calculate
bounds on any mobility statistic, as in Sections~\ref{sec:method} and
\ref{sec:mob}. Table~\ref{tab:double_censor} shows the expected child
rank by parent education for the high and low mobility scenario, as
well as bounds on the rank-rank gradient and on $\mu_0^{50}$.  Taking
censoring in the child distribution into account widens the bounds on
all parameters. The effect is proportionally the greatest on the
interval mean measure, because it was so precisely estimated
before---the bounds on $\mu_0^{50}$ approximately double in width when
censoring of son data is taken into account.

These bounds are very conservative, as the worst case scenario is
unlikely to reflect the true uncensored joint parent-child rank
distribution, due to the number and sharpness of kinks in the CDFs in
Panel A of Figure~\ref{fig:son_solution}. A curvature constraint on
the CDF would move the set of feasible solutions closer to the high
mobility scenario. We next draw on additional data on children, which
suggests that the best case mobility scenario is close to the true
joint distribution.

\subsection{Estimating the Child Distribution Within Censored Bins}

\begin{figure}[H]

  \caption{Best- and Worst-Case Son CDFs \cnewline by Father Education (1960-69
    Birth Cohort)}
  \label{fig:son_solution}

  \begin{center}
    \begin{tabular}{c}

      Panel A: Lowest Feasible Mobility
      \\
      \includegraphics[scale=0.80]{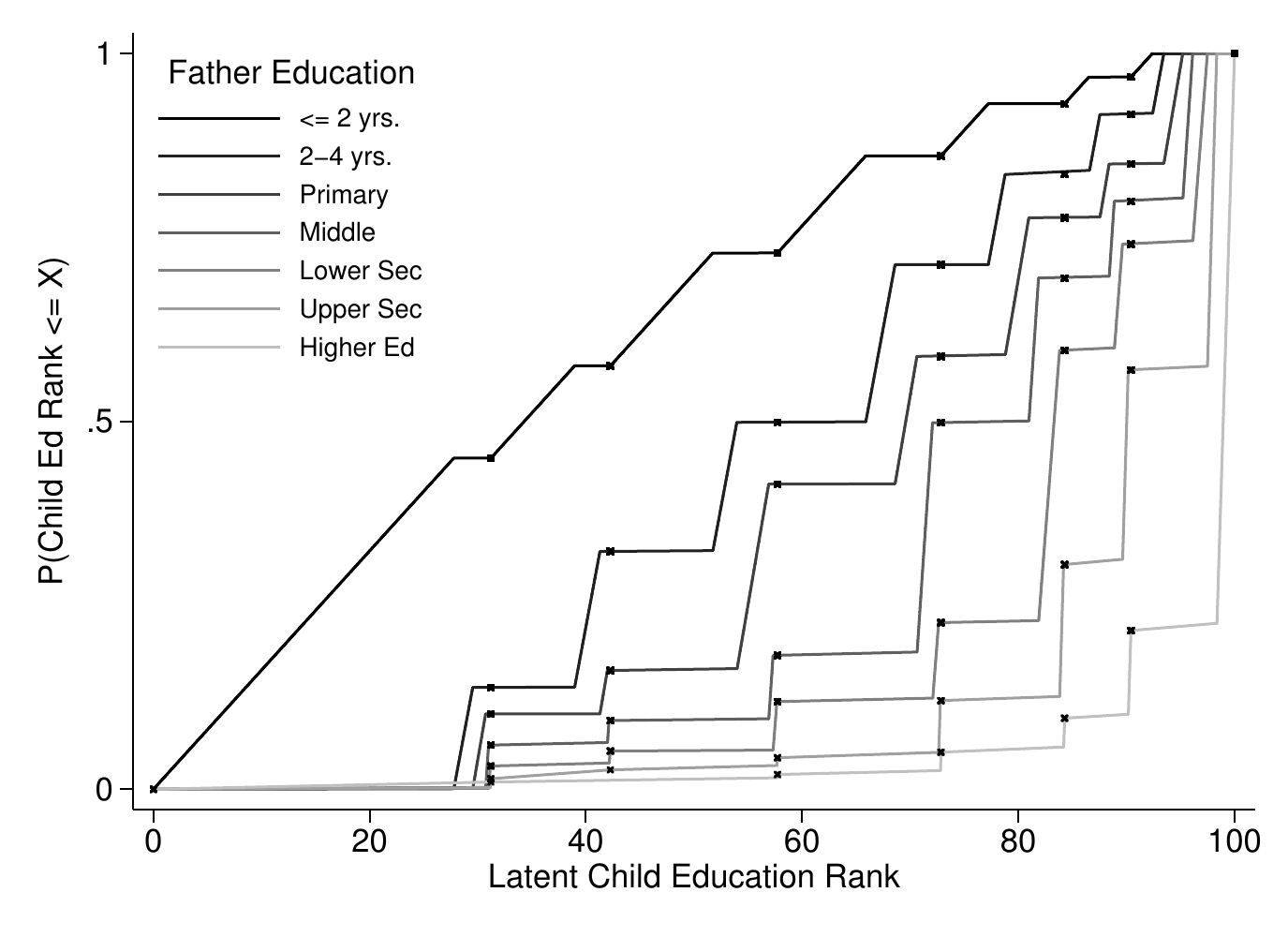} \\
      Panel B: Highest Feasible Mobility
      \\
      \includegraphics[scale=0.80]{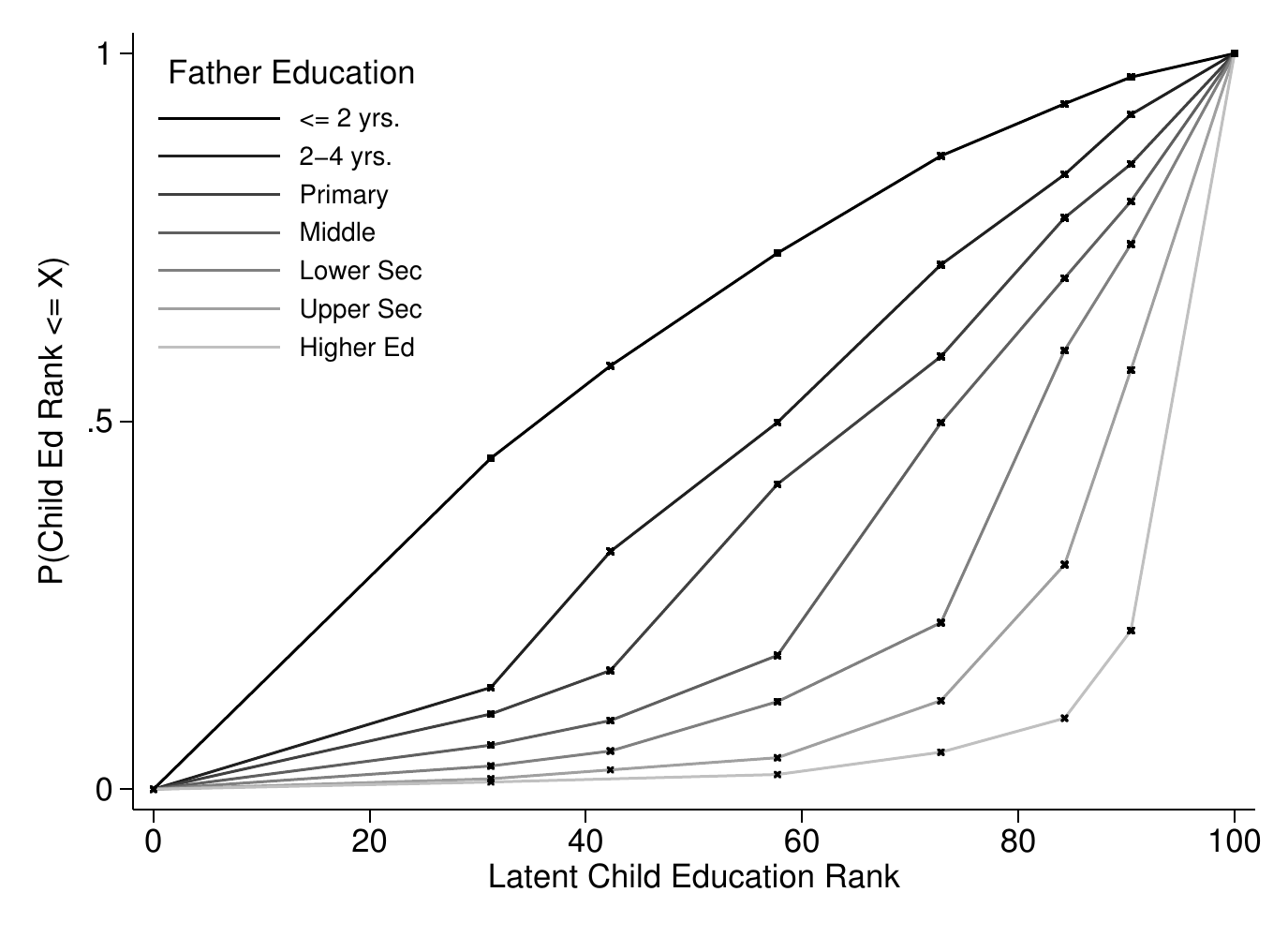} \\

      \hline
    \end{tabular}
  \end{center}

  { \footnotesize Figure~\ref{fig:son_solution} shows bounds on the
    CDF of child education rank, separately for each father
    education group. The lines index father types. Each point on a
    line shows the probability that a child of a given father type
    obtains an education rank less than or equal to the value on the X
    axis in the national education distribution. The large markers
    show the points observed in the data.}

\end{figure}

Because we have additional data on children, we can estimate the shape
of the child CDF within parent-child education bins using rank data
from other outcome variables that are not censored. Under the
assumption that latent education rank is correlated with other
measures of socioeconomic rank, this exercise sheds light on whether
Panel A or Panel B in Figure~\ref{fig:son_solution} better describes
the true latent distribution.

\begin{figure}[H]
  \caption{Son Outcome Rank CDF \cnewline
    by Father Education (1960-69 Birth
    Cohort) \cnewline {\small Joint Education/Wage Estimates}} 
  \label{fig:son_wage_cdf}
  \begin{center}
    \includegraphics[scale=0.80]{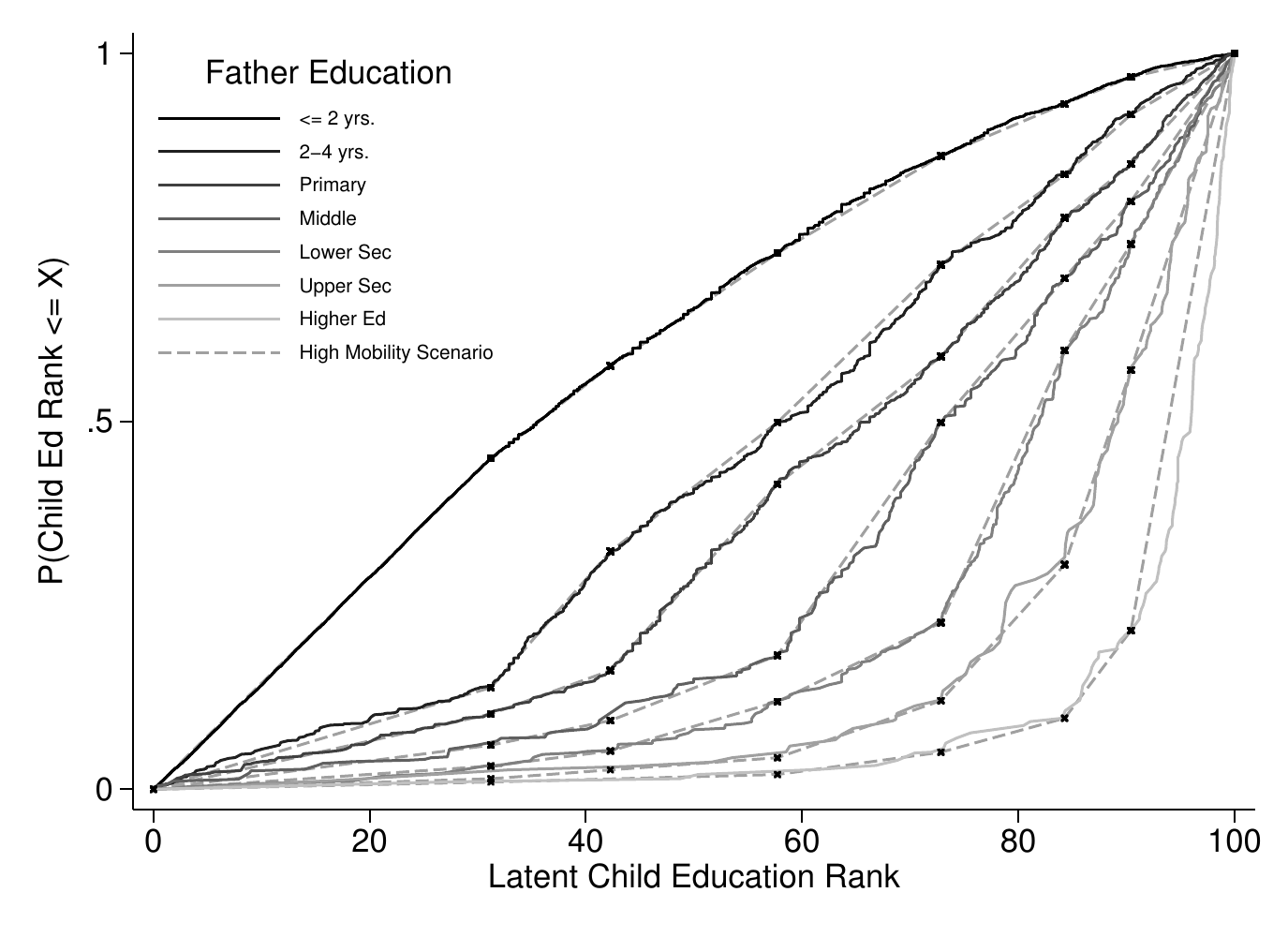}
  \end{center}
      {\small Figure~\ref{fig:son_wage_cdf} plots separate son rank
        CDFs separately for each father
        education group, for sons born in the 1960s in India. Sons
        are ranked first in terms of education, and then in terms of
        wages. Sons not reporting
        wages are dropped. For each father type, the graph shows a
        child's probability of attaining less than or equal to the
        rank given on the X axis.}
\end{figure}

Figure~\ref{fig:son_wage_cdf} shows the result of this exercise using wage data
from men in the 1960s birth cohort. To generate this figure, we
calculate children's ranks first according to education, and then
according to wage ranks within each education bin.\footnote{We limit
  the sample to the ~50\% of men who report wages. Results are similar
  if we use household income, which is available for all
  men. Household income has few missing observations, but in the many
  households where fathers are coresident with their sons, it is
  impossible to isolate the son's contribution to household income
  from the father's, which biases mobility estimates downward.} The
solid lines depict this uncensored rank distribution for each father
education; the dashed gray lines overlay the estimates from the high
mobility scenario in Panel B of Figure~\ref{fig:son_solution}.

If parent education strongly predicted child wages \textit{within}
each child education bin, we would see a graph like Panel A of
Figure~\ref{fig:son_solution}. The data clearly reject this
hypothesis. There is some additional curvature in the expected
direction in some bins, particularly among the small set of
college-educated children, but the distribution of child cumulative
distribution functions is strikingly close to the high mobility
scenario, where father education has little predictive power over
child outcomes after child education is taken into account. The last
row of Table~\ref{tab:double_censor} shows mobility estimates using
the within-bin parent-child distributions that are predicted by child
wages; the mobility estimates are nearly identical to the high
mobility scenario.  This result supports the assumption made in
Section~\ref{sec:mob} that latent child rank within a child rank bin
is uncorrelated with parent rank.

Note that there is no comparable exercise that we can conduct to
improve upon the situation when parent ranks are interval censored,
because we have no information on parents other than their education,
as is common in mobility studies. If we had additional information on
parents, we could conduct a similar exercise. The closest we can come
to this is by observing the parent-child rank distribution in
countries with more granular parent ranks, as we did in
Section~\ref{sec:method}. The results in that section suggest that
interval censoring of parent ranks does indeed mask important features
of the mobility distribution.

An additional factor that makes censoring in the child distribution a
smaller concern is the fact that children are more educated than
parents in every cohort, and thus the size of the lowest education bin
is smaller for children than for parents. This result is likely to be
true in many other countries where education is rising. Of course, in
other contexts, we may lack additional information about the
distribution of the $x$ and $y$ variable within bins, and researchers may
prefer to work with conservative bounds as described in~\ref{sec:best_worst}.

\newpage
\singlespacing

\begin{table}[H]
  \caption{Mobility Estimates under Double-Censored CEF }
  \label{tab:double_censor} 
  \begin{tabular}{lcc}
  \hline\hline
  &  Upward Interval          &  Rank-Rank \\
  & Mobility ($\mu_0^{50}$)  &  Gradient ($\beta$)        \\
  Low mobility scenario &   [32.33, 35.90]          &  [0.55, 0.80]  \\
  High mobility scenario &  [35.86, 38.80]      &  [0.45, 0.67]  \\
  Wage imputation scenario &  [35.79, 38.70]     &  [0.46, 0.67]  \\
  \hline
\end{tabular}

\end{table}  
\footnotesize{Table \ref{tab:double_censor} presents bounds on
  $\mu_0^{50}$ and the rank-rank gradient $\beta$ under
  three different sets of assumptions about child rank distribution
  within child rank bins. The low mobility scenario assumes children
  are ranked by parent education within child bins. The high mobility
  scenario assumes parent rank does not affect child rank after
  conditioning on child education bin. The wage imputation predicts the
  within-bin child rank distribution using child wage ranks and parent
  education.}

\newpage
\section{Appendix D: Data Sources} 
\label{sec:app_data}
\subsection{Data on Mortality in the United States}
\label{sec:app_mort_data}

For comparability, we follow the data
construction procedure used in \citeasnoun{Case2017}. We are
grateful that these authors shared software for data construction on their paper's
website to simplify this process.

Death records come from the CDC WONDER database. We have deaths
counts by race, gender, and education from 1992--2015, as well as
information on cause of death. To obtain mortality rates by year, we obtain
the number of people in each age-race-gender-education cell from the
Current Population Survey. 

The death records contain the universe of deaths in the U.S. The CPS
only interviews people who are not institutionalized --- e.g., not in
a prison or health institution. As a result, the denominator used by
\citeasnoun{Case2017} is slightly smaller than the true
denominator. To account for people who are institutionalized, we
obtain the number of institutionalized people missing from the CPS in
the U.S. Census for 1990 and 2000, and the American Communities Survey
for 2005--2015. For non-Census years prior to 2005, we linearly impute
the number of institutionalized people in each
age-race-gender-education cell; e.g., for 1995, we take the midpoint
of the observed number of institutionalized people in 1990 and
2000. For instance, among women ages 50--54 in 1992, just under 0.4\%
with a high school degree or less are institutionalized. Among that
group, mortality falls from 460.8 to 459.0 once we include
institutionalized people in the denominator.

The mortality records are characterized by some data with missing
education. We follow standard practice in assuming that the education
data are missing at random; we assign the missings the educations of
the observed educations in the age-race-gender cell whose deaths we
observe in that year. \citeasnoun{Case2017} drop several states that
inconsistently report education. After 2005, state identifiers are
available only in restricted access data, so we do not yet take this
step, and we apply the imputation procedure described above for all
people. We have applied for the restricted data from the National
Center for Health Statistics, and the revision of this paper will use
only the states with constant data. To predict whether these
exclusions are likely to affect our results, we calculated bounds on
mortality change from 1992-2004 for all states, and for the subset of
states with consistent reporting. Estimates of mortality change
differed by at most 0.2\%, suggesting that exclusion of these states
in all periods will also minimally affect results.

\subsection{Intergenerational Mobility: Matched Parent-Child Data from
  India}
\label{sec:app_mob_data}

To estimate intergenerational educational mobility in India, we draw
on two databases that report matched parent-child educational
attainment. The first is an administrative census dataset describing
the education level of all parents and their coresident
children. Because coresidence-based intergenerational mobility
estimates may be biased, we supplement this with 
a representative sample of non-coresident father-son pairs. We focus
on fathers and sons because we do not have data on non-coresident
mothers and/or daughters. This section describes the two datasets.

The Socioeconomic and Caste Census (SECC) was conducted in 2012, to
collect demographic and socioeconomic information determining
eligibility for various government programs.\footnote{It is often
  referred to as the 2011 SECC, as the initial plan was for the survey
  to be conducted between June and December 2011. However, various
  delays meant that the majority of surveying was conducted in
  2012. We therefore use 2012 as the relevant year for the SECC.} The
data was posted on the internet by the government, with each village
and urban neighborhood represented by hundreds of pages in PDF
format. Over a period of two years, we scraped over two million files,
parsed the embedded data into text, and translated the text from
twelve different Indian languages into English.\footnote{Additional
  details of the SECC and the scraping process are described in
  \citeasnoun{Asher2016}.} The individual-level data that we use
describe age, gender, and relationship with household head. Assets and
income are reported at the household rather than the individual level,
and thus cannot be used to estimate mobility. The SECC provides the
education level of every parent and child residing in the same
household. Sons who can be matched to fathers through coresidence
represent about 85\% of 20-year-olds and 7\% of
50-year-olds. Education is reported in seven categories.\footnote{The
  categories are (i) illiterate; (ii) literate without primary (iii)
  primary; (iv) middle; (v) secondary (vi) higher secondary; and (vii)
  post-secondary.} To ease the computational burden of the analysis,
we work with a 1\% sample of the SECC, stratified across India's 640
districts.

We supplement the SECC with data from the 2011-2012 round of the India
Human Development Survey (IHDS). The IHDS is a nationally
representative survey of 41,554 households in 1,503 villages and 971
urban neighborhoods across India. Crucially, the IHDS solicits
information on the education of fathers of household heads, even if
the fathers are not resident, allowing us to fill the gaps in the SECC
data. Since the SECC contains data on all coresident fathers and sons,
our main mobility estimates use the IHDS strictly for non-coresident
fathers and sons. IHDS contains household weights to make the data
nationally representative; we assign constant weights to SECC, given
our use of a 1\% sample. By appending the two datasets, we can obtain
an unbiased and nationally representative estimate of the joint
parent-child education distribution.\footnote{We verified that IHDS
  and SECC produce similar point estimates for the coresident
  father-son pairs that are observed in both datasets. Point estimates
  from the IHDS alone (including coresident and non-coresident pairs)
  match our point estimates, albeit with larger standard errors.} IHDS
reports neither the education of non-coresident mothers nor of women's
fathers, which is why our estimates are restricted to fathers and
sons.

IHDS records completed years of education. To make the two data
sources consistent, we recode the SECC into years of education, based
on prevailing schooling boundaries, and we downcode the IHDS so that
it reflects the highest level of schooling completed, \textit{i.e.},
if someone reports thirteen years of schooling in the IHDS, we recode
this as twelve years, which is the level of senior secondary
completion.\footnote{We code the SECC category ``literate without
  primary'' as two years of education, as this is the number of years
  that corresponds most closely to this category in the IHDS data,
  where we observe both literacy and years of education. Results are
  not substantively affected by this choice.} The loss in precision by
downcoding the IHDS is minimal, because most students exit school at
the end of a completed schooling level.

We estimate changes in mobility over time by examining the joint
distribution of fathers' and sons' educational attainment for sons in
different birth cohorts. All outcomes are measured in 2012, but
because education levels only rarely change in adulthood, these
measures capture educational investments made decades earlier. We use
decadal cohorts reflecting individuals' ages at the time of
surveying. To allay concerns that differential mortality across more
or less educated fathers and sons might bias our estimates, we
replicated our analysis on the \textit{same} birth cohorts using the
IHDS 2005. By estimating mobility on the same cohort at two separate
time periods, we identified a small survivorship bias for the 1950-59
birth cohort (reflecting attrition of high mobility dynasties), but
zero bias for the cohorts from the 1960s forward. The fact that
mobility in the 1950s is biased slightly downward only strengthens our
conclusions about zero mobility change (see
Figure~\ref{fig:mob_time_stats}).

\end{appendix}

\end{document}